\documentclass[aps,prb,reprint,amsmath,amssymb,longbibliography,floatfix]{revtex4-2}
\usepackage[T1]{fontenc}
\usepackage{bm}
\usepackage{graphicx}
\usepackage{xcolor}
\usepackage{tikz}
\usetikzlibrary{arrows.meta,positioning,calc,decorations.pathreplacing}
\usepackage[colorlinks=true,linkcolor=blue,citecolor=blue,urlcolor=blue]{hyperref}
\newcommand{\ii}{\mathrm{i}}
\newcommand{\R}{\bm R}
\newcommand{\rvec}{\bm r}
\newcommand{\avec}{\bm{\mathcal A}}
\newcommand{\Id}{\mathbb I}
\newcommand{\Xmat}{\mathsf X}
\newcommand{\Zmat}{\mathsf Z}
\newcommand{\Sframe}{\mathsf S}
\newcommand{\Gframe}{\mathsf G}
\newcommand{\Cframe}{\mathsf C}
\newcommand{\Acell}{\mathcal V}
\definecolor{learn}{HTML}{2a78d6}
\definecolor{fixed}{HTML}{52514e}
\definecolor{jas}{HTML}{eb6834}
\definecolor{feat}{HTML}{1baf7a}

\begin{document}
\title{Continuum neural wavefunctions for anyons on a torus: superfluidity in the semion gas}
\author{Ahmed Khalifa}
\affiliation{Department of Physics, Carnegie Mellon University, Pittsburgh, PA 15213, USA}
\author{Shubhayu Chatterjee}
\affiliation{Department of Physics, Carnegie Mellon University, Pittsburgh, PA 15213, USA}
\begin{abstract}
Mobile anyons in topologically ordered phases provide a setting to investigate how fractional exchange statistics shapes collective behavior at finite density. 
We introduce AnyFormer: a neural variational framework for continuum anyons on a torus that explicitly incorporates statistical interactions and global winding constraints. 
We construct a smooth orbital frame that enforces the matrix-valued torus boundary conditions, and supplement it with an analytic pair factor that supplies the regular collision behavior. 
The orbital geometry and many-body amplitudes are optimized in parallel by variational Monte Carlo. 
We apply AnyFormer to semions without an external field or interaction potential, isolating unique correlations generated solely by statistics. 
Our framework closely matches exact two-semion energies and yields lower variational energies than finite-cutoff exact diagonalization for $N = 4$ and $6$ semions.
Scaling up to $N = 16$ semions, well beyond the limits of continuum diagonalization, we find a largely uniform liquid with pronounced correlation holes and an approximately linear structure factor at the smallest accessible wavevectors.
The pair circulation reverses near the mean particle spacing, revealing chiral relative motion.
We then identify paired superfluid behavior through spatially persistent coherence of compact semion pairs, carried by a counter-rotating pair channel.
The ansatz is not biased toward pairing, so this result emerges from energy minimization alone.
AnyFormer thus provides a scalable variational neural-net framework for microscopic studies of correlations, chiral motion, and collective phases of bulk anyonic matter.
\end{abstract}
\maketitle

\section{Introduction}
\label{sec:introduction}

In two spatial dimensions, quantum mechanics allows for unconventional particles known as anyons, that exhibit exchange statistics beyond the familiar bosonic and fermionic cases~\cite{LeinaasMyrheim1977,Wilczek1982,Wu1984}.
In particular, a counterclockwise exchange of two Abelian anyons adiabatically multiplies their quantum wavefunction by $e^{i \pi \alpha}$, where $0 < \alpha < 1$.  
In condensed matter systems, such Abelian anyons naturally emerge as quasiparticle excitations of topologically ordered phases. 
Prominent examples include charge-neutral semions ($\alpha = 1/2$) in chiral spin liquid phases of frustrated magnets~\cite{KalmeyerLaughlin87,WWZ89,LaughlinZou90,bauer2014chiral} and fractionally charged $e/3$ Laughlin anyons ($\alpha = 1/3$) in fractional quantum Hall phases~\cite{Laughlin1983,Arovas1984,Halperin84,WenZee92}.  
To connect the exchange statistics of anyons to their collective behavior at finite density, it is crucial to develop a microscopic description of interacting, mobile anyons~\cite{Callin1993}.
A central question is whether statistics alone can turn such a gas into a superfluid, as predicted by early theories of anyon superconductivity~\cite{Laughlin88,FetterHannaLaughlin,ChenWilczekWittenHalperin1989}. 

Recent experimental developments in two-dimensional Van der Waals heterostructures and quantum simulators have accelerated the need to describe quantum states at finite anyon density.
The discovery of fractional Chern insulators in moir\'e transition metal dichalcogenides (TMDs)~\cite{cai2023signatures,park2023observation,zeng2023thermodynamic,XuPRX2023} and multilayer graphene~\cite{lu2024fractional,xu2025signatures}, where electrically charged anyons are not confined by a strong external magnetic field and can hence disperse~\cite{schleith2026anyon,yan2026anyon,iyer2026dispersion}, opens up the possibility to study mobile anyons in quantum materials. 
In parallel, the rapid development of quantum simulation platforms with carefully engineered Hamiltonians allow anyons to emerge as low-energy quasiparticle excitations in correlated bosonic systems~\cite{lu2009demonstrating,google2023non,iqbal2024non}. 
These discoveries have stimulated widespread theoretical efforts to characterize the many-body physics of anyons at finite density~\cite{divic2025anyon, shi2025doping,shi2026superconductivity,shi2025non,senthil2026fractionalized,han2026single,fan2026hidden,Taige2026,mehta2026coloring,cui2026anyonic,zhang2026color,li2026boundstatesanyonsgeometric,Nosov_2026,ShiPRX,WagnerNeupert2026}.

The canonical starting point to study many-body phases of interacting bosons or fermions is to formulate their continuum non-interacting ground states with periodic boundary conditions, topologically equivalent to a torus in two dimensions. 
However, for anyons, even this \textit{non-interacting} problem is complicated for two reasons. 
First, even in the absence of any interaction potential $V(r)$ between two anyons, their braiding statistics makes its way into the Hamiltonian via a statistical interaction mediated by Chern-Simons gauge fields~\cite{WenZee92,iengo1992anyon}.
This couples motion of anyons even at large separation, leading to a non-trivial spectral problem.
Second, the torus introduces an additional global constraint: transporting anyons around its two non-contractible loops do not commute.
This global transport algebra cannot be represented by an ordinary scalar-periodic wavefunction, so the wavefunction must carry additional structure that encode the braid algebra~\cite{Einarsson1990,HosotaniHo1992}.
The inherent complexity of the toroidal many-anyon Hilbert space thus restricts direct access to many-anyon states on the torus. 

Owing to its fundamental importance, several alternative approaches have been developed to theoretically tackle the problem of anyons at finite density.
Analytical field-theoretic descriptions often model anyons as charge-flux composites~\cite{Wilczek82}, and approximate the fluctuating statistical gauge field by an average effective flux generated by the other anyons~\cite{FetterHannaLaughlin,ChenWilczekWittenHalperin1989}.
Such approaches permit controlled perturbative limits near bosonic ($\alpha \to 0^+$) or fermionic ($\alpha \to 1^{-}$) statistics~\cite{correggi2019vortex,girardot2021semiclassical,LundholmRougerie2015,Lundholm2024}, self-consistent magnetic states such as nonlinear Landau levels~\cite{LundholmRougerie2015,AtaeiEtAl2025} or specific large-N limits~\cite{komargodski2026anyon}. 
Their applicability to generic statistics and parameter regimes, however, is limited.
Alternately, numerical exact diagonalization (ED) studies have focused on few-anyon spectral problems~\cite{Callin1993}, often confining the anyonic wave-functions using potential wells~\cite{Sporre1991,Sporre1992}.
Extending these calculations to delocalized anyonic states on a torus, and scalability with anyon number remain challenging. 
Similar scalability limitations arise when the anyons are allowed to emerge as excitations by solving for excited states of correlated bosonic or fermionic systems via ED or density matrix renormalization group (DMRG)~\cite{Taige2026,wang2026anyon,valentini2026single}. 
These challenges exemplify the importance of developing complementary techniques to study the fate of anyons at finite density on a torus. 

Motivated by the success of neural networks~\cite{CarleoTroyer2017} for continuum electronic systems~\cite{Pfau2020,vonglehn2023,geier2025,LuoDaiFu2024,abouelkomsan2025topologicalorderdeepstate,abouelkomsan2026firstprinciplesaifindscrystallization,hu2026hyperdet,zhang2026correlated,khalifa2026continuum}, here we introduce AnyFormer: an attention-based variational neural network ansatz for anyons on a torus. 
As a minimal setting beyond the reach of perturbation theory in the small deviation in statistics from bosons/fermions, we specialize to semions with $\alpha=1/2$ on a torus, without an external field or additional
interaction potential, isolating correlations generated by statistics.
Our AnyFormer ansatz combines a smooth two-component orbital frame that encodes the required torus transport structure with periodic neural-network coefficients that capture many-body correlations. 
The orbital geometry and neural coefficients are learned jointly in the training process, while an analytic pair factor enforces the appropriate short-distance behavior. 
We formulate a variational Monte Carlo (VMC) optimization protocol applicable across particle numbers, and validate it against an exact two-semion solution and against continuum ED for four and six semions.

We then apply AnyFormer to systems of up to $N = 16$ semions, well beyond the reach of ED benchmarks. 
The optimized states form a liquid with a largely uniform density, a pronounced correlation hole, and a static structure factor that is approximately linear at the smallest accessible wavevectors.
The liquid is chiral: its pair circulation reverses near the mean spacing, and at $N=12,16$ the circulation develops damped spatial oscillations.

This raises the central question of our work: is the semion gas a superfluid?
A pair of semions is a local bosonic composite, and its coherence probes the paired order proposed in gauge-theory descriptions of the anyon gas~\cite{FetterHannaLaughlin,ChenWilczekWittenHalperin1989}.
We therefore formulate a compact-pair correlator with the required torus statistical string and resolve its internal angular channels.
Through spatially persistent off-diagonal coherence of compact semion pairs, we identify paired superfluid behavior generated by statistics alone.
The coherence at large accessible separations is concentrated in the counter-rotating $\ell=-2$ channel, whereas the more populated $\ell=0$ contact channel decorrelates.
Crucially, this pairing is not built in.
Our ansatz fixes only the statistics, the torus boundary conditions, and the collision behavior; it is not biased toward pairing and learns its many-body correlations from energy minimization alone.
These results connect paired superfluid behavior to the microscopic chiral structure of the optimized semion gas and establish a route toward variational studies of bulk anyonic matter.

The rest of the paper is organized as follows.
In Sec.~\ref{sec:model}, we specify the continuum semion Hamiltonian and its torus boundary conditions, which inform our choice of variational wavefunctions in Sec.~\ref{sec:ansatz}. Sec.~\ref{sec:ansatz} constructs the learned orbital-frame ansatz, and Sec.~\ref{sec:vmc} describes its VMC optimization and evaluation. We validate the method against exact solutions and continuum ED, and compare training and variational energies, in Sec.~\ref{sec:validation}. We then develop the physical picture from liquid structure through chiral pair motion to paired superfluidity in Sec.~\ref{sec:beyond-ed}. Section~\ref{sec:conclusion} concludes with a summary of our key findings and a discussion of future directions.

\section{Continuum semion gas on a torus}
\label{sec:model}
\label{sec:hamiltonian}

We begin with an introduction to the many-anyon Hamiltonian and Hilbert space on a torus.
Specifically, we consider $N$ identical point semions of mass $m$ on a rectangular
torus with sides $L_x,L_y$ and area $\Acell=L_xL_y$. 
We study their kinetic energy without any
external field or pair interaction potential $V(r)$.
The statistical connection, global transport,
and collision domain specify the key facets of the model used to develop our variational neural state.

Let us denote the coordinate of the i$^{th}$ anyon by $\rvec_i=(x_i,y_i)$, $z_i=x_i+\ii y_i$, and define $\R=(\rvec_1,\ldots,\rvec_N)$.
In the anyon gauge, the wavefunction $\Psi_{\mathrm A}(\R)$ acquires a phase $e^{\ii\pi\alpha}$ under a counterclockwise exchange of any two anyons, with $\alpha=1/2$ for semions.
This leads to a multi-valued wavefunction for a many-semion state. 
The kinetic Hamiltonian acting on such multi-valued wavefunctions is $H_{\mathrm A}=-\kappa\sum_i\nabla_i^2$, where $\kappa=\hbar^2/(2m)$.

Statistical transmutation factors this phase out of the wavefunction:
\begin{equation}
 \Psi_{\mathrm A}(\R)=e^{\ii\chi(\R)}\Psi(\R),
 \qquad \avec_i=\nabla_i\chi.
 \label{eq:statistical-transmutation}
\end{equation}
Choosing $\chi$ to increase by $\pi/2$ under a counterclockwise exchange
leaves $\Psi$ permutation symmetric. Differentiation gives
$-\ii\nabla_i\Psi_{\mathrm A}=e^{\ii\chi}D_i\Psi$, so conjugating the
anyon-gauge operator, $H=e^{-\ii\chi}H_{\mathrm A}e^{\ii\chi}$, gives
the bosonic-gauge Hamiltonian
\begin{equation}
 H=\kappa\sum_{i=1}^N D_i\cdot D_i,
 \qquad D_i=-\ii\nabla_i+\avec_i.
 \label{eq:hamiltonian}
\end{equation}
This is the torus anyon Hamiltonian~\cite{HosotaniHo1992}.

On the plane, $\chi=\tfrac12\sum_{i<j}\arg(z_i-z_j)$. On the torus,
we replace the relative angle by the phase of the Jacobi theta function
\begin{equation}
 f(z)=\vartheta_1\!\left(\frac{\pi z}{L_x}\,\middle|\,
                   \tau\right),\qquad \tau=\ii L_y/L_x,
 \label{eq:theta}
\end{equation}
which vanishes linearly at $z=0$ and at all of its periodic images. (The
first argument of $\vartheta_1$ has periods $\pi$ and $\pi\tau$.) We take
$\chi=\tfrac12\sum_{i<j}\arg f(z_i-z_j)$ on a collision-free simply
connected region, choosing each argument continuously there.
Differentiating this phase, each pair $z=z_i-z_j$ contributes to particle
$i$ the connection
\begin{equation}
 \bm a(z)=\frac12
 \begin{pmatrix}
  \operatorname{Im}[f'(z)/f(z)]\\
  \operatorname{Re}[f'(z)/f(z)]
 \end{pmatrix},\qquad
 \avec_i=\sum_{j\ne i}\bm a(z_i-z_j).
 \label{eq:pair-connection}
\end{equation}
The prime denotes the derivative with respect to the dimensional
coordinate $z$. Near a coincidence, $\bm a\simeq(-y,x)/(2r^2)$, with counterclockwise circulation $\pi$: half a flux quantum and a phase $-1$ for a full pair loop. 
Since $\avec_i=\nabla_i\chi$, the connection
is locally pure gauge away from collisions, and the statistical flux is entirely
concentrated on the particles. 
Because $f$ is odd, $\bm a(-z)=-\bm a(z)$. 
The connection is dimensionally an inverse length, acts as the identity on the components introduced below, and carries no adjustable coupling. 
Supplemental Material, Sec.~\ref{app:consistency} verifies the sign of $D_i$ against the stated exchange phase.

\label{sec:boundary}

The noncontractible torus cycles require additional transport data.
Transport around the $y$ cycle gives a position-dependent Aharonov--Bohm phase. Translating another particle around the $x$ cycle changes this phase by $-1$, so the two cycle transports anticommute.
This algebra requires a multicomponent wavefunction~\cite{Einarsson1990}.
For semions its minimal representation has two components,
\begin{equation}
 \Psi(\R)=\begin{pmatrix}\Psi_1(\R)\\\Psi_2(\R)\end{pmatrix},
 \qquad \Psi(P\R)=\Psi(\R),
 \label{eq:spinor}
\end{equation}
where $P$ is any permutation of the particles. 
At a fixed configuration, $\Psi_1$ and $\Psi_2$ are amplitudes for two global winding states of the same $N$ particles. 
They retain information about transport around the torus handles that the instantaneous particle positions alone do not specify. 
In the basis chosen below, carrying any one particle around the $x$ cycle interchanges these amplitudes; the $y$ cycle gives them opposite signs, in addition to a common position-dependent phase.
Thus there is one two-component amplitude for the whole configuration, rather than an extra two-state label for each particle. 
The probability density for finding the semions in configuration $\R$ is
$\Psi^\dagger(\R)\Psi(\R)=|\Psi_1(\R)|^2+|\Psi_2(\R)|^2$.

\begin{figure}[!htbp]
\centering
\begin{tikzpicture}[>=Stealth]
 \draw[thick,fill=black!3] (0,0) rectangle (2.8,2.1);
 \node[below,font=\small] at (1.4,-0.02) {$L_x$};
 \node[left,font=\small] at (-0.02,1.05) {$L_y$};
 \foreach \p/\q in {0.55/1.5,1.45/0.55,2.2/1.45} {\fill[jas] (\p,\q) circle (2.2pt);}
 \fill[learn] (2.45,0.95) circle (2.4pt);
 \draw[->,thick,learn] (2.55,0.95) -- (3.25,0.95);
 \node[learn,right,align=left,font=\footnotesize] at (3.25,0.95)
   {$x$ edge: $\Psi\to e^{\ii\beta_x}\Xmat\Psi$\\ $(\Psi_1\leftrightarrow\Psi_2)$};
 \fill[feat] (1.0,1.8) circle (2.4pt);
 \draw[->,thick,feat] (1.0,1.9) -- (1.0,2.5);
 \node[feat,right,align=left,font=\footnotesize,anchor=south west] at (0.0,2.5)
   {$y$ edge: $\Psi\to e^{\ii\beta_y+\ii\pi s_i/L_x}\Zmat\Psi$
    \ $(\Psi_2\to-\Psi_2$, AB phase$)$};
 \fill[jas] (3.55,0.15) circle (2.2pt); \fill[jas] (3.95,0.15) circle (2.2pt);
 \draw[<->,thin] (3.55,-0.08) -- node[below,font=\scriptsize]{$r$} (3.95,-0.08);
 \node[right,font=\footnotesize] at (4.1,0.15) {close pair: $\Psi\propto r^{1/2}$};
\end{tikzpicture}
\caption{The three requirements on a semion wavefunction in the bosonic
gauge. Crossing an $x$ edge swaps the components. Crossing a $y$ edge flips
the sign of the second component and adds the Aharonov--Bohm (AB) phase of
the flux carried by the other particles. Exchanging two particles leaves
$\Psi$ unchanged, since the exchange phase sits in $\avec$. A close pair
vanishes as $r^{1/2}$ (Sec.~\ref{sec:jastrow}).}
\label{fig:rules}
\end{figure}
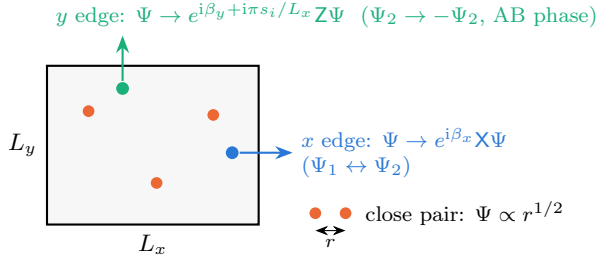

The sign from anti-commutation when transported around the two cycles of the torus is carried by the Pauli-like matrices
\begin{equation}
 \Xmat=\begin{pmatrix}0&1\\1&0\end{pmatrix},\qquad
 \Zmat=\begin{pmatrix}1&0\\0&-1\end{pmatrix},\qquad
 \Xmat\Zmat=-\Zmat\Xmat.
 \label{eq:transport-matrices}
\end{equation}
We work with unwrapped coordinates on the plane covering the torus. Moving
particle $i$ once around either cycle gives
\begin{subequations}\label{eq:boundary}
\begin{align}
 \Psi(\R+L_x\hat e_{ix})&=e^{\ii\beta_x}\Xmat\Psi(\R),\\
 \Psi(\R+L_y\hat e_{iy})&=
 e^{\ii\beta_y+\ii\pi s_i/L_x}\Zmat\Psi(\R),
\end{align}
\end{subequations}
with $s_i=\sum_{j\ne i}(x_i-x_j)=Nx_i-X$ and $X=\sum_jx_j$. Here
$\hat e_{i\mu}$ shifts one coordinate of one particle. The twists
$\beta_x,\beta_y$ specify additional Aharonov--Bohm phases around the
handles. A phase redefinition can move these twists from the boundary
rules into a constant vector potential, but cannot in general remove
them from both. Supplemental Material, Sec.~\ref{app:consistency}, gives
the explicit transformation.
Figure~\ref{fig:rules} summarizes these rules.

Crossing an $x$ edge exchanges the two
components. Crossing a $y$ edge flips the sign of the second component and
adds a position-dependent phase. As a function of $x_i$, that phase is
exactly the Landau-gauge boundary phase of a charged particle on a torus
pierced by $(N-1)/2$ flux quanta, the half quanta carried by the other $N-1$
particles. Its dependence on the other $x_j$ records where those fluxes sit.
The matrices also ensure consistency when a single particle crosses both
periods. Moving particle $i$ first by $L_x$ and then by $L_y$ must give
the same amplitude at the lifted configuration as performing these
coordinate shifts in the opposite order. Interchanging $\Xmat$ and
$\Zmat$ gives a minus sign, while evaluating the $y$-cycle phase after
the $x$ shift gives the factor $e^{\ii\pi(N-1)}$. These signs cancel
precisely for even $N$. The different-particle shifts obey the same
consistency requirement (Supplemental Material, Sec.~\ref{app:consistency}).

The singular statistical connection makes
the behavior at a coincidence essential to the kinetic energy.
We choose the regular, Friedrichs condition, with no contact parameter
or finite core radius.
Consider an isolated pair with relative separation
$\rvec_i-\rvec_j=r(\cos\varphi,\sin\varphi)$, while the other particles
remain separated. The connection approaches
$\bm a=\hat{\bm e}_\varphi/(2r)$, shifting the relative angular derivative
from $-\ii\partial_\varphi$ to $-\ii\partial_\varphi+1/2$.
With reduced mass $m/2$, the leading relative kinetic operator is
\begin{equation}
 H_{\rm rel}\simeq2\kappa\left[
 -\partial_r^2-\frac1r\partial_r
 +\frac{(-\ii\partial_\varphi+1/2)^2}{r^2}\right].
 \label{eq:pair-relative-operator}
\end{equation}
Exchanging the particles sends $\varphi\to\varphi+\pi$. Since the
bosonic-gauge amplitude is unchanged, an angular harmonic
$e^{\ii\ell\varphi}$ must satisfy $e^{\ii\ell\pi}=1$, so only even
integers $\ell$ occur. In each harmonic, the shifted angular term is the
centrifugal barrier $(\ell+1/2)^2/r^2$. Balancing this term against the
radial derivatives gives the regular behavior
\begin{equation}
 r^{\nu_\ell}e^{\ii\ell\varphi},\qquad
 \nu_\ell=|\ell+1/2|,\qquad \ell\in2\mathbb Z.
 \label{eq:pair-channels}
\end{equation}
The leading channel, $\ell=0$, vanishes as $r^{1/2}$; the next,
$\ell=-2$, starts as $r^{3/2}e^{-2\ii\varphi}$. These are the regular
solutions selected by the Friedrichs collision condition. A nonzero
contact limit would instead give a kinetic density proportional to
$1/r^2$, with a logarithmically divergent integral
$\int_0^\epsilon r\,dr/r^2$.

\section{The learned orbital-frame ansatz}
\label{sec:ansatz}

\subsection{Transport, smoothness, and completeness}
\label{sec:why-frame}

Our starting idea is to let known two-component functions carry the
boundary transport, while a neural network supplies ordinary periodic
scalar amplitudes. We denote the neural variational parameters by $\theta$.
If a given function $\bm s(\R)$ obeys
Eqs.~\eqref{eq:boundary}, then
\begin{equation}
 \bm f_\theta(\R)=\bm s(\R)c_\theta(\R)
 \label{eq:single-section}
\end{equation}
obeys the same rules for any periodic scalar $c_\theta$. Choosing both
factors to be permutation symmetric also preserves bosonic exchange
symmetry. The prescribed transport is then enforced analytically, and
the network learns the remaining configuration dependence.

Mathematically, $\bm s$ is a \emph{section} of the rank-two vector bundle
specified by the boundary rules~\cite{nakahara2018geometry}: it assigns a two-component vector to each
configuration, with vectors at identified boundaries related by the
prescribed transport matrices. A single product fixes
$f_{\theta,2}/f_{\theta,1}=s_2/s_1$ wherever that ratio is defined, and
inherits every zero of $\bm s$. To vary both components independently,
we use several known sections with independent scalar coefficients,
\begin{equation}
 \bm f_\theta(\R)=\sum_{k=1}^{K_N}\bm s_k(\R)c_{\theta,k}(\R)
                 =\Sframe_0(\R)\bm c_\theta(\R).
 \label{eq:frame-expansion}
\end{equation}
The sections are stored as the columns of the $2\times K_N$ matrix
$\Sframe_0$, and every coefficient is complex, periodic, and permutation
symmetric. A collection spanning the two-component space at every
configuration is called a \emph{frame}; redundant columns are allowed~\cite{nakahara2018geometry}.

Correct transport alone does not determine a useful choice of frame.
The kinetic energy differentiates the given functions as well as their
neural coefficients. Applied to Eq.~\eqref{eq:frame-expansion}, the
product rule gives $D_i\bm f_\theta=(D_i\Sframe_0)\bm c_\theta
-\ii\Sframe_0\nabla_i\bm c_\theta$.
A rapidly varying frame therefore asks the coefficients to compensate
for its variation when representing a low-energy state. We seek a frame
that is smooth across the torus boundaries and distributes the required
winding over the cell, with controlled spatial variation.

We also require uniform conditioning of its two physical components,
\begin{equation}
 \Sframe_0(\R)\Sframe_0(\R)^\dagger=\Id_2.
 \label{eq:tightness}
\end{equation}
This identity, called \emph{tightness}~\cite{CasazzaChen2017}, says that the sections resolve
both components at every configuration. Individual sections may
vanish, but the collection always spans both components. 
Any smooth, symmetric target $\bm f_*$ with the prescribed boundary transport can then be
reconstructed using a coefficient function
\begin{equation}
 \bm c_*(\R)=\Sframe_0(\R)^\dagger\bm f_*(\R).
 \label{eq:right-inverse}
\end{equation}
These coefficients are periodic because the transport of $\Sframe_0$
and $\bm f_*$ cancels, and they are permutation-symmetric because both factors are
symmetric. 
Since the coefficients are functions of the full configuration, a finite number of sections in a frame does not restrict the representable space of smooth targets. 
In practice, however, a finite neural network must approximate these coefficients, and the choice of frame can strongly affect the complexity of that approximation. 
We now construct a smooth, tight starting frame from single-particle orbitals and a center-of-mass frame.

\subsection{The analytic frame and pair factor}
\label{sec:winding}

We first construct an analytic frame $\Sframe_0(\R)$ that satisfies the torus boundary conditions of Eq.~\eqref{eq:boundary}.
To separate the single-particle and collective transport, we factor the boundary phase as
\begin{equation}
 e^{\ii\pi(Nx_i-X)/L_x}
       =e^{\ii\pi N x_i/L_x}\,e^{-\ii\pi X/L_x}.
 \label{eq:transport-factorization}
\end{equation}
Accordingly, we seek a factorized frame
\begin{equation}
 \Sframe_0(\R)=\Cframe(X,Y)\otimes\bm p(\R),
 \label{eq:factorized-frame}
\end{equation}
with $X=\sum_jx_j$ and $Y=\sum_jy_j$.
Here $\otimes$ denotes the Kronecker product with $\bm p$ treated as a row
vector: $(\Sframe_0)_{\alpha,(a,k)}=\Cframe_{\alpha a}p_k$, where
$\alpha=1,2$ labels the transport component, $a$ a column of $\Cframe$,
and $k$ a scalar function.
The matrix $\Cframe$ carries
the collective, two-component transport, while the scalar functions
$p_k$ carry the single-particle magnetic phases.

The first phase factor involves particle $i$ alone. It is the boundary phase of
one particle in a \emph{uniform} field of $n_\phi=N/2$ flux quanta. This is
the familiar average-field picture of anyons~\cite{ChenWilczekWittenHalperin1989},
in which each particle sees the half quanta of the others smeared over the
torus. The second factor depends only on the center of mass. It carries the
remaining $-1/2$ flux quantum, together with both components and the
anticommuting matrices. The center-of-mass matrix must therefore satisfy
\begin{subequations}\label{eq:com-transport}
\begin{align}
 \Cframe(X+L_x,Y)&=\Xmat\Cframe(X,Y),\\
 \Cframe(X,Y+L_y)&=e^{-\ii\pi X/L_x}\Zmat\Cframe(X,Y),
\end{align}
\end{subequations}
and the scalar functions must obey, for each particle,
\begin{subequations}\label{eq:scalar-transport}
\begin{align}
 p_k(\R+L_x\hat e_{ix})&=e^{\ii\beta_x}p_k(\R),\\
 p_k(\R+L_y\hat e_{iy})&=
   e^{\ii\beta_y+\ii\pi N x_i/L_x}p_k(\R).
\end{align}
\end{subequations}
The product then has exactly the transport~\eqref{eq:boundary}.

These are precisely the boundary conditions of a symmetric many-body
wavefunction of particles in the same uniform auxiliary field
$B_*=2\pi n_\phi/\Acell=\pi N/\Acell$, with twists $\beta_x,\beta_y$.
For one particle, the $y$-cycle phase is $e^{\ii\beta_y+\ii B_*L_yx}$;
in the many-body function, moving particle $i$ gives that same phase
with $x$ replaced by $x_i$. This holds for correlated states as well
as product states. Symmetric products of single-particle magnetic
orbitals therefore supply building blocks for the many-body scalars
$p_k$. Landau-level wavefunctions provide a natural starting point for
the orbital texture, while the
center-of-mass frame supplies the matrix-valued transport.

The following construction defines $\Sframe_0$ at the physical
configuration $\R$. We subsequently allow the network to adapt this
starting geometry in Sec.~\ref{sec:learned-frame}.

\label{sec:com-frame}

In this factorization the two transport components are carried by the
center-of-mass frame. This parallels
the fractional quantum Hall effect on a torus, where the topological
degeneracy is a property of the center-of-mass motion~\cite{Haldane1985}.
We therefore need a two-component function of $(X,Y)$ alone that obeys
Eqs.~\eqref{eq:com-transport} and never vanishes.

The magnetic phase in Eq.~\eqref{eq:com-transport} corresponds to a
uniform auxiliary field $B_C=-\pi/\Acell$, or $-1/2$ flux quantum through
the $L_x\times L_y$ cell. The two components accommodate this half flux:
the extra sign from the magnetic phase around the cell is canceled by
$\Xmat\Zmat=-\Zmat\Xmat$. Since an $X$ lap swaps the components and
$\Xmat^2=\Id_2$, two laps restore each component. We can therefore unfold
the two-component problem into a scalar one on the doubled cell
$2L_x\times L_y$, carrying one flux quantum in magnitude.

As a smooth starting choice, take a lowest-Landau-level orbital on this
doubled cell. Its magnetic length fixes the Gaussian width,
$w_C^2=1/|B_C|=\Acell/\pi$. In Landau gauge, each orbital mode is a plane
wave in $X$ times a Gaussian in $Y$. A single Gaussian is not compatible
with identifying $Y$ and $Y+L_y$. We add its translates by every $jL_y$,
with the corresponding magnetic phases, so crossing the boundary simply
relabels the terms in the sum. This gives
\begin{equation}
 F(X,Y)=\sum_{j\in\mathbb Z}
 e^{-(Y-jL_y)^2/(2w_C^2)}e^{-\ii\pi jX/L_x}.
 \label{eq:com-orbital}
\end{equation}
Equivalently, this is a Jacobi theta function~\cite{DLMFTheta} times a nonzero Gaussian:
\begin{equation}
 F(X,Y)=e^{-Y^2/(2w_C^2)}
 \vartheta_3\!\left(\frac{\pi(X+\ii Y)}{2L_x}\,\middle|\,\tau_C\right),
 \label{eq:com-theta}
\end{equation}
where $\tau_C=\ii L_y/(2L_x)$ and
$\vartheta_3(z|\tau)=\sum_{j\in\mathbb Z}e^{\ii\pi\tau j^2+2\ii jz}$.
Each term is $2L_x$ periodic in $X$. Reindexing the sum under
$Y\to Y+L_y$ gives $F(X,Y+L_y)=e^{-\ii\pi X/L_x}F(X,Y)$.
We fold this scalar orbital back into two components on the original cell,
\begin{equation}
 \bm\Phi(X,Y)=\frac1{\sqrt2}
 \begin{pmatrix}F(X,Y)\\F(X+L_x,Y)\end{pmatrix}.
 \label{eq:com-spinor}
\end{equation}
An $X$ lap exchanges the entries, giving $\Xmat$. A $Y$ lap multiplies
the first entry by $e^{-\ii\pi X/L_x}$ and the second by its negative,
giving $\Zmat$ and the common phase. Thus the folded orbital obeys
Eqs.~\eqref{eq:com-transport} exactly.

The zeros of $\vartheta_3$ occur at
$z=\pi(m+\tfrac12)+\pi\tau_C(n+\tfrac12)$ for integers $m,n$
\cite{DLMFTheta}. Thus $F$ has exactly one zero on the doubled cell,
at $(X,Y)=(L_x,L_y/2)$ modulo $(2L_x,L_y)$. The two
entries of $\bm\Phi$ sample points separated by $L_x$, so they cannot
vanish together. Its normalized version
$\bm\varphi=\bm\Phi/\|\bm\Phi\|$ is therefore smooth everywhere.

One section still locks the ratio of components. Two ``boosts'', which
multiply by a unit-modulus phase and a constant matrix, map sections to
sections:
\begin{equation}
 \begin{gathered}
 C_0=[\bm\varphi,\,\mathsf B_x\bm\varphi,\,\mathsf B_y\bm\varphi],\\
 \mathsf B_x=e^{\ii\pi X/L_x}\Zmat,\qquad
 \mathsf B_y=e^{\ii\pi Y/L_y}\Xmat.
 \end{gathered}
 \label{eq:com-three-sections}
\end{equation}
For any unit spinor, $C_0C_0^\dagger$ has eigenvalues exactly $1$ and $2$
(Supplemental Material, Sec.~\ref{app:com}). Pointwise normalization makes the frame tight,
$\Cframe=(C_0C_0^\dagger)^{-1/2}C_0$, with $\Cframe\Cframe^\dagger=\Id_2$ at
every point.

\label{sec:orbital-frame}

Having constructed the center-of-mass matrix $\Cframe$, we now move on to construct the scalar functions $p_k(\R)$.
Guided by the magnetic boundary conditions in Eq.~\eqref{eq:scalar-transport}, we use two lowest-Landau-level (LLL) functions for the scalar factor, $\phi_A$ and $\phi_B$, each
carrying the full $n_\phi$ flux quanta. We set $\beta_x=\beta_y=0$
for the calculations reported here, as specified in Sec.~\ref{sec:vmc}.
Each has $n_\phi$ vortices, points
where its amplitude vanishes and its phase winds. 
We arrange the vortices
of $\phi_A$ in a near-regular array and obtain $\phi_B$ by a translation
with a phase that preserves the LLL and its torus boundary rules. The two
vortex arrays interleave. Where one function vanishes the other remains
nonzero, so the pair has no common zero. Supplemental Material, Sec.~\ref{app:orbitals}
gives the construction. Two functions here do not mean two flux quanta:
as $N$ increases, each function carries more vortices.

\begin{figure}[t]
\centering
\includegraphics[width=0.812\columnwidth]{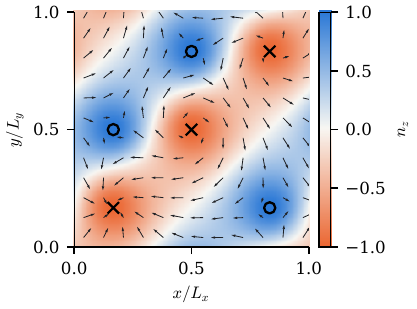}
\caption{Analytic starting orbital texture for $N=6$. Color shows the
orbital pseudospin component $n_z$, and arrows show $(n_x,n_y)$.
Crosses and circles mark the three vortices of each orbital function.
The disjoint zero sets give a smooth normalized orbital vector.}
\label{fig:texture}
\end{figure}

The locally normalized orbital vector is
\begin{equation}
 \bm v(\rvec)=\frac{(\phi_A(\rvec),\phi_B(\rvec))}
 {\sqrt{\rho(\rvec)}},\qquad
 \rho=|\phi_A|^2+|\phi_B|^2.
 \label{eq:local-normalization}
\end{equation}
It has two entries at every even $N\ge4$, and unit norm everywhere.
Normalization uses the combined density. The relative weights and phase define an \emph{orbital
pseudospin} direction
$\bm n=\bm v^\dagger\bm\sigma\bm v$, where $\bm\sigma$ denotes the three
Pauli matrices. This is an auxiliary way of describing the two orbital
amplitudes, rather than a physical particle spin. The direction lies on
a unit sphere: a zero of $\phi_A$ is its south pole and a zero of
$\phi_B$ its north pole. Between the vortices it changes smoothly.
The sphere is covered $n_\phi$ times as one traverses the torus; this
winding is a skyrmion texture. Figure~\ref{fig:texture} shows the
six-particle starting texture.

Three features motivate this compact choice. First, it carries exactly
the scalar transport of Eq.~\eqref{eq:scalar-transport}. Second, the
interleaved zeros make it smooth everywhere. Third, remaining within the
LLL gives the smallest possible average twisting for that flux.
The flux requires the orbital direction to wind a fixed number of times.
The first-order relation between the $x$ and $y$ derivatives of LLL
functions makes their gradient cost attain the lower bound set by this
winding. More precisely, the phase-independent derivative cost is
\begin{equation}
 g=\sum_{\mu=x,y}\left(\|\partial_\mu\bm v\|^2
          -|\bm v^\dagger\partial_\mu\bm v|^2\right)
   =\tfrac14|\nabla\bm n|^2.
 \label{eq:metric}
\end{equation}
This measures how quickly the orbital direction changes, after removing
changes of its common phase. Its torus average is
$B_*=2\pi n_\phi/\Acell$, the auxiliary field strength, and its integral
attains the flux lower bound of Belavin and Polyakov
(Supplemental Material, Sec.~\ref{app:orbitals})~\cite{BelavinPolyakov1975}. This property
does not depend on the number of orbital functions. The natural length
is the magnetic length $\ell_*=(\Acell/\pi N)^{1/2}$, proportional to the
mean particle spacing. Interleaving the vortices also aims to spread this
variation over the cell, rather than concentrating it into sharp regions.

The two orbital entries are distinct from the physical transport
components, which come from $\Cframe$. For $N=2$, the auxiliary flux is
a single quantum and the LLL is one dimensional: every LLL function is
proportional to the same orbital and vanishes at its zero. Two such
functions would therefore leave $\rho=0$ there and the normalized orbital
vector undefined. We instead use a first excited Landau-level partner
that is nonzero at the LLL zero, keeping $\rho>0$ (Supplemental Material,
Sec.~\ref{app:orbitals}). Everything else is unchanged, including the count of
$3(N+1)=9$ sections, so the same ansatz construction applies at every
even $N\ge2$.

\label{sec:symmetric-products}

Each particle samples a unit orbital vector $\bm v_i=\bm v(\rvec_i)$.
To form an exchange-symmetric product of these vectors, we can place $k=0,\ldots,N$
particles in the second orbital $v_1$ while keeping the rest of the particles in the first orbital $v_0$, leading to $N + 1$ occupation amplitudes. 
They can be obtained by multiplying a polynomial in a bookkeeping variable $t$,
\begin{equation}
 \prod_{i=1}^N[v_0(\rvec_i)+v_1(\rvec_i)t]
      =\sum_{k=0}^N a_k(\R)t^k,
 \label{eq:polynomial}
\end{equation}
and applying the usual occupation-state normalization,
\begin{equation}
 \widetilde p_k=\frac{a_k}{\sqrt{\binom Nk}},\qquad
 \mathcal N^2=\sum_{k=0}^N|\widetilde p_k|^2,\qquad
 p_k=\frac{\widetilde p_k}{\mathcal N}.
 \label{eq:symmetrized-orbitals}
\end{equation}
Every term carries the same
boundary phase, so each $p_k$ is symmetric and obeys the scalar transport.
The count is $M_N=N+1$.
In pseudospin language, symmetrizing $N$ spin-$\tfrac12$ orbital vectors
gives a collective spin $N/2$ with these $N+1$ components. The individual
orbital directions are its Majorana stars~\cite{Majorana1932}.

The normalization $\mathcal N^2$ is the weight of the product in the
symmetric subspace. It equals one for parallel orbital directions and
falls as they spread over the sphere. It never vanishes: each factor in
Eq.~\eqref{eq:polynomial} is a nonzero polynomial, so their product is
nonzero. A quantitative bound is $(N+1)e^{-N}\le\mathcal N^2\le1$
(Supplemental Material, Sec.~\ref{app:products}). This proves regularity
at each fixed $N$ and gives an explicit size-dependent lower bound.

Combining the scalar amplitudes with the three COM sections gives
\begin{equation}
 (\Sframe_0)_{ak}(\R)=\bm C_a(X,Y)p_k(\R).
 \label{eq:full-frame}
\end{equation}
Here $a=1,2,3$ and $k=0,\ldots,N$, so $K_N=3(N+1)$ physical sections
suffice for the starting frame.
Because $\sum_k|p_k|^2=1$ and $\Cframe\Cframe^\dagger=\Id_2$,
$\Sframe_0\Sframe_0^\dagger=\Id_2$. The frame is smooth, tight, exchange
symmetric, and has exact transport. The reconstruction in
Eq.~\eqref{eq:right-inverse} therefore applies with this compact frame:
configuration-dependent coefficients provide the remaining many-body
freedom.

\label{sec:jastrow}
\label{sec:collisions}

The analytic frame and smooth coefficient functions need not vanish
when particles meet. We enforce the leading semion cusp with the pair factor
\begin{equation}
 J_\eta(\R)=\prod_{i<j}
       \left(\frac{d_{ij}^2}{d_{ij}^2+\eta^2}\right)^{1/4},
 \qquad \eta>0.
 \label{eq:jastrow}
\end{equation}
Here $d_{ij}$ is the chord distance
\begin{equation}
 d_{ij}^2=\sum_{\mu=x,y}\left[
 \frac{L_\mu}{\pi}\sin\!\left(\frac{\pi(r_{i\mu}-r_{j\mu})}{L_\mu}\right)
                         \right]^2,
\end{equation}
whose square root approaches the separation $r_{ij}$ at short distance but, unlike the
minimum-image distance, is smooth and periodic. Each pair factor behaves as
$(r_{ij}/\eta)^{1/2}$ when the pair is close and tends to a constant far
apart. The exponent is fixed by the statistics. The range $\eta$ only sets
how fast the factor recovers. It is not an excluded radius and does not
change $H$, and it may be fixed or optimized. Multiplying a smooth bounded
frame expansion by this factor gives a state in the Friedrichs form domain.
Supplemental Material, Sec.~\ref{app:collisions}, gives the domain argument
and discusses subleading pair channels and their representation by
additional analytic factors.

\subsection{Learned geometry and neural amplitudes}
\label{sec:learned-frame}

The analytic frame and pair factor supply the required transport and
collision behavior. We now let the network adapt the orbital geometry
and learn the coefficients that combine the frame sections. Replacing
$\Sframe_0$ by a learned frame $\Gframe_\theta$ preserves the same
transport and tightness. Adapting the geometry gives a finite network
another way to express correlations, alongside learning the coefficients.

The network assigns a real two-dimensional auxiliary displacement
$\bm u_{\theta,i}(\R)$ to each particle and forms
\begin{equation}
 \bm q_i=\rvec_i+\bm u_{\theta,i}(\R),\qquad
 \bm Q_\theta=(\bm q_1,\ldots,\bm q_N).
 \label{eq:internal-coordinates}
\end{equation}
These auxiliary displacements are smooth and periodic in every physical coordinate.
They are also \emph{permutation equivariant}: exchanging two particle
labels exchanges their displacements. 
Thus particles in different local
environments can sample different parts of the orbital texture without introducing distinguishable particle labels. 
The auxiliary coordinates play a role similar to backflow coordinates~\cite{FeynmanCohen1956}: a particle's orbital value responds to its neighbors. 
All sections use the same displacement map.

Simply using $\Sframe_0(\bm Q_\theta)$ would give the wrong $y$-cycle phase, because that phase depends on the $x$ coordinates. The required correction
has an explicit form. With $\bar y=N^{-1}\sum_i y_i$, define
\begin{align}
 \varphi_\theta(\R)&=-\frac{\pi N}{\Acell}
             \sum_i(y_i-\bar y)u_{\theta,i,x}(\R),
 \label{eq:endpoint-phase}\\
 \Gframe_\theta(\R)&=e^{\ii\varphi_\theta(\R)}
                         \Sframe_0(\bm Q_\theta).
 \label{eq:learned-frame}
\end{align}
This phase realigns the orbital winding with the physical boundary rule.
When particle $i$ crosses the $y$ boundary, the internally evaluated frame
gains an extra phase
$\pi[Nu_{\theta,i,x}-\sum_j u_{\theta,j,x}]/L_x$.
The change in $\varphi_\theta$ is exactly its negative. The $x$ boundary
rule needs no correction. Exchange symmetry is preserved because the sum
in Eq.~\eqref{eq:endpoint-phase} is unchanged by relabeling particles.
Supplemental Material, Sec.~\ref{app:learned-frame} gives the short check.

The distinction between physical and
auxiliary positions is essential.
The pair factor $J_\eta(\R)$ remains evaluated at the physical positions,
as does the statistical connection $\avec_i(\R)$.
Sampling and expectation-value integrals use the physical coordinates
$\R$ with measure $d\R=\prod_i d^2r_i$. Only the
smooth frame is evaluated at $\bm Q_\theta$. The internal map need not be
invertible, and two particles may have the same auxiliary position. The analytic frame is smooth even there. Such an auxiliary
coincidence neither imposes a physical collision nor forces a node.
There is no change of integration coordinates or volume factor.

The learned frame remains tight since
\begin{equation}
 \Gframe_\theta\Gframe_\theta^\dagger
 =\Sframe_0(\bm Q_\theta)\Sframe_0(\bm Q_\theta)^\dagger=\Id_2.
 \label{eq:learned-tightness}
\end{equation}
The learned frame evaluates the same $N+1$ scalar amplitudes at
$\bm Q_\theta$, so its section count stays $3(N+1)$ throughout training.
Zero displacement recovers the starting frame exactly. Learning therefore
begins with the analytic orbital construction and can adapt it continuously
while preserving transport and the ability to resolve both components.
The frame derivative in the product rule now acts on $\Gframe_\theta$.
It includes both the change of the internal coordinates
and the explicit phase correction, with the physical connection
$\avec_i(\R)$. All are differentiated along with the coefficients.
The starting orbitals have minimal twisting, while a learned deformation
can increase or decrease their derivative cost. The variational energy,
which includes every product-rule term, determines which deformations
are useful. Supplemental Material, Sec.~\ref{app:learned-frame} gives
the explicit chain rule.

\begin{figure*}[t]
\centering
\resizebox{\textwidth}{!}{%
\begin{tikzpicture}[>=Stealth, x=1cm,y=1cm,
  every node/.style={font=\small,execute at begin node={\hyphenpenalty=10000\relax}},
  block/.style={draw=fixed!55,rounded corners=2pt,fill=white,
    align=center,inner sep=5pt,text width=24mm,minimum height=12mm},
  train/.style={block,draw=learn,fill=learn!7},
  analytic/.style={block,draw=feat!80!black,fill=feat!6},
  flow/.style={->,line width=.65pt,draw=fixed},
  data/.style={->,line width=.65pt,draw=feat!80!black},
  note/.style={font=\scriptsize,align=center,text=fixed}]
 \node[anchor=west,font=\footnotesize\bfseries] at (-.15,7.65)
   {(a) Shared particle encoder};
 \node[block,text width=18mm] (R) at (1,6.5)
   {Positions\\$\R=\{\rvec_i\}$};
 \node[analytic] (fourier) at (3.8,6.5)
   {Periodic features\\$\gamma(\rvec_i)$\\
    $\langle\gamma(\rvec_i-\rvec_j)\rangle_{j\ne i}$};
 \node[train,text width=31mm,minimum height=23mm] (enc) at (7.5,6.5)
   {Shared embedding\\$\downarrow$\\
    Self-attention + residual\\$\tanh$\quad($L$ layers)};
 \node[block,text width=21mm,minimum height=23mm] (h) at (11,6.5)
   {Particle\\features\\[3pt]
    $\begin{bmatrix}\text{---}\;\bm h_1\;\text{---}\\
       \vdots\\\text{---}\;\bm h_N\;\text{---}\end{bmatrix}$};
 \draw[flow] (R) -- (fourier);
 \draw[flow] (fourier) -- (enc);
 \draw[flow] (enc) -- (h);
 \node[note,anchor=west] at (12.55,6.5)
   {One hidden vector\\per particle;\\shared parameters};

 \node[anchor=west,font=\footnotesize\bfseries] at (-.15,4.8)
   {(b) Learned frame and coefficients};
 \node[analytic,text width=25mm] (pool) at (7.5,4.25)
   {Mean pooling\\$\overline{\bm h}=N^{-1}\sum_i\bm h_i$};
 \node[train,text width=24mm] (move) at (11,4.25)
   {Displacement readout\\$u_{\theta,i,\mu}=sL_\mu\tanh(\cdots)$};
 \node[analytic,text width=25mm] (Q) at (14.9,4.25)
   {Internal coordinates\\$\bm Q_\theta=\R+\bm u_\theta$};
 \draw[flow] (h.south) -- (move.north);
 \draw[flow] (h.west) -- ++(-.35,0) |- (pool.north);
 \draw[flow] (move) -- (Q);

 \node[train,text width=29mm,minimum height=18mm] (c) at (7.5,2.15)
   {Concatenate + dense $\tanh$\\
    Real and imaginary readouts\\$\bm c_\theta\in\mathbb C^{K}$};
 \node[analytic,text width=31mm,minimum height=18mm] (G) at (14.9,2.15)
   {Analytic frame + phase\\
    $\Sframe_0(\bm Q_\theta)=\Cframe\otimes\bm p$\\
    $\Gframe_\theta=e^{\ii\varphi_\theta}\Sframe_0(\bm Q_\theta)$};
 \draw[flow] (pool) -- (c);
 \draw[data] (Q) -- (G);
 \draw[data] (G.west) -- node[above,note]
   {Orbital weights $\{|p_k(\bm Q_\theta)|^2\}$} (c.east);
 \node[note,anchor=west,text width=42mm] at (.1,2.65)
   {Blue: trainable maps\\Green: prescribed operations\\
    $K=3(N+1)$ generating sections};

 \node[block,draw=jas,fill=jas!7,text width=29mm] (J) at (3.8,-.05)
   {Physical pair factor\\$J_\eta(\R)\sim r^{1/2}$};
 \node[block,text width=31mm,minimum height=15mm] (psi) at (11,-.05)
   {Contract and multiply\\$\Psi_\theta=J_\eta\Gframe_\theta\bm c_\theta$\\
    two global components};
 \node[analytic,text width=25mm,minimum height=15mm] (proj) at (14.9,-.05)
   {Coherent shift average\\$\Psi_\theta^{(M)}=P_M\Psi_\theta$};
 \draw[flow,draw=jas] (R.south) -- (1,5.35) -- (-.5,5.35)
   -- (-.5,-.05) -- (J.west);
 \draw[flow,draw=jas] (J.east) -- (psi.west);
 \draw[flow] (c.south) -- ++(0,-.1) -| (psi.north);
 \draw[data] (G.south) -- ++(0,-.1) -| ([xshift=9mm]psi.north);
 \draw[flow] (psi) -- (proj);
 \node[note,anchor=north,text width=47mm] at (14.9,-1.04)
   {Reevaluate the complete graph at each\\collectively shifted configuration};
\end{tikzpicture}
}
\caption{Computational graph of the learned orbital-frame ansatz.
(a) Periodic position and pair features pass through one shared attention
encoder to give a hidden vector for each particle.
(b) A particlewise readout produces displacements; their internal coordinates
sample the analytic frame, and a common phase restores physical transport.
The same hidden vectors are mean-pooled and augmented by the orbital weights
$|p_k(\bm Q_\theta)|^2$ before the complex coefficient readout.
The pair factor uses the physical positions. Contracting the frame with its
coefficients and multiplying by $J_\eta$ gives the two-component amplitude.
The final momentum projection coherently averages evaluations of this entire
graph at collectively shifted positions, with shared parameters
[Eq.~\eqref{eq:projected-ansatz}]. Blue blocks contain trainable maps;
green blocks are prescribed operations.}
\label{fig:ansatz}
\end{figure*}
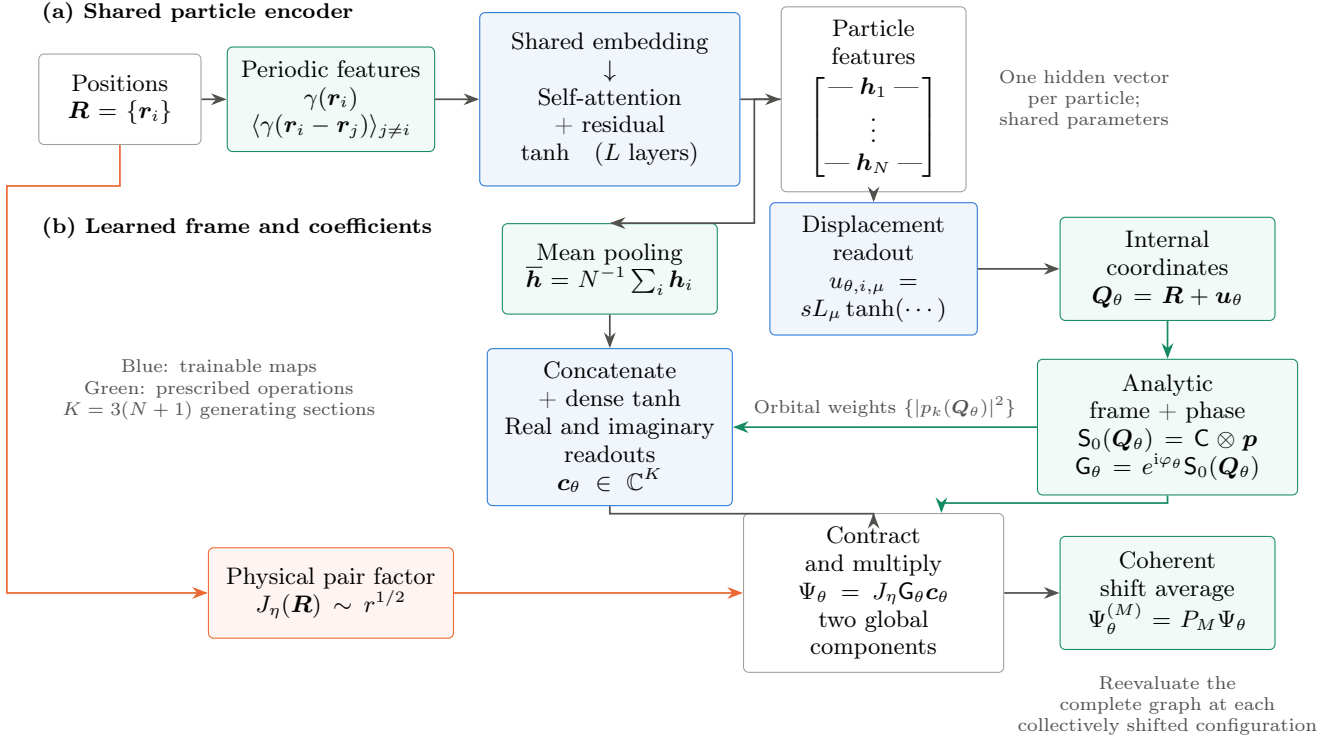

Combining the pair factor with the learned frame and neural coefficients
completes the trial wavefunction,
\begin{equation}
 \Psi_\theta(\R)=J_\eta(\R)\,
                  \Gframe_\theta(\R)\,\bm c_\theta(\R).
 \label{eq:ansatz}
\end{equation}
The frame supplies smooth two-component transport, the coefficients
learn amplitudes, phases, and nodes, and the pair factor supplies the
leading collision behavior. Including this last factor in the product
rule gives
\begin{align}
 D_i\Psi_\theta={}&-\ii(\nabla_iJ_\eta)\Gframe_\theta\bm c_\theta
 +J_\eta(D_i\Gframe_\theta)\bm c_\theta\nonumber\\
 &-\ii J_\eta\Gframe_\theta\nabla_i\bm c_\theta.
 \label{eq:ansatz-derivative}
\end{align}
These choices enforce statistics, transport,
and finite collision energy for every parameter value. The frame and
coefficients share one neural encoder, as described below.
Figure~\ref{fig:ansatz} shows the complete structure.
Equation~\eqref{eq:ansatz} is the building block for the momentum-projected
state in Sec.~\ref{sec:projection}.

For a target $\Psi_*$ whose remainder $\bm f_*=\Psi_*/J_\eta$ is smooth,
the reconstruction in Eq.~\eqref{eq:right-inverse} applies equally to the
learned frame, with $\bm c_*=\Gframe_\theta^\dagger\bm f_*$. Its smoothness
at contact depends on the target's collision channels (Supplemental
Material, Sec.~\ref{app:collisions}).

\label{sec:network}

The displacement and coefficients come from one permutation-equivariant
attention network~\cite{PhysRevB.110.035108}. For each direction $\mu=x,y$ and harmonic
$h=1,\ldots,H_f$, collect $\sin(2\pi h r_\mu/L_\mu)$ and
$\cos(2\pi h r_\mu/L_\mu)$ into $\gamma(\rvec)$. Particle $i$ receives
features of its own position and its separations from the others:
\begin{equation}
 \begin{aligned}
 \bm f_i^{\mathrm{in}}&=\left[\gamma(\rvec_i),\,
       \frac1{N-1}\sum_{j\ne i}\gamma(\rvec_i-\rvec_j)\right],\\
 \bm h_i^{(0)}&=\tanh\!\left(W_{\mathrm{in}}\bm f_i^{\mathrm{in}}
                                      +\bm b_{\mathrm{in}}\right).
 \end{aligned}
 \label{eq:main-network-input}
\end{equation}
Square brackets denote concatenation, and $\tanh$ acts entrywise.
All matrices $W$ and biases $\bm b$ are learned. The Fourier inputs
make every output periodic; the hidden vectors $\bm h_i^{(0)}$ have
width $w$ and describe each particle's environment.

Each attention layer $\ell=1,\ldots,L$ refines these descriptions.
Head $m=1,\ldots,H_a$ forms query, key, and value vectors
$\bm q_i^{(\ell,m)}$, $\bm k_i^{(\ell,m)}$, and
$\bm v_i^{(\ell,m)}$ by affine maps of $\bm h_i^{(\ell-1)}$.
These are auxiliary features, distinct from the internal particle
positions $\bm q_i$. With head width $d_h=w/H_a$, the update is
\begin{equation}
 \begin{aligned}
 A_{ij}^{(\ell,m)}&=
 \frac{\exp[\bm q_i^{(\ell,m)}\!\cdot\bm k_j^{(\ell,m)}/\sqrt{d_h}]}
 {\sum_{j'=1}^N\exp[\bm q_i^{(\ell,m)}\!\cdot\bm k_{j'}^{(\ell,m)}/\sqrt{d_h}]},\\
 \bm o_i^{(\ell,m)}&=\sum_{j=1}^N A_{ij}^{(\ell,m)}\bm v_j^{(\ell,m)},\\
 \bm h_i^{(\ell)}&=\tanh\!\left[\bm h_i^{(\ell-1)}
             +W_O^{(\ell)}\bm o_i^{(\ell)}+\bm b_O^{(\ell)}\right].
 \end{aligned}
 \label{eq:main-attention-update}
\end{equation}
The weights sum to one over $j$, so each head returns a learned weighted
average over particles. Here $\bm o_i^{(\ell)}$ concatenates all head
outputs. The last line adds their contribution to the previous hidden
vector. Parameters are shared across particles, with no particle-index
encoding; relabeling particles therefore only relabels the hidden vectors.

A shared readout converts each final hidden vector into a displacement,
\begin{equation}
 u_{\theta,i,\mu}=sL_\mu\tanh\!\left[
                  (W_u\bm h_i^{(L)}+\bm b_u)_\mu\right].
 \label{eq:main-movement-head}
\end{equation}
The dimensionless scale $s$ bounds each component by $sL_\mu$.
We then evaluate the analytic frame at $\bm Q_\theta$ and apply the phase
in Eq.~\eqref{eq:endpoint-phase}. Initializing $W_u$ and $\bm b_u$ to
zero recovers the analytic starting frame exactly.

The coefficient readout uses the whole configuration. Mean pooling gives
a permutation-invariant hidden vector, augmented by the deformed frame's
orbital weights:
\begin{equation}
 \begin{aligned}
 \bar{\bm h}&=\frac1N\sum_i\bm h_i^{(L)},\\
 \bm g_\theta&=\tanh\!\left\{
 W_g\left[\bar{\bm h},\{|p_k(\bm Q_\theta)|^2\}_{k=0}^N\right]
                                      +\bm b_g\right\},\\
 \bm c_\theta&=(W_R\bm g_\theta+\bm b_R)
                 +\ii(W_I\bm g_\theta+\bm b_I).
 \end{aligned}
 \label{eq:main-coefficient-readout}
\end{equation}
Here brackets denote vector concatenation. The pooled hidden layer has
learned real weights $W_g\in\mathbb R^{w\times(w+N+1)}$ and bias
$\bm b_g\in\mathbb R^w$. The learned real matrices
$W_R,W_I\in\mathbb R^{K_N\times w}$ and biases
$\bm b_R,\bm b_I\in\mathbb R^{K_N}$ produce the real and imaginary
parts, respectively, of the $K_N=3(N+1)$ complex coefficients.
The orbital weights are symmetric and periodic because their common
boundary phase cancels. They tell the network which orbital occupations
dominate at the internal configuration. The complex coefficients allow
interference and movable nodes. Absolute positions are retained, so
translation invariance is not imposed; exact transport comes from the
learned frame for every parameter value. The calculation is sequential:
encode the physical positions, move the frame, then mix its sections.
Supplemental Material, Sec.~\ref{app:network}, gives the full layer
equations, dimensions, and a pedagogical account of the shared encoder
and its two readouts.

\subsection{Momentum selection and computational cost}
\label{sec:projection}

\label{sec:com}

Each pair connection is odd, so $\sum_i\avec_i=0$. Consequently, the total
momentum $\bm P=-\ii\hbar\sum_i\nabla_i$ is conserved. The internal
coordinates adapt the orbital geometry, but do not by themselves give a
definite total momentum. A learned state can still depend on where the
whole configuration sits on the torus. Such dependence mixes momentum
sectors, called \emph{momentum leakage} here. Its energy contains a
nonnegative center-of-mass part,
\begin{equation}
 E_{\mathrm{COM}}[\Psi]=\frac{\kappa}{N}
 \frac{\int\|\sum_iD_i\Psi\|^2\,d\R}{\int\|\Psi\|^2\,d\R},
 \label{eq:com-energy}
\end{equation}
which equals $\kappa|\bm K|^2/N$ in a state of momentum $\hbar\bm K$
(Supplemental Material, Sec.~\ref{app:com}). The remainder is the relative
energy. Reporting both separates improved correlations from residual
collective motion. We suppress unwanted momentum components by averaging
translated copies of the complete learned wavefunction with the phases
appropriate to the target momentum.

Set $\beta_x=\beta_y=0$, as in our calculations. Write
$\R+\bm a=(\rvec_1+\bm a,\ldots,\rvec_N+\bm a)$ for a collective
translation. A momentum eigenstate transforms by
$\Psi_{\bm K}(\R+\bm a)=e^{\ii\bm K\cdot\bm a}\Psi_{\bm K}(\R)$.
The continuous projection onto $\bm K=0$ is thus
\begin{equation}
 (\mathcal P_0\Psi_\theta)(\R)=\frac1{\Acell}
       \int_{T^2}\Psi_\theta(\R+\bm a)\,d^2a.
 \label{eq:continuous-projection}
\end{equation}
All copies add coherently in the zero-momentum sector; the oscillating
phases of other sectors average to zero. A nonzero target $\bm K_*$ is
selected by inserting $e^{-\ii\bm K_*\cdot\bm a}$ in the average.
The zero-momentum choice removes the center-of-mass energy in
Eq.~\eqref{eq:com-energy}. The relative-motion energies can also differ
between sectors, so the lowest sector is identified by comparing their
variational energies, rather than from the center-of-mass term alone.

In practice we use an $M\times M$ grid of collective shifts,
\begin{equation}
 \bm a_{ab}=\left(\frac{aL_x}{M},\frac{bL_y}{M}\right),
 \qquad a,b=0,\ldots,M-1,
\end{equation}
and train the projected ansatz
\begin{equation}
 \Psi_\theta^{(M)}(\R)=\frac{J_\eta(\R)}{M^2}
 \sum_{a,b=0}^{M-1}
 \Gframe_\theta(\R+\bm a_{ab})\,
 \bm c_\theta(\R+\bm a_{ab}).
 \label{eq:projected-ansatz}
\end{equation}
The pair factor comes outside because it depends only on separations.
Every shifted copy reevaluates the encoder, displacement, orbital frame,
phase correction, and coefficients. We average amplitudes, so interference
removes unwanted momenta before computing the density or energy. Sampling
and differentiation use this projected state throughout training.

The finite grid removes every sector except those with both integer
momentum labels divisible by $M$, where $K_\mu L_\mu=2\pi n_\mu$.
It therefore suppresses leakage while retaining possible higher-momentum
copies. To suppress the dominant residual momentum sectors, we choose
the smallest $M\ge2$ that shares no factor with $N/2$:
$M=2,3,2,3,2,5,3$ for $N=2,4,6,8,10,12,16$.
The residual center-of-mass energy, and
refinement of the shift grid, check the sector selection. All collective
translates have the same particle boundary rules: separations and
$Nx_i-X$ are unchanged by a collective shift. Their sum preserves the
exact transport, exchange symmetry, and physical collision behavior.
The tightness identity applies to each learned frame inside the sum;
projection is performed on the whole state, so it needs no separately
normalized projected frame.
Projected-state derivatives are obtained by differentiating the coherent
shift sum itself, including the physical pair factor.

The center-of-mass energy provides a direct bound on residual momentum
weight. If the evaluation grid is $Q\times Q$, every surviving nonzero
momentum on the square has magnitude at least $2\pi Q/L$.
For normalized weights $w_{\bm n}$ in the orthogonal momentum sectors,
$E_{\rm COM}=(\kappa/N)\sum_{\bm n}w_{\bm n}|\bm K_{\bm n}|^2$.
Each nonzero sector therefore contributes at least
$\kappa(2\pi Q/L)^2/N$ per unit weight, giving the bound on their total
norm fraction,
\begin{equation}
 w_{\rm leak}\le\frac{N E_{\rm COM}}{\kappa(2\pi Q/L)^2}.
 \label{eq:momentum-leakage-bound}
\end{equation}
We use $Q=2M$ for the reported energies and observables. This diagnostic
complements the change in energy when the evaluation grid is refined.

\label{sec:scaling}

The adopted pseudospin-texture frame has $M_N=N+1$ scalar amplitudes
and $K_N=3(N+1)$ physical sections at all even $N\ge4$.
The two-orbital symmetric products are built by polynomial multiplication
with a number of operations proportional to $N^2$.
At fixed series accuracy and aspect ratio, the orbital construction also
has quadratic forward cost. At fixed neural width and depth, attention
has quadratic particle-number dependence because it couples every pair
of particles. The movement head adds two real outputs per particle and
$2(w+1)$ parameters at hidden width $w$; it requires no extra encoder.
For example, the $N=6$ two-orbital model has twenty-one complex coefficient
outputs and $41\,836$ real parameters at width $64$, including the $130$
movement parameters.

An $M\times M$ projection evaluates $M^2$ shifted copies with shared
parameters. This increases evaluation work, not the parameter count.

\section{Variational Monte Carlo}
\label{sec:vmc}
\label{sec:energy}

For the numerical results we use the unit square, $L_x=L_y=1$,
$\beta_x=\beta_y=0$, and $\kappa=1$. Energies are quoted in units of
$\kappa/(L_xL_y)$. The Hamiltonian has no intrinsic length scale, so
at fixed $N$ and aspect ratio its energies scale as $\kappa/(L_xL_y)$;
the intensive length scale is set by the density $n=N/\Acell$.

We minimize the normalized kinetic-energy expectation of the full
complex two-component wavefunction,
\begin{equation}
 E[\Psi]=\frac{\langle\Psi|H|\Psi\rangle}{\langle\Psi|\Psi\rangle},
 \qquad E_0=\inf_{\Psi\ne0}E[\Psi],
 \label{eq:variational}
\end{equation}
where the infimum runs over states satisfying the specified transport,
exchange symmetry, and collision condition.
Metropolis sampling~\cite{Metropolis1953} uses the Born density
$\|\Psi^{(M)}_\theta\|^2$, with all learned factors reevaluated inside
the translation sum. Writing $\Psi=J_\eta\bm f$ and $u=\ln J_\eta$,
integration by parts gives the first-derivative estimator
\begin{gather}
 E[\Psi]=\big\langle e_{\rm W}\big\rangle_{\|\Psi\|^2},\nonumber\\
 e_{\rm W}=\kappa\sum_i\left[
 \frac{\|D_i\bm f\|^2}{\|\bm f\|^2}
 -\nabla_i^2u-|\nabla_i u|^2\right].
 \label{eq:weak-energy}
\end{gather}
For a close pair at separation $r$, the cusp $J_\eta\propto r^{1/2}$
gives $u\simeq\tfrac12\ln r$. The pair contributions to
$|\avec_i|^2$ and $|\nabla_i u|^2$ are then both $1/(4r^2)$.
They enter Eq.~\eqref{eq:weak-energy} with opposite signs, so the leading
$1/r^2$ terms cancel. We combine these terms analytically before numerical
evaluation to avoid subtracting large, nearly equal quantities.
The energy gradient differentiates both numerator and normalization,
including the displacement and complex coefficient parameters. Updates
use Kronecker-factored approximate curvature
(K-FAC)~\cite{MartensGrosse2015,Pfau2020}.

The training protocol shares the attention architecture and optimizer,
with lengths set by the mean particle spacing and the projection grid
chosen as in Sec.~\ref{sec:projection}. Energies and observables are
measured in separate VMC runs with frozen parameters and independent
random streams. We estimate statistical errors by
blocking~\cite{FlyvbjergPetersen1989} and examine the energy tails.
Supplemental Material, Sec.~\ref{app:vmc}, gives the sampling moves,
gradient formulation, hyperparameters, and error analysis. All reported production states are trained through $\mathcal P_M$.
Energy and observable evaluations use a finer filter on the same frozen checkpoint:
$\mathcal P_{2M}$ through $N=12$ and $\mathcal P_{4M}=\mathcal P_{12}$
at $N=16$. The width and walker count of each state are specified
in Sec.~\ref{sec:state-selection}.

The optimization histories and frozen variational energies are compared
in Fig.~\ref{fig:training}, alongside the validation and scaling results
in Sec.~\ref{sec:validation}.

\section{Validation and variational energies}
\label{sec:validation}

\subsection{Exact and continuum-diagonalization benchmarks}

We validate the representation at $N=2,4,6$ before studying the larger
systems. For two semions, center-of-mass and relative motion separate, and
an exact solution in a convergent Bessel series gives the ground energy
(Supplemental Material, Sec.~\ref{app:n2-exact}).
For $N=4,6$ we use continuum exact diagonalization (ED) in the zero-momentum
sector. Symmetric Fourier configurations with total squared integer
wavenumber at most $\Lambda$ multiply fixed Gaussian-periodized
two-component sections and a regular collision factor. A collective
translation average selects total momentum zero. The expansion coefficients
follow from a generalized eigenvalue problem for the continuum kinetic and
overlap matrices. Supplemental
Material, Sec.~\ref{app:ed}, gives the basis, the additional pair channel
used at $N=4$, and the independent integration checks.

With exact matrix integrals, finite-basis ED gives variational upper
bounds. Numerical integration introduces a separate error, assessed by
evaluating frozen eigenvectors on independent quadratures. We therefore
compare the neural energies with the exact two-particle result and the
best available finite-cutoff estimates at $N=4,6$, retaining their different
levels of numerical precision.

\begin{table}[b]
\begin{ruledtabular}
\begin{tabular}{rclll}
$N$ & $M$ & Reference & Neural & $E_{\rm COM}$\\
\hline
2 & 2 & $13.673\,332\,280$ (exact) & $13.6850(22)$ & $0.0005$\\
4 & 3 & $72.43$ (ED, $\Lambda=6$, $+T$) & $71.826(19)$ & $0.019$\\
6 & 2 & $\simeq176.3$ (ED, $\Lambda=9$) & $170.036(46)$ & $0.189$\\
\end{tabular}
\end{ruledtabular}
\caption{Validation on the unit square at $\beta_x=\beta_y=0$, in units of
$\kappa/(L_xL_y)$. The neural states are trained through $\mathcal P_M$ for
$2\times10^4$ updates and evaluated through $\mathcal P_{2M}$, using width
64 and 2048 walkers. Parentheses give one blocked standard error from 4000
frozen-parameter measurements. The $N=4$ ED entry is a fitted finite-basis
Ritz value; the $N=6$ entry is the mean of two independent positive-form
evaluations of a frozen ED state. ED convergence and integration checks
are given in Supplemental Material, Sec.~\ref{app:ed}.}
\label{tab:validation}
\end{table}

Table~\ref{tab:validation} shows close agreement with the exact two-semion
energy and neural energies below the available finite-cutoff ED estimates
at $N=4,6$. The exact solution also tests the pair-correlation and current
estimators (Supplemental Material, Sec.~\ref{app:n2-exact}).

Doubling the momentum-filter grid leaves the $N=4$ energy unchanged within
statistical errors. At $N=6$, it lowers both the total and center-of-mass
energies. The total reduction is larger, showing that the removed momentum
sectors also had higher relative-motion energy.

\subsection{Scaling and variational energies}
\label{sec:state-selection}

All states use the learned two-orbital
frame, a random start, and $2\times10^4$ updates through $\mathcal P_M$
($4\times10^4$ at $N=10,12$). At $N=16$, training continues to
$1.2\times10^5$ updates, followed by $2\times10^4$ further updates
with a reduced K-FAC norm constraint. All observables use
$\mathcal P_{2M}$ for $N\le12$ and $\mathcal P_{12}$ at $N=16$
on the same selected checkpoint. We use width 64 at $N=8,10,12$ and width 256 at $N=16$.
Table~\ref{tab:beyond-ed} specifies the larger states used for the
physical observables in Sec.~\ref{sec:beyond-ed}. Additional states test the dependence of pair coherence on
training, width, and momentum filtering in Supplemental Material,
Sec.~\ref{app:pair-coherence}. Smaller systems enter the comparisons of size dependence.

\begin{table}[b]
\begin{ruledtabular}
\begin{tabular}{rrrlll}
$N$ & $w$ & $g$ & $E$ & $E_{\rm COM}$ & $E/(N\pi n\kappa)$\\
\hline
8 & 64 & 6 & $301.584(68)$ & $0.372$ & $1.5000$\\
10 & 64 & 4 & $491.44(10)$ & $1.074$ & $1.5643$\\
12 & 64 & 10 & $704.32(22)$ & $0.998$ & $1.5569$\\
16 & 256 & 12 & $1266.91(109)$ & $1.339$ & $1.5753$\\
\end{tabular}
\end{ruledtabular}
\caption{States beyond ED used for energy and observable results.
Training uses $\mathcal P_M$, with $M=3,2,5,3$ for
$N=8,10,12,16$. The table gives the frozen evaluation grid $g$
of $\mathcal P_g$ for all observables; $w$ is the hidden width.
The tabulated evaluations use
2048 walkers at $N=8,10$, 1024 at $N=12$, and 512 at $N=16$. Energies are in units of
$\kappa/(L_xL_y)$; parentheses are one blocked standard error. The last column
uses the density $n=N/(L_xL_y)$.}
\label{tab:beyond-ed}
\end{table}

\begin{figure}[t]
\centering
\includegraphics[width=\columnwidth]{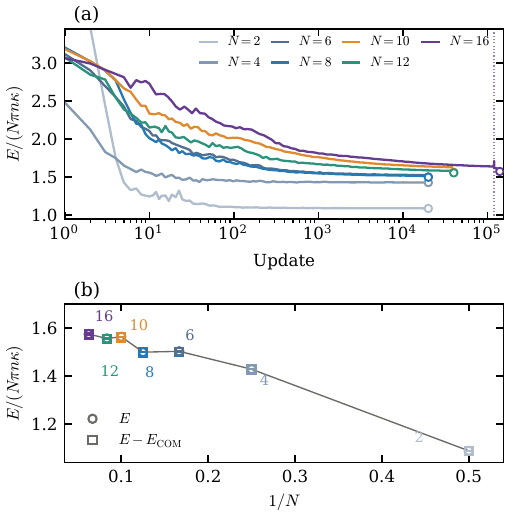}
\caption{Optimization and variational energies in units of $N\pi n\kappa$.
(a) Training energies averaged over logarithmically spaced update intervals.
The $N=10,12$ states run to $4\times10^4$ updates; $N=16$ runs to
$1.2\times10^5$, followed by $2\times10^4$ with the K-FAC norm constraint
reduced from $10^{-3}$ to $10^{-4}$ (dotted vertical line).
Open circles mark the frozen evaluations in Tables~\ref{tab:validation}
and \ref{tab:beyond-ed}: $\mathcal P_{2M}$ through $N=12$ and
$\mathcal P_{12}$ at $N=16$.
(b) The same frozen energies against $1/N$, with and without the
center-of-mass part. Point labels give $N$; error bars show two statistical
standard errors. Colors identify the same particle numbers in both panels.}
\label{fig:training}
\end{figure}

Average-field estimates provide an energy
scale~\cite{ChenWilczekWittenHalperin1989}. Each semion carries half a
flux quantum of statistical flux. Smearing this flux over the cell gives
a uniform field of flux density $\pi n$, with two particles per flux
quantum and cyclotron energy $\hbar\omega_c=2\pi n\kappa$, where
$\kappa=\hbar^2/(2m)$. Starting from bosons, all particles can occupy
the lowest Landau level of this field, which gives
$E/N=\hbar\omega_c/2=\pi n\kappa$. Starting from fermions, two
particles per flux quantum must fill the two lowest Landau levels, which
gives $E/N=\hbar\omega_c=2\pi n\kappa$.
The dimensionless ratio $E/(N\pi n\kappa)$ lies between these values at
every size [Fig.~\ref{fig:training}(b)]. For $N=10$--16 it is
$1.56$--$1.58$, with $1.575$ at $N=16$ (Table~\ref{tab:beyond-ed}).

Finer projection lowers the energies at the larger sizes. Increasing the
network width and extending training further improve the $N=16$ state.
The residual center-of-mass energies in Table~\ref{tab:beyond-ed} bound
the remaining momentum weight through Eq.~\eqref{eq:momentum-leakage-bound}.

Figure~\ref{fig:training} shows the optimization histories and
energies of these selected states. Most of the reduction occurs within
the first $10^3$ updates, followed by a slower decline at the larger
sizes. The open endpoint markers show
the further reduction from the finer frozen-evaluation filter. The
particle-number comparison places the same energies on the density scale
used to compare their physical correlations.

\section{Structure and paired superfluidity of the semion gas}
\label{sec:beyond-ed}

Having established the energies of the variational states, we turn to
their physical structure. We first characterize their density
correlations, then examine their chiral pair motion, and finally
identify paired superfluid behavior through channel-resolved
pair coherence. The selected states and evaluation grids are specified
in Tables~\ref{tab:validation} and \ref{tab:beyond-ed}.

\subsection{Liquid structure and density correlations}
\label{sec:structure}

The one-particle density is largely uniform at all four beyond-ED sizes.
We characterize the spatial organization around this homogeneous density
through the pair correlation and static structure factor, using the mean
spacing $a=(\Acell/N)^{1/2}$ to compare different particle numbers.

\begin{figure}[t]
\centering
\includegraphics[width=\columnwidth]{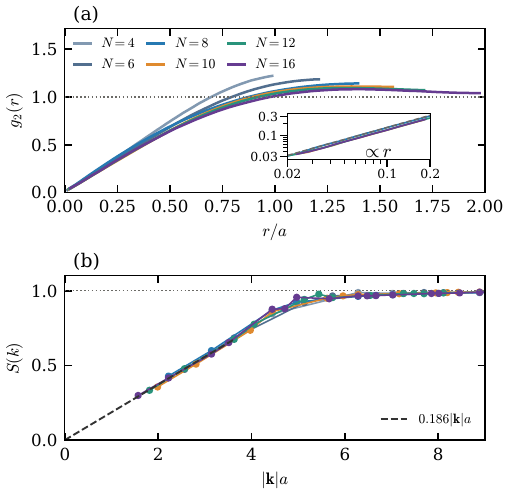}
\caption{Density correlations of the semion liquid, with mean spacing
$a=(\Acell/N)^{1/2}$. (a) Radial pair correlation at $N=4$--16;
the logarithmic inset shows the contact law at $N=8,10,12,16$
over $0.02\le r/a\le0.2$.
(b) Shell-averaged static structure factor for the same particle numbers.
The dashed line fits the $N\ge8$ points with $|\bm k|a<3.6$.
All data use the selected states in Tables~\ref{tab:validation} and
\ref{tab:beyond-ed}; colors are common to both panels.}
\label{fig:liquid-structure}
\end{figure}

We normalize the pair correlation to one for independent uniform positions,
\begin{equation}
 g_2(\bm d)=\frac{\Acell}{N(N-1)}
 \left\langle\sum_{i\ne j}
 \delta_{T^2}(\bm d-\rvec_i+\rvec_j)\right\rangle.
 \label{eq:pair-correlation}
\end{equation}
Here $\delta_{T^2}$ is the periodic delta function and the expectation
sums both physical components. Its integral over the separation cell is
$\Acell$. Each beyond-ED state has a pronounced correlation hole despite
the absence of a repulsive potential [Fig.~\ref{fig:liquid-structure}(a)].
At contact, the $\ell=0$ semion channel gives $g_2\propto r^{2\alpha}=r$,
as imposed by the cusp factor [Fig.~\ref{fig:liquid-structure}(a), inset].
The optimized profiles determine the crossover out of
the contact regime and the extent of the correlation hole.

In spacing units the short-distance profiles are similar.
At intermediate distances the size dependence is clearer
[Fig.~\ref{fig:liquid-structure}(a)]. The first maximum is resolved inside
$r<L/2$ from $N=10$ onward, near $1.3$--$1.4a$, and becomes weaker
as particle number increases.
The modest maxima and suppressed contact probability describe weak
spatial order.

For $\rho_{\bm k}=\sum_i e^{\ii\bm k\cdot\rvec_i}$, with
$k_\mu=2\pi n_\mu/L_\mu$ and integer $n_\mu$, define
\begin{equation}
 S(\bm k)=\frac{\langle|\rho_{\bm k}|^2\rangle}{N},\qquad
 S_{\rm c}(\bm k)=S(\bm k)-\frac{|\langle\rho_{\bm k}\rangle|^2}{N}.
 \label{eq:structure-factor}
\end{equation}
For reference, $N$ particles placed independently and uniformly have
$\langle\rho_{\bm k}\rangle=0$ but $\langle|\rho_{\bm k}|^2\rangle=N$
at nonzero $\bm k$, so $S=1$. At $\bm k=0$, $\rho_{\bm 0}=N$ exactly,
so $S(0)=N$ and $S_{\rm c}(0)=0$. We average $S(\bm k)$ over shells of
wavevectors with equal $|\bm k|$. The smallest nonzero shell,
$|\bm k|=2\pi/L$, shows increasingly suppressed density
fluctuations as longer wavelengths become accessible in spacing units.
The shell averages approximately collapse against $|\bm k|a$
[Fig.~\ref{fig:liquid-structure}(b)]. A fit through the origin to the $N\ge8$
points with $|\bm k|a<3.6$ gives $S\simeq0.19|\bm k|a$.
The smallest available $|\bm k|a$ is $\pi/2$ at $N=16$;
larger systems would test the continuation of this trend toward zero
wavevector.

Together, these observables describe a liquid with a short-distance
correlation hole and suppressed collective density fluctuations. We next
resolve the chiral relative motion accompanying this spatial structure.

\subsection{Chiral pair motion}
\label{sec:pairs-current}

\begin{figure*}[t]
\centering
\includegraphics[width=\textwidth]{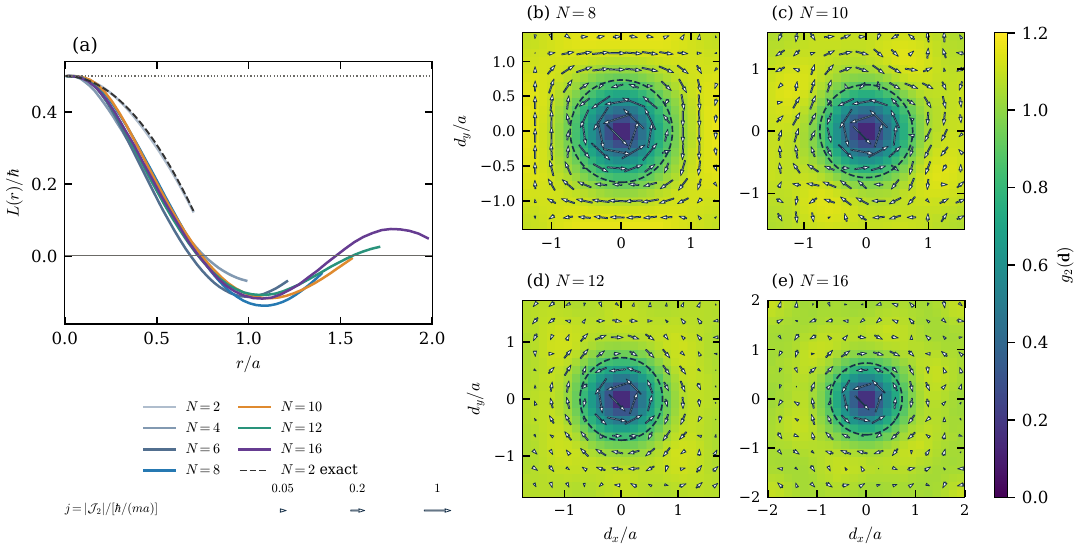}
\caption{Chiral pair motion. (a) Pair relative angular momentum with
two-standard-error bands, including the exact two-semion result.
(b--e) Pair-correlation maps for the selected $N=8,10,12,16$ states on
their full minimum-image cells in spacing units. Arrows show the joint
pair current, with every second map bin displayed. Arrow lengths are
proportional to $\ln(1+j/0.05)$, where
$j=|\bm{\mathcal J}_2|/[\hbar/(ma)]$, on a common scale; reference arrows
show representative magnitudes. Directions and measured maps are
unchanged by this nonlinear length scale. Dashed circles mark the first
zero of the radial $L(r)$ profile. No rotation or reflection averaging
is applied. Colors in (a) match the particle-number colors in
Fig.~\ref{fig:liquid-structure}.}
\label{fig:pair-motion}
\end{figure*}

The covariant current resolves the relative motion accompanying the
density correlations. Figure~\ref{fig:pair-motion} combines its radial
angular-momentum profile with the spatial current maps.
At a configuration $\R$, the velocity of particle $i$ is
\begin{equation}
 \bm v_i(\R)=\frac{\hbar}{m}
 \frac{\operatorname{Re}[\Psi^\dagger D_i\Psi]}{\|\Psi\|^2}.
 \label{eq:local-current-velocity}
\end{equation}
The derivative uses the physical positions and the statistical connection,
and the numerator sums both components. Averaging $\bm v_i-\bm v_j$ at a
fixed pair separation removes common drift. Its azimuthal part defines
the pair relative angular momentum,
\begin{equation}
 L(r)=\frac m2\,r\,
 \big\langle(\bm v_i-\bm v_j)\cdot\hat{\bm e}_\varphi\big\rangle_r,
 \label{eq:pair-angular-momentum}
\end{equation}
where $\hat{\bm e}_\varphi$ points counterclockwise and $m/2$ is the
reduced mass. The joint pair current is
$\bm{\mathcal J}_2(\bm d)=g_2(\bm d)\overline{\bm v}_{\rm rel}(\bm d)$.
For an isolated close pair, $L\to(\ell+\alpha)\hbar=\hbar/2$.
The sign records the chosen semion chirality: time reversal conjugates
the exchange phase $+\ii$ to $-\ii$ and reverses the current.
The exact two-semion state shows what statistics alone does to an
isolated pair [dashed curve in Fig.~\ref{fig:pair-motion}(a);
Supplemental Material, Fig.~\ref{fig:n2-exact}(c)]. The pair circulates
counterclockwise, with $L=\hbar/2$ at contact, and $L(r)$ decreases
monotonically with separation while remaining positive, down to
$0.12\hbar$ at $r=L/2$.

All four beyond-ED states counter-rotate at larger separation
[Fig.~\ref{fig:pair-motion}(a)]. The first zero lies near
$0.73$--$0.75a$, followed by a negative minimum near $r=a$.
The reversal also occurs at $N=4,6$, whereas the exact two-semion
current remains positive.
As particle number increases, the reversal
remains near the mean spacing while moving inward relative to the box
size. The reversal is therefore set by the particle density rather than
by the cell size: beyond about three quarters of the mean spacing, the
other particles reverse the counterclockwise circulation that
statistics imposes on a close pair.

The density correlations and currents characterize the internal spatial
structure and relative motion of particle pairs. To identify paired
superfluidity, we now translate the pair center and resolve its internal
angular channels, testing which pairs remain coherent between separated
positions.

\subsection{Paired superfluidity and chiral pair coherence}
\label{sec:pair-coherence}

\begin{figure*}[t]
\centering
\includegraphics[width=\textwidth]{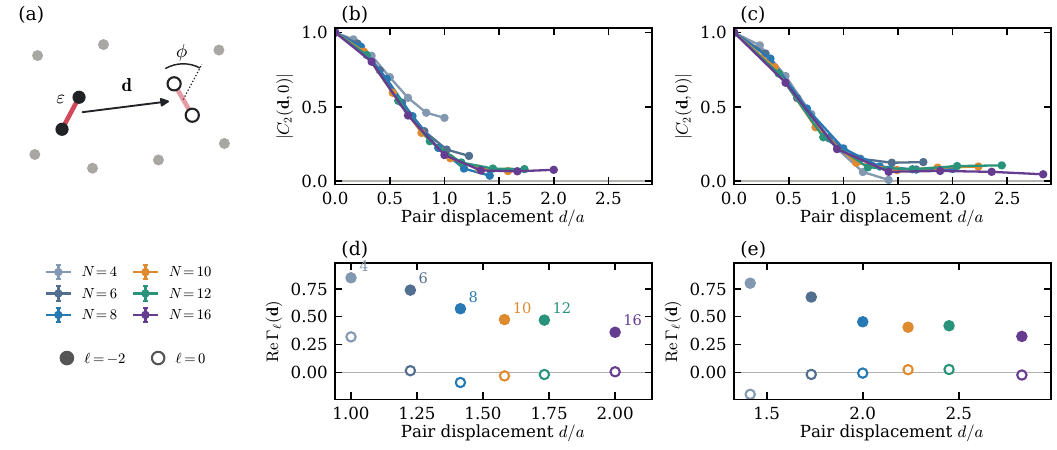}
\caption{Pair coherence of the optimized semion gas. (a) The measured
move: a compact pair (filled) with internal vector $\bm\varepsilon$ is
displaced by $\bm d$ and its internal vector is rotated by $\phi$ (open),
while the spectators (gray) stay fixed. The red segment marks the
statistical branch cut, joined inside the pair. (b,c) Magnitude
$|C_2(\bm d,0)|$ for pairs with internal separation $0.45$--$0.65a$,
translated along $x$ and the diagonal. The faint horizontal line marks
zero. (d,e) Channel-resolved coherence
$\operatorname{Re}\Gamma_\ell(\bm d)$ at the farthest points along $x$ (d) and the diagonal (e); filled symbols are $\ell=-2$, open symbols $\ell=0$.
Each point is labeled by particle number in (d); colors are common to
panels (b)--(e). Distances increase with $N$, so these panels compare coherence
across each cell rather than a fixed-distance size extrapolation.
The counter-rotating channel remains coherent as the contact channel
loses coherence. All data use the states in Tables~\ref{tab:validation}
and \ref{tab:beyond-ed}. At $N=16$, all observables use
$\mathcal P_{12}$ on the selected checkpoint after 120\,000 plus
20\,000 updates.
Error bars show two blocked statistical standard errors; magnitude errors
in (b,c) use a delete-one-group jackknife, with bars clipped at zero.}
\label{fig:pair-coherence}
\end{figure*}

A paired condensate is characterized by off-diagonal order in its pair
density matrix~\cite{Yang1962}: the amplitude to remove a pair at one
position and reinsert it at another remains finite as the two positions
separate. For semions, the natural object is a compact pair. Taking one
semion around another gives the phase $-1$, so taking a pair around a
third semion, or exchanging two pairs, gives $+1$. A compact pair is
therefore a local boson and can carry an ordinary order parameter,
whereas a single semion cannot. At fixed particle number the pair
amplitude itself vanishes, so we measure the pair density matrix. In
first quantization, it is the overlap between the state and a copy in
which one compact pair has been displaced, with all other particles, the
spectators, held fixed. We sample particle configurations from
$\|\Psi\|^2$, take each pair whose separation lies in a fixed short
range, move it rigidly [Fig.~\ref{fig:pair-coherence}(a)], and average
the ratio of the displaced to the original wavefunction.

This ratio must compare physical amplitudes. Only gauge-invariant
correlators are physical: by Elitzur's theorem, operators that are not
gauge invariant have vanishing expectation values~\cite{Elitzur1975}.
In the Chern--Simons description of the statistical
interaction~\cite{WenZee92,iengo1992anyon}, a pair operator is
therefore physical only together with its statistical string. In first
quantization, the physical amplitudes are the anyon-gauge amplitudes
$\Psi_{\mathrm A}=e^{\ii\chi}\Psi$ of
Eq.~\eqref{eq:statistical-transmutation}, whereas the network evaluates
the bosonic-gauge state $\Psi$. Moving the pair changes its statistical
phase with every spectator, so the bosonic-gauge overlap must be
multiplied by $e^{\ii[\chi(\R')-\chi(\R)]}$, where $\R'$ is the moved
configuration. The factor is not unique, however. Because $\Psi_{\mathrm A}$
is multivalued, its phase at $\R'$ depends on how $\chi$ is continued
from $\R$.

We fix this choice by treating the pair as a single object. For each
spectator $k$, we write the pair's contribution to the planar
$\chi$ as
\begin{equation}
 \begin{aligned}
 &\tfrac12\left[\arg(z_i-z_k)+\arg(z_j-z_k)\right]\\
 &\qquad=\arg(c-z_k)
   +\frac12\sum_{m=i,j}\operatorname{Arg}\frac{z_m-z_k}{c-z_k},
 \end{aligned}
 \label{eq:pair-phase-split}
\end{equation}
where $c$ is the pair center in complex notation. The first term has an
integer coefficient. Seen from spectator $k$, the two half flux quanta of
the pair add to one flux quantum, whose phase factor
$(c-z_k)/|c-z_k|$ is single valued. The
second term depends on the internal geometry of the pair, and its ratios
are close to one for a compact pair. Evaluating them on the principal
branch ($\operatorname{Arg}$) leaves a single branch cut, on the short
segment joining the two members. This is the local-pair prescription of
Ref.~\cite{ZhangIdealAnyonGas2026}. The resulting phase is a function of
the configuration alone, so the overlap does not depend on the path of
the move. Continuing the phase of each member separately along the move
would instead add a factor $-1$ whenever a spectator passes between
the members. Such crossings are frequent in a liquid, and their signs
would suppress the correlator even for coherent pairs. We use this
member-by-member continuation only as a control.

On the torus, we replace $z$ by the theta function $f(z)$ of
Eq.~\eqref{eq:theta} in Eq.~\eqref{eq:pair-phase-split}, as in $\chi$.
The product of the resulting phases over all spectators is the pair
string $\mathcal E_{ij}(\R)$, given explicitly in Supplemental Material,
Sec.~\ref{app:pair-coherence}.

For a pair $(i,j)$, let $\widetilde\R$ be the sampled configuration with
particle $j$ replaced by its periodic copy nearest to $i$. The internal
vector $\bm\varepsilon=\rvec_i-\rvec_j$ is then the shortest separation
of the two members, and $\bm c$ is their midpoint. The moved
configuration $\R'$ places the members at
$\bm c+\bm d\pm\mathsf R_\phi\bm\varepsilon/2$ and leaves the
spectators in place; $\mathsf R_\phi$ rotates the internal vector by
$\phi$. We select compact pairs with a shell indicator: $\omega(r)=1$ if
the internal separation $r$ lies in a chosen shell and $\omega(r)=0$
otherwise. We use two shells, $0.25$--$0.45a$ and $0.45$--$0.65a$, with
mean spacing $a=(\Acell/N)^{1/2}$. In terms of the bosonic-gauge state,
the pair correlator is
\begin{equation}
 \begin{aligned}
 C_2(\bm d,\phi)&=\frac1{\mathcal Z}\int d\R\sum_{i<j}\omega(r_{ij})\\
 &\qquad\times\Psi(\widetilde\R)^\dagger\Psi(\R')\,
 \frac{\mathcal E_{ij}(\R')}{\mathcal E_{ij}(\widetilde\R)},
 \end{aligned}
 \label{eq:pair-coherence}
\end{equation}
with $\mathcal Z=\int d\R\sum_{i<j}\omega(r_{ij})\|\Psi(\R)\|^2$.
Only pairs inside the shell contribute, so a configuration with no such
pair adds nothing to either the numerator or $\mathcal Z$.
Here $\Psi$ is the full evaluated two-component state, the inner product
sums both components, and the integrals run over all particle positions
in the cell. This is the pair density matrix averaged over pair centers
and over the internal separations in the shell, with $C_2(\bm0,0)=1$.

We evaluate both wavefunctions in the unwrapped coordinates of
Sec.~\ref{sec:model}, without mapping the moved pair back into the cell.
The result does not depend on which periodic copy of each particle is
used. Changing the copy of a particle applies the boundary rules of
Eq.~\eqref{eq:boundary} to both $\Psi(\widetilde\R)$ and $\Psi(\R')$.
The matrices $\Xmat$ and $\Zmat$ are unitary and cancel in
$\Psi(\widetilde\R)^\dagger\Psi(\R')$, but the Aharonov--Bohm phases
depend on the particle positions and differ between the two
configurations. The theta functions in $\mathcal E_{ij}$ change by the
opposite phase.

We sample $\R$ from $\|\Psi\|^2$. Then $C_2$ is the average of
$\sum_{i<j}\omega(r_{ij})F_{ij}$ divided by the average of
$\sum_{i<j}\omega(r_{ij})$, with the single-configuration estimator
\begin{equation}
 F_{ij}(\bm d,\phi;\R)=
 \frac{\Psi(\widetilde\R)^\dagger\Psi(\R')}
      {\|\Psi(\widetilde\R)\|^2}
 \frac{\mathcal E_{ij}(\R')}{\mathcal E_{ij}(\widetilde\R)}.
 \label{eq:pair-overlap}
\end{equation}
We average the complex overlap before examining its real part or
magnitude.

The magnitude of the total pair correlator initially falls over a distance of order the
mean spacing and then retains a finite signal across the larger cells
[Fig.~\ref{fig:pair-coherence}(b,c)]. At $N=10,12$ the outer-shell
correlator levels off near $0.07$--$0.10$ at the largest separations,
extending to $2.45a$ along the $N=12$ diagonal. A nonzero signal also
persists at $N=16$, where the diagonal reaches $2.83a$.

Resolving internal motion identifies the pairs that carry this coherence.
Rotating the internal vector during the move compares the relative
wavefunction of the pair at different orientations, so the Fourier
components of $C_2$ in $\phi$ separate its relative-angular-momentum
channels. We rotate the bosonic-gauge amplitude: $\mathcal E_{ij}$ omits
the internal term $\tfrac12\arg f(z_i-z_j)$ of $\chi$. A rotation by
$\pi$ then exchanges the members and leaves $\Psi$ unchanged, so $C_2$
has period $\pi$ in $\phi$, and only the even $\ell$ of
Eq.~\eqref{eq:pair-channels} appear. Define
\begin{equation}
 \begin{aligned}
 W_\ell(\bm d)&=\frac1\pi\int_0^\pi d\phi\,
                  e^{-\ii\ell\phi}C_2(\bm d,\phi),\\
 \Gamma_\ell(\bm d)&=W_\ell(\bm d)/W_\ell(\bm0).
 \end{aligned}
 \label{eq:coherence-channels}
\end{equation}
At $\bm d=\bm0$, $W_\ell$ is the fraction of compact pairs in channel
$\ell$, and $\Gamma_\ell$ is the coherence of that channel relative to
its population. Restoring the omitted internal phase shifts each label
by the statistical $1/2$, so the physical relative angular momentum is
$(\ell+1/2)\hbar$.
Most compact pairs occupy $\ell=0$, the $+\hbar/2$ contact channel.
The counter-rotating $\ell=-2$ channel, with label $-3\hbar/2$, has
only about $9$--$11\%$ of the inner-shell weight and $19$--$21\%$ of
the outer-shell weight. Nevertheless, it carries the persistent coherence.
At the farthest points of the $N=10,12$ cells,
$\operatorname{Re}\Gamma_{-2}\simeq0.4$--$0.5$, while
$|\operatorname{Re}\Gamma_0|\lesssim0.03$
[Fig.~\ref{fig:pair-coherence}(d,e)]. The normalized $\ell=-2$
coherence agrees between the two internal shells to about $0.03$,
showing that the same chiral channel remains coherent across a range
of compact-pair sizes.

This channel resolution connects the persistent pair coherence to the
chiral relative motion in Sec.~\ref{sec:pairs-current}. The channel
weights give the mean relative angular momentum of a shell's pairs,
$\hbar[1/2+\sum_\ell\ell W_\ell(\bm0)]$. For compact pairs it agrees with
the shell average of the current-based $L(r)$ of
Eq.~\eqref{eq:pair-angular-momentum} to about $0.003\hbar$ at $N=4$ and
$0.01\hbar$ at $N=6$. The small difference comes from the connection
terms of nearby spectators.

Together, these measurements identify paired superfluid behavior and
resolve the chiral internal motion of the coherent pairs. This behavior
emerges in a variational state learned without a prescribed pairing
structure.

\section{Discussion and conclusion}
\label{sec:conclusion}

We have constructed a neural variational representation of continuum
semions on a torus that preserves the statistical connection, global
transport, and collision domain. Its compact orbital frame
contains $3(N+1)$ generating sections. Configuration-dependent neural
coefficients retain the smooth many-body freedom, while learned orbital
coordinates adapt the frame without changing the physical boundary
conditions. A common phase correction and coherent translation average
allow this geometry to be optimized in the target momentum sector.

Validation at two, four, and six semions anchors the many-body
calculation. The two-particle energy closely matches the exact
solution, and the neural energies at four and six particles lie below
our best available finite-cutoff ED estimates (Table~\ref{tab:validation}).
The same construction reaches up to 16 semions, beyond the reach of
our continuum diagonalization. Our central physical finding is paired
superfluid behavior in the semion gas: compact-pair coherence persists across
the accessible cells and is concentrated in the chiral $\ell=-2$
channel. The more populated contact channel loses coherence, while the
normalized coherence of the counter-rotating channel is nearly independent
of the chosen internal-separation shell. These correlations emerge from
energy optimization with no prescribed pairing structure.
The correlation hole, suppressed long-wavelength density fluctuations,
and circulation reversal near $0.73$--$0.75$ mean spacings characterize
the accompanying liquid. At $N=12,16$, the circulation returns to positive
values at larger separation, giving a damped oscillatory radial profile.

Following the same correlator to larger cells at fixed density, with
variational accuracy controlled at each size, will resolve whether the
persistent pair coherence approaches a nonzero limit or decays slowly,
and will quantify the long-range pair order. A complementary calculation of the static transverse
current response at finite wavevector would determine the superfluid
density and connect the observed pair coherence to phase rigidity.

Our work represents a first attempt at constructing neural variational states for anyons on a torus.
Further methodological improvements can focus on the joint choice of orbital frame and neural architecture. Can one identify an optimal starting frame, in both its geometry and number of generating sections, that reaches lower variational energies at a fixed computational cost? How does accuracy depend on network depth, and which combinations of width and depth are most effective for a given frame size and particle number? Exploring these choices systematically could yield more efficient variational states and extend calculations of energy densities and correlations toward the thermodynamic limit.

Our work also sets the stage to study several open problems via anyonic neural quantum states.
Within our framework, both short-range interaction potentials and long-range Coulomb repulsion for charged anyons can be incorporated immediately, allowing us to study how interactions reshape the paired liquid and whether they stabilize competing anyonic crystals~\cite{komargodski2026anyon}.
Our work can also be extended to characterize correlations in multi-component anyon systems with mutual anyonic statistics, e.g., $\mathbb{Z}_2$ spin liquids where the spinon and vison excitations are self-bosons but mutual semions.
Finally, a long term goal would be to generalize our learned frame ansatz to non-Abelian anyons on a torus~\cite{NayakEtAl2008}, thereby opening the door to study the effect of non-Abelian statistics on anyonic correlations at finite density.

\begin{acknowledgments}
We thank Eslam Khalaf for insightful conversations and ongoing
collaboration on related topics, and T. Senthil for inspiring discussions. 
We thank Ya-Hui Zhang for privately communicating his results before
publication.
The computations in this work were done using Bridges-2 at PSC and SDSC
Expanse cluster through allocation PHY250361 from the Advanced
Cyberinfrastructure Coordination Ecosystem: Services \& Support (ACCESS)
program~\cite{access}, which is supported by U.S. National Science Foundation
grants No. 2138259, No. 2138286, No. 2138307, No. 2137603, and No. 2138296.
\end{acknowledgments}

\vspace{5mm}
\noindent\textit{Note added.} While our manuscript was in preparation, we became
aware of related work by Perez Fadon \textit{et al.}~\cite{PerezFadonEtAl2026}.
They develop a neural-network based variational ansatz for continuum anyons in harmonic and Coulomb traps.
Here we pursue a complementary direction: we address the continuum anyon problem on a torus with a neural
ansatz that exactly enforces global transport and provides microscopic
access to bulk anyon physics.
We also note the concurrent work of
Y.-H.~Zhang~\cite{ZhangIdealAnyonGas2026}, which studies the ideal anyon
gas on the sphere using composite-fermion and parton variational
wavefunctions. For semions, their conclusion of a paired superfluid is
consistent with our results, which were obtained with a different
variational approach.

\onecolumngrid
\clearpage
\setcounter{section}{0}
\setcounter{equation}{0}
\renewcommand{\thesection}{S\arabic{section}}
\renewcommand{\theequation}{S\arabic{equation}}
\renewcommand{\theHsection}{supp.\arabic{section}}
\renewcommand{\theHequation}{supp.\arabic{equation}}
\setcounter{figure}{0}
\setcounter{table}{0}
\renewcommand{\thefigure}{S\arabic{figure}}
\renewcommand{\thetable}{S\arabic{table}}
\renewcommand{\theHfigure}{supp.\arabic{figure}}
\renewcommand{\theHtable}{supp.\arabic{table}}
\section*{Supplemental Material}
\par\medskip

\section{Consistency of the boundary rules and of the connection}
\label{app:consistency}

\paragraph{Corner algebra and even $N$.}
Translating $x_j$ by $L_x$ changes the phase in the $y_i$ rule of
Eq.~\eqref{eq:boundary} by $\exp[\ii\pi(N\delta_{ij}-1)]$. For $i\ne j$ this
is $-1$ for every $N$. For $i=j$ it is $(-1)^{N-1}$, which equals $-1$ only
for even $N$. The anticommutation $\Xmat\Zmat=-\Zmat\Xmat$ compensates a
factor $-1$ for every pair $i,j$. Both orders of the coordinate
translations therefore agree for even $N$, while for odd $N$ the
same-particle corner fails.

\paragraph{Boundary twists and gauge choice.}
Let $\bm b=(\beta_x/L_x,\beta_y/L_y)$ and
$\chi_\beta(\R)=\bm b\cdot\sum_i\rvec_i$. The rephasing
\begin{equation}
 \widetilde\Psi=e^{-\ii\chi_\beta}\Psi,\qquad
 \widetilde{\avec}_i=\avec_i+\bm b
 \label{eq:twist-rephasing}
\end{equation}
removes the scalar factors $e^{\ii\beta_\mu}$ from both boundary rules.
With $\widetilde D_i=-\ii\nabla_i+\widetilde{\avec}_i$, however,
\begin{equation}
 \widetilde D_i\widetilde\Psi=e^{-\ii\chi_\beta}D_i\Psi.
 \label{eq:twist-covariance}
\end{equation}
Thus the same Hamiltonian expectation is represented by untwisted
boundaries and an additional constant vector potential. Its curl is
zero, but its phases around the handles retain the twist information.
For generic $\beta_\mu$, $e^{-\ii\chi_\beta}$ is not a single-valued
periodic scalar gauge transformation on the torus. Setting both the
boundary twists and this constant potential to zero specifies a physical
boundary choice; it is not a gauge fixing of arbitrary twists.

\paragraph{Sign of the covariant derivative.}
The inverse of the statistical transmutation in
Eq.~\eqref{eq:statistical-transmutation} recovers the locally free
anyon-gauge derivative. On a simply connected collision-free chart,
\begin{equation}
 \Psi_{\mathrm A}=e^{\ii\chi}\Psi
 \quad\Longrightarrow\quad
 -\ii\nabla_i\Psi_{\mathrm A}=e^{\ii\chi}D_i\Psi.
 \label{eq:singular-gauge}
\end{equation}
Near a pair, $f(z)\propto z$, so a counterclockwise
exchange raises $\chi$ by $\pi/2$. The anyon-gauge wavefunction
$\Psi_{\mathrm A}$ thus acquires $+\ii$, and a full pair loop gives $-1$.
Reference~\onlinecite{HosotaniHo1992} uses a fermionic reference
wavefunction. Its exchange factor $-e^{-\ii\pi/2}$ describes the same
physical semion phase. Our bosonic representation and boundary convention
are fixed by Eqs.~\eqref{eq:boundary} and \eqref{eq:singular-gauge}.

\paragraph{Quasiperiodicity of the connection.}
The theta-function identities give
\begin{equation}
 \bm a(z+L_x)=\bm a(z),\qquad
 \bm a(z+\ii L_y)=\bm a(z)-\frac{\pi}{L_x}\hat{\bm x}.
\end{equation}
Translating $y_i$ therefore changes the connection acting on particle $k$ by
\begin{equation}
 \avec_k(\R+L_y\hat e_{iy})=
 \avec_k(\R)-\frac{\pi}{L_x}(N\delta_{ik}-1)\hat{\bm x}.
 \label{eq:connection-boundary}
\end{equation}
This is exactly minus the gradient of the phase in Eq.~\eqref{eq:boundary}.
Hence $D_k\Psi$ obeys the same unitary transport as $\Psi$, and the kinetic
density is single valued. The connection is locally a gradient away from
coincidences. Its singular circulation represents point statistical flux,
not a smooth external field.

\section{Collision domain and pair channels}
\label{app:collisions}

The coincidence set $\Delta$ contains all configurations with
$\rvec_i=\rvec_j$ modulo a torus period for at least one pair. Positions are
continuous, and the configuration space is $F_N(T^2)=(T^2)^N\setminus\Delta$,
with exchange symmetry imposed on labeled coordinates. The kinetic
expectation in Eq.~\eqref{eq:variational} is defined through the
nonnegative form
\begin{equation}
 q[\Psi]=\kappa\sum_i\int\|D_i\Psi(\R)\|^2\,d\R,
 \qquad E[\Psi]=\frac{q[\Psi]}{\langle\Psi|\Psi\rangle}.
 \label{eq:quadratic-form}
\end{equation}
We define it first on smooth physical sections supported away from
$\Delta$ and then close it. The constant
factor associated with indistinguishability cancels in the energy ratio.
This domain choice is part of our model; it is not inferred from the
differential expression of Ref.~\onlinecite{HosotaniHo1992} alone.

For an isolated pair the singular angular operator is
$(-\ii\partial_\varphi+1/2)^2/r^2$. The regular radial solutions give the
powers $\nu_\ell=|\ell+1/2|$ of Eq.~\eqref{eq:pair-channels}, with leading
power $\nu_0=1/2$. A wavefunction with nonzero value at coincidence has a
kinetic density $\propto|\Psi|^2/(4r^2)$ near the pair and a logarithmically
divergent form energy.

With smooth bounded frame and coefficient derivatives, both
$D_i\Psi_\theta$ and the collision cutoff approximants of
$\Psi_\theta$ have finite form norm, so the pair factor places the ansatz in
the Friedrichs form domain. It does not prescribe all eigenfunction
asymptotics. The $\ell=-2$ channel starts as $r^{3/2}e^{-2\ii\varphi}$,
whereas a smooth even Cartesian remainder multiplied by $r^{1/2}$ starts that
angular channel at $r^{5/2}$. This affects the resolution required of the
coefficient functions near a collision. In local relative coordinates
$z=re^{\ii\varphi}$, the missing remainder is proportional to
$\bar z^2/r$. Smooth, exchange-even approximants are
\begin{equation}
 b_\epsilon(z)=\frac{\bar z^2}{\sqrt{r^2+\epsilon^2}},
 \qquad \epsilon>0.
 \label{eq:channel-regularization}
\end{equation}
After multiplication by $J\sim r^{1/2}$ and a smooth local cutoff,
these converge to $r^{3/2}e^{-2\ii\varphi}$ in the local kinetic-form
norm as $\epsilon\to0$. Thus increasing coefficient resolution can
approximate this channel in energy. Supplying $\bar z^2/r$ explicitly
in an additional analytic factor represents its leading power directly,
reducing the near-collision resolution required of the network.
For a simultaneous collision, let $k$ particles approach a common point
with all pair separations proportional to a size $\varrho$, keeping their
relative shape fixed and the other particles separated. Their $k(k-1)/2$
internal pair factors each contribute $\varrho^{1/2}$, so
$J\sim\varrho^{k(k-1)/4}$; for three particles, $J\sim\varrho^{3/2}$.
A bounded smooth remainder makes the full ansatz vanish at least this
fast. The physical collective exponent follows from the $k$-particle
relative Schr\"odinger problem, rather than the isolated-pair cusp
alone~\cite{PerezFadonEtAl2026}. Representing a channel that vanishes more
slowly requires coefficients $\Psi/J$ that grow as the cluster shrinks.
Approximating this growth can increase the resolution required of the
network; an adapted collective factor can supply it analytically.

\section{Center of mass}
\label{app:com}

\paragraph{Momentum and COM energy.}
Since $\sum_i\avec_i=0$, collective translations commute with $H$. A state
with $\bm P\Psi=\hbar\bm K\Psi$ obeys
\begin{equation}
 K_\mu L_\mu=N\beta_\mu+2\pi n_\mu,\qquad n_\mu\in\mathbb Z.
\end{equation}
The $N$ particle derivatives decompose orthogonally into their sum and
relative parts. The sum contributes Eq.~\eqref{eq:com-energy}, which equals
$\kappa|\bm K|^2/N$ for a definite momentum.

\paragraph{The COM section.}
From Eq.~\eqref{eq:com-orbital}, $F(X+2L_x,Y)=F(X,Y)$ and
$F(X,Y+L_y)=e^{-\ii\pi X/L_x}F(X,Y)$. Consequently
$\bm\Phi(X+L_x)=\Xmat\bm\Phi$ and
$\bm\Phi(Y+L_y)=e^{-\ii\pi X/L_x}\Zmat\bm\Phi$, which are
Eqs.~\eqref{eq:com-transport}. The boosts $\mathsf B_x$ and $\mathsf B_y$
preserve these laws.

\paragraph{Gram eigenvalues.}
For a unit spinor $\bm v$ with Bloch vector $\bm m$,
\begin{equation}
 \bm v\bm v^\dagger+\Zmat\bm v\bm v^\dagger\Zmat+\Xmat\bm v\bm v^\dagger\Xmat
 =\tfrac12\left[3+(m_x,-m_y,m_z)\cdot\bm\sigma\right],
\end{equation}
with eigenvalues $1$ and $2$. The boost phases have unit modulus and do not
change this. The inverse square root of $G=C_0C_0^\dagger$ is therefore the
fixed polynomial
\begin{equation}
 G^{-1/2}=\left(2-\frac1{\sqrt2}\right)\Id_2
              +\left(\frac1{\sqrt2}-1\right)G,
\end{equation}
which is regular everywhere and requires no eigenvector convention.

\section{Transport and derivatives of the learned frame}
\label{app:learned-frame}

Write $U_x=\sum_j u_{\theta,j,x}$ and
$Q_x=\sum_j q_{j,x}$. Under $y_i\to y_i+L_y$, periodicity of the
displacement means that only $q_{i,y}$ changes, by the same period. The
analytic frame therefore transforms as
\begin{equation}
 \Sframe_0(\bm Q_\theta+L_y\hat e_{iy})
 =e^{\ii\beta_y+\ii\pi(Nq_{i,x}-Q_x)/L_x}
       \Zmat\Sframe_0(\bm Q_\theta).
\end{equation}
Meanwhile Eq.~\eqref{eq:endpoint-phase} gives
\begin{equation}
 \Delta\varphi_\theta=-\frac{\pi}{L_x}
                  (Nu_{\theta,i,x}-U_x).
\end{equation}
Since $Nq_{i,x}-Q_x=Nx_i-X+Nu_{\theta,i,x}-U_x$, these extra phases
cancel exactly. Under an $x_i$ period, $\varphi_\theta$ is unchanged and
the analytic frame supplies $e^{\ii\beta_x}\Xmat$. Permutation equivariance
of the displacement makes both $\Sframe_0(\bm Q_\theta)$ and
$\varphi_\theta$ invariant under particle relabeling. These arguments
use coordinate values continued across the boundary, retaining the
analytic frame's quasiperiodic phase.

The ordinary chain rule, for a Cartesian direction $\mu=x,y$, is
\begin{equation}
 \begin{split}
 \partial_{i\mu}\Gframe_\theta=e^{\ii\varphi_\theta}
 \bigg[&\ii(\partial_{i\mu}\varphi_\theta)\Sframe_0(\bm Q_\theta)\\
 &+\sum_{j,\nu}(\delta_{ij}\delta_{\mu\nu}
       +\partial_{i\mu}u_{\theta,j,\nu})
       \partial_{q_{j,\nu}}\Sframe_0(\bm Q_\theta)\bigg].
 \end{split}
 \label{eq:learned-chain-rule}
\end{equation}
Here $\partial_{i\mu}$ differentiates with respect to the physical
coordinate $r_{i\mu}$ and the Kronecker delta equals one for matching
indices and zero otherwise. The matrix of coordinate derivatives in
parentheses can lose rank without affecting the tightness of the frame.
No inverse coordinate map appears.

For clarity, setting $\bm b=\Sframe_0(\bm Q_\theta)\bm c_\theta$ gives
\begin{equation}
 D_i(e^{\ii\varphi_\theta}\bm b)=e^{\ii\varphi_\theta}
 \left[-\ii\nabla_i\bm b+
       (\avec_i(\R)+\nabla_i\varphi_\theta)\bm b\right].
\end{equation}
Thus the phase contributes to kinetic and parameter derivatives even
though it cancels from the unprojected density. Between translated copies
it also affects interference. Keeping $J_\eta(\R)$ and differentiating
the full smooth amplitude allows the same integration-by-parts kinetic
form and isolated-pair pole cancellation as for the analytic starting
frame. The smoothness and collision-domain argument in
Supplemental Material, Sec.~\ref{app:collisions} applies to both the learned amplitude and
its finite projection sum.

\section{Local orbital frame}
\label{app:orbitals}

\paragraph{Orbitals.}
Let $B_*=2\pi n_\phi/\Acell$ and $\ell_*=B_*^{-1/2}$. A basis of the
auxiliary LLL is
\begin{align}
 \phi_k(x,y)={}&\sum_{j\in\mathbb Z}
 \exp\!\left[-\frac{B_*}{2}(y-L_y\lambda_{jk})^2\right]\nonumber\\
 &\times\exp\!\left[\frac{2\pi\ii n_\phi}{L_x}
                  \lambda_{jk}x+\ii\beta_y j\right],
 \label{eq:lll-orbitals}\\
 \lambda_{jk}={}&j+\frac{k}{n_\phi}
                    +\frac{\beta_x}{2\pi n_\phi},
 \qquad k=0,\ldots,n_\phi-1.\nonumber
\end{align}
Each orbital has the single-particle boundary law of
Eq.~\eqref{eq:scalar-transport}.

\paragraph{Two interleaved orbital functions.}
Put $m=n_\phi$ in this paragraph. Choose a near-regular index-$m$ point
lattice on the torus; ``index $m$'' means that it has $m$ sites per cell.
In the implementation its sites are
$\rvec_j=(jL_x/m,(s j\bmod m)L_y/m)$ for $j=0,\ldots,m-1$, with the
integer $s$ chosen to maximize their minimum separation. Shift all sites
so that their sum equals, modulo torus periods,
\begin{equation}
 \left(\frac{mL_x}{2}-\frac{\beta_yL_x}{2\pi},\,
       \frac{mL_y}{2}+\frac{\beta_xL_y}{2\pi}\right)
 .
\end{equation}
This is the zero-sum condition for an LLL function with these boundary
twists. The coefficients in $\phi_A=\sum_{l=0}^{m-1}c_l\phi_l$ are found
by requiring zeros at these sites and normalizing the coefficient vector.
This selects a single orbital function, rather than retaining the full
LLL as local features.

Its partner is the magnetic translate
\begin{equation}
 \phi_B(x,y)=e^{\ii B_*d_y x}\phi_A(x-d_x,y-d_y),
 \qquad d_\mu\in (L_\mu/m)\mathbb Z.
 \label{eq:magnetic-partner}
\end{equation}
The phase in this translation keeps the boundary twists and the LLL
condition. The shift is chosen to maximize the minimum separation
between the two vortex sets. As two distinct translates of the same
index-$m$ lattice, their zero sets are disjoint, ensuring $\rho>0$.
For $N=6$ and zero twists, the three zeros of $\phi_A$ are
$(1/6,1/6)$, $(1/2,1/2)$, and $(5/6,5/6)$ on the unit square, with
$\bm d=(0,1/3)$ for its partner. These are the constants used in
Fig.~\ref{fig:texture} and the trained compact model.
For $N=2$, $m=1$ allows only one independent LLL
function. The first-excited partner
$\phi_1^{\mathrm{ex}}=-\sqrt{2/B_*}\,\partial_y\phi_0$ is nonzero at the
simple zero of $\phi_0$, so the sum of their squared magnitudes remains
positive there.

\paragraph{Normalization density and metric.}
For the two selected LLL functions,
Eq.~\eqref{eq:metric} satisfies
$g=B_*+\tfrac12\nabla^2\log\rho$, and hence
$\langle g\rangle_{T^2}=B_*$. Here $\rho$ is the sum of squared magnitudes
of these two functions. Disjoint zeros ensure a regular denominator;
interleaving aims to keep
the derivative cost spread over the cell. The geometry checks measure
this distribution as well as its fixed mean.

\paragraph{Why this is minimal.}
With the projected derivatives
$\mathcal D_\mu\bm v=\partial_\mu\bm v-\bm v(\bm v^\dagger\partial_\mu\bm v)$,
$g=\|\mathcal D_x\bm v\|^2+\|\mathcal D_y\bm v\|^2$, and the Berry curvature
of $\bm v$ is
$\mathcal F=2\operatorname{Im}\langle\mathcal D_x\bm v,\mathcal D_y\bm v\rangle$.
Pointwise, $g\ge2\|\mathcal D_x\bm v\|\|\mathcal D_y\bm v\|\ge|\mathcal F|$.
Because $\bm v$ carries the boundary phases of $n_\phi$ flux quanta,
$\int_{T^2}\mathcal F\,d^2r=\pm2\pi n_\phi$. Hence
$\int g\,d^2r\ge2\pi n_\phi$, with equality for the LLL textures used
here. For the two-component orbital vector, $g=\tfrac14|\nabla\bm n|^2$ and
$\mathcal F$ is the skyrmion density, so this is the Belavin--Polyakov
bound~\cite{BelavinPolyakov1975}. 

These properties control the local orbital geometry. The complete physical
derivatives also act on the COM frame, the product normalization, the
collision factor and the learned coefficients.

\section{Normalization of the symmetric products}
\label{app:products}

The amplitudes $\widetilde p_k$ of Eq.~\eqref{eq:symmetrized-orbitals}
are the components of the symmetric projection of
$\bm v_1\otimes\cdots\otimes\bm v_N$, in the normalized two-orbital
occupation basis. This projection removes the parts that change under
particle permutations, so its squared norm is at most one. To obtain a
lower bound, average over unit two-component vectors $\bm z$, with
uniform probability measure $d\mu(\bm z)$ on their pseudospin sphere:
\begin{equation}
 \mathcal N^2=(N+1)\int\prod_i|\bm z^\dagger\bm v_i|^2\,d\mu(\bm z).
\end{equation}
For each unit $\bm v_i$, the average of
$\log|\bm z^\dagger\bm v_i|^2$ is $-1$. The inequality that the average
of an exponential is at least the exponential of its average then gives
$\mathcal N^2\ge(N+1)e^{-N}$, proving the stated lower bound.

\section{Network equations}
\label{app:network}

The network constructs the two learned ingredients of
Eq.~\eqref{eq:ansatz}: the displacement $\bm u_\theta(\R)$ that changes
where the analytic frame is evaluated, and the complex coefficients
$\bm c_\theta(\R)$ that combine its columns. In terms of the three
two-component COM columns $\bm C_a$ and the $N+1$ scalar occupation
amplitudes $p_k$, the resulting wavefunction is
\begin{equation}
 \Psi_\theta(\R)=J_\eta(\R)e^{\ii\varphi_\theta(\R)}
 \sum_{a=1}^3\sum_{k=0}^N
 \bm C_a(X_Q,Y_Q)\,p_k(\bm Q_\theta)\,c_{\theta,ak}(\R),
 \label{eq:network-ansatz}
\end{equation}
where $\bm Q_\theta=\R+\bm u_\theta(\R)$,
$X_Q=\sum_i q_{i,x}$, and $Y_Q=\sum_i q_{i,y}$.
The pair factor, COM columns, scalar amplitudes, and compensating
phase~\eqref{eq:endpoint-phase} have the analytic forms given in the
main text. The following layers specify how the network produces
$\bm u_\theta$ and $\bm c_\theta$ from the physical positions $\R$;
they introduce no additional wavefunction factor.

\paragraph{Periodic inputs and the first layer.}
For harmonics $h=1,\ldots,H_f$, collect $\sin(2\pi hr_\mu/L_\mu)$ and
$\cos(2\pi hr_\mu/L_\mu)$ into a vector $\gamma(\rvec)$ with $4H_f$ entries.
These smooth periodic functions describe positions consistently across
the edges of the torus. Particle $i$ receives the $8H_f$-component vector
\begin{equation}
 \bm f_i^{\mathrm{in}}=\left[\gamma(\rvec_i),\,
       \frac1{N-1}\sum_{j\ne i}\gamma(\rvec_i-\rvec_j)\right],
 \label{eq:network-input}
\end{equation}
where square brackets denote concatenation. The first part records its
position, and the second describes its environment through an average
over the other particles. The input layer is explicitly
\begin{equation}
 \bm h_i^{(0)}=\tanh\!\left(W_{\mathrm{in}}\bm f_i^{\mathrm{in}}
                          +\bm b_{\mathrm{in}}\right),
 \qquad
 W_{\mathrm{in}}\in\mathbb R^{w\times8H_f},\quad
 \bm b_{\mathrm{in}}\in\mathbb R^w.
 \label{eq:input-layer}
\end{equation}
An affine map multiplies its input by a learned matrix and adds a learned
bias; $\tanh$ then acts on each entry, giving a smooth nonlinear
transformation. The resulting hidden vector $\bm h_i^{(0)}\in\mathbb R^w$
is a learned description of particle $i$ in its environment, rather than
a physical wavefunction component. Its width $w$ is the number of internal
features. The same matrix and bias act on every particle.

\paragraph{Successive attention layers.}
The encoder refines these descriptions through
$\bm h_i^{(0)}\to\bm h_i^{(1)}\to\cdots\to\bm h_i^{(L)}$.
The superscript counts successive transformations, not particles or time
steps. At layer $\ell=1,\ldots,L$, each of $H_a$ attention heads
$m=1,\ldots,H_a$ maps the previous hidden vector into three vectors of
width $d_h=w/H_a$:
\begin{equation}
 \begin{aligned}
 \bm q_i^{(\ell,m)}&=W_Q^{(\ell,m)}\bm h_i^{(\ell-1)}+\bm b_Q^{(\ell,m)},\\
 \bm k_i^{(\ell,m)}&=W_K^{(\ell,m)}\bm h_i^{(\ell-1)}+\bm b_K^{(\ell,m)},\\
 \bm v_i^{(\ell,m)}&=W_V^{(\ell,m)}\bm h_i^{(\ell-1)}+\bm b_V^{(\ell,m)}.
 \end{aligned}
 \label{eq:attention-projections}
\end{equation}
These are called query, key, and value vectors. The query of particle $i$
is compared with the key of particle $j$ to determine how much of $j$'s
value vector is passed to $i$:
\begin{equation}
 A_{ij}^{(\ell,m)}=
 \frac{\exp\!\left[\bm q_i^{(\ell,m)}\cdot\bm k_j^{(\ell,m)}/\sqrt{d_h}\right]}
      {\sum_{j'=1}^N\exp\!\left[\bm q_i^{(\ell,m)}\cdot\bm k_{j'}^{(\ell,m)}/\sqrt{d_h}\right]},
 \qquad
 \bm o_i^{(\ell,m)}=\sum_{j=1}^N A_{ij}^{(\ell,m)}\bm v_j^{(\ell,m)}.
 \label{eq:attention-weights}
\end{equation}
The weights are positive and sum to one over $j$, so each head returns
a configuration-dependent weighted average. All particles, including $i$,
enter this sum. Different heads can learn different ways of combining
the information. These attention queries are unrelated to the internal
coordinates $\bm q_i$, and $A_{ij}^{(\ell,m)}$ is unrelated to the
statistical connection $\avec_i$.

The head outputs are concatenated back into a vector of width $w$,
mixed by another affine map, and added to the previous hidden vector:
\begin{align}
 \bm o_i^{(\ell)}&=
 [\bm o_i^{(\ell,1)},\ldots,\bm o_i^{(\ell,H_a)}],\nonumber\\
 \bm h_i^{(\ell)}&=\tanh\!\left[
 \bm h_i^{(\ell-1)}+W_O^{(\ell)}\bm o_i^{(\ell)}+\bm b_O^{(\ell)}\right],
 \qquad W_O^{(\ell)}\in\mathbb R^{w\times w}.
 \label{eq:attention-update}
\end{align}
This addition, often called a residual connection, lets each layer
modify the existing description using information from the other
particles. The parameters differ between layers and heads, but are
shared across particles; no particle-index labels are supplied.
Relabeling particles therefore only relabels the hidden vectors.
Repeating Eq.~\eqref{eq:attention-update} $L$ times gives the common
encoder output $\{\bm h_i^{(L)}\}$ used by both readouts below.

\paragraph{From hidden vectors to the learned frame.}
A readout is a map from hidden features to an output needed by the
ansatz. The displacement readout acts separately on each particle:
\begin{equation}
 u_{\theta,i,\mu}=sL_\mu
     \tanh\!\left[(W_u\bm h_i^{(L)}+\bm b_u)_\mu\right],
 \label{eq:movement-head}
\end{equation}
where $W_u\in\mathbb R^{2\times w}$, $\bm b_u\in\mathbb R^2$, and
$s=0.1$ is the dimensionless displacement scale. Thus
$|u_{\theta,i,\mu}|<sL_\mu$, and a particle's internal displacement can
depend on the entire physical configuration through its hidden vector.
The matrix and bias start at zero, giving $\bm Q_\theta=\R$ initially.
The bound controls scale; exact transport needs smooth, periodic,
permutation-equivariant displacements.

Once these displacements are known, the analytic construction gives
\begin{equation}
 \bm Q_\theta=\R+\bm u_\theta(\R),\qquad
 \Gframe_\theta(\R)=e^{\ii\varphi_\theta(\R)}\Sframe_0(\bm Q_\theta).
 \label{eq:network-frame}
\end{equation}
In particular, $\varphi_\theta$ is calculated from $\bm u_\theta$ by
Eq.~\eqref{eq:endpoint-phase}; it is not a separate neural output.
Only the frame uses the internal coordinates. The pair factor and
physical statistical connection remain evaluated at $\R$.

\paragraph{From hidden vectors to complex coefficients.}
The coefficient vector describes the whole configuration, so its input
must be invariant under particle relabeling. Mean pooling forms
$\bar{\bm h}=N^{-1}\sum_i\bm h_i^{(L)}$. We concatenate this vector
with the $N+1$ orbital weights $|p_k(\bm Q_\theta)|^2$ to give the
coefficient readout information about the frame it will combine:
\begin{equation}
 \bm g_\theta=\tanh\!\left\{
 W_g\left[\bar{\bm h},\,
                \{|p_k(\bm Q_\theta)|^2\}_{k=0}^N\right]+\bm b_g\right\},
 \qquad W_g\in\mathbb R^{w\times(w+N+1)}.
 \label{eq:pooled-layer}
\end{equation}
The orbital weights are symmetric and periodic because the common
boundary phase of the $p_k$ cancels in their absolute squares.
The vector $\bm g_\theta\in\mathbb R^w$ is another hidden representation.
Two linear readouts produce the real and imaginary parts of the
$K_N=3(N+1)$ complex coefficients:
\begin{equation}
 \bm c_\theta=(W_R\bm g_\theta+\bm b_R)
              +\ii(W_I\bm g_\theta+\bm b_I).
 \label{eq:complex-readout}
\end{equation}
Here $W_R,W_I\in\mathbb R^{K_N\times w}$ and
$\bm b_R,\bm b_I\in\mathbb R^{K_N}$. Each coefficient labels a COM
column and an occupation amplitude, not an individual particle.
Both readouts use the same encoder. The calculation is sequential:
encode $\R$, obtain $\bm u_\theta$, evaluate $p_k$ and the frame at
$\bm Q_\theta$, and then obtain $\bm c_\theta$. The hidden vectors
are computed once from $\R$; the appended orbital weights account for
the dependence of the coefficient readout on the learned coordinates.

For the width-64 architecture, $H_f=2$, $w=64$, $L=2$, and $H_a=4$.
Thus the input layer maps 16 features to 64 hidden entries per particle;
two attention updates preserve this width, with 16 entries per head.
The displacement readout maps each final hidden vector to two real
numbers, while pooling and the complex readout give one vector of
$3(N+1)$ complex coefficients for the whole configuration. For $N=6$
these are 21 coefficients multiplying a $2\times21$ frame, whose
contraction gives the two physical components in
Eq.~\eqref{eq:network-ansatz}. The layer count $L$ refers to the
attention updates; the input layer and readouts are specified separately.
All learned matrices and biases above belong to $\theta$ and are
optimized together by the variational energy. Momentum projection
reevaluates this entire construction at each shifted configuration
with the same parameters and combines the resulting amplitudes.

\section{Exact ground state of two semions}
\label{app:n2-exact}

For $N=2$ the problem separates, and its ground state can be written in
closed form up to one quantization condition. 

\paragraph{Reduction.}
With $\bm w=\rvec_1-\rvec_2$, $\bm R_c=(\rvec_1+\rvec_2)/2$ and
$\avec_1=-\avec_2=\bm a(\bm w)$,
\begin{equation}
 H=\tfrac12(-\ii\nabla_{R_c})^2+2(-\ii\nabla_w+\bm a)^2 .
\end{equation}
The boundary rules give $\Psi=e^{\ii\bm K\cdot\bm R_c}\bm\Phi(\bm w)$ with
$\bm K=2\pi(n_x,n_y)$, and the relative function obeys
\begin{equation}
 \bm\Phi(w+1)=s_x\Xmat\bm\Phi(w),\quad
 \bm\Phi(w+\ii)=s_ye^{\ii\pi\operatorname{Re}w}\Zmat\bm\Phi(w),\quad
 \bm\Phi(-w)=\bm\Phi(w),
\end{equation}
with $s_\mu=(-1)^{n_\mu}$. The four sign sectors are unitarily equivalent,
since $\Zmat\bm\Phi$ flips $s_x$ and $\Xmat\bm\Phi$ flips $s_y$. Hence
$E=|\bm K|^2/2+2\lambda$, with $\lambda$ an eigenvalue of one relative
problem $(-\ii\nabla_w+\bm a)^2\bm\Phi=\lambda\bm\Phi$. The ground state has
$\bm K=0$. Its center-of-mass boosts lie $2\pi^2$ higher, and the first internal
excitation is at $E=44.6066$.

\paragraph{Local solutions.}
On the cell $C=\{|\operatorname{Re}w|,|\operatorname{Im}w|\le1/2\}$ around
the collision, write $w=re^{\ii\varphi}$ and
$\chi=\varphi/2+\eta(w)$, with $\eta=\tfrac12\arg[f(w)/w]$ smooth and
single valued on $C$ [$f$ of Eq.~\eqref{eq:theta}]. In the anyon gauge each
component solves the free Helmholtz equation and changes sign around $w=0$.
The solutions that are regular in the Friedrichs sense and even under
exchange are $J_{n+1/2}(k r)e^{\ii(m_n+1/2)\varphi}$, with $m_n=n$ for even
$n$ and $m_n=-(n+1)$ for odd $n$, $n=0,1,2,\ldots$. The local powers
$\nu=n+1/2$ are those of Eq.~\eqref{eq:pair-channels}. Half-integer Bessel
functions are elementary, for example $J_{1/2}(x)=\sqrt{2/\pi x}\,\sin x$.
In the bosonic gauge the ground state is, for $w\in C$,
\begin{equation}
 \Phi_c(w)=e^{-\ii\eta(w)}\sum_{n\ge0}a_{c,n}
 \frac{J_{n+1/2}(kr)}{J_{n+1/2}(k/2)}e^{\ii m_n\varphi},\qquad c=1,2,
 \label{eq:n2-series}
\end{equation}
continued to the whole plane by the relative boundary rules. The series
converges geometrically on $C$: its radius of convergence is the distance
$1$ to the nearest image collision, while the cell corners lie at $0.707$.

\paragraph{Quantization and result.}
The wavenumber $k$ is fixed by requiring that the value and the normal
derivative of $\bm\Phi$ match across opposite edges of $C$ under the
boundary rules. We solve this with the method of particular solutions,
locating the minimum of the Betcke--Trefethen subspace angle with interior
normalization~\cite{BetckeTrefethen2005}. With $M=40$ terms per component,
\begin{equation}
 k=2.614\,701\,921\,8,\qquad
 E_0=2k^2=13.673\,332\,280\pm10^{-9}.
\end{equation}
Between 32 and 56 terms per component, $k$ changes by less than
$10^{-10}$. The ground state is non-degenerate in each
sector: at $k$ the second subspace angle is $0.73$, while the first is about
$10^{-8}$. The coefficients are real. The leading ones are
\begin{center}
\begin{tabular}{rrrrr}
$n$ & $\nu$ & $m_n$ & $a_{1,n}$ & $a_{2,n}$\\
\hline
0 & 1/2 & 0 & $0.898935$ & $0.372351$\\
1 & 3/2 & $-2$ & $-0.184385$ & $0.445145$\\
2 & 5/2 & 2 & $0.028556$ & $-0.068941$\\
3 & 7/2 & $-4$ & $0.013241$ & $0.005485$\\
4 & 9/2 & 4 & $-0.011300$ & $-0.004681$\\
5 & 11/2 & $-6$ & $0.000563$ & $-0.001359$
\end{tabular}
\end{center}
They are normalized so that $\int_C\|\bm\Phi\|^2=1$, which makes
$g_2(\bm d)=\|\bm\Phi(\bm d)\|^2$ in the convention of
Eq.~\eqref{eq:pair-correlation}. Two relations hold numerically to
$3\times10^{-8}$ without being imposed: $c_{2,2j}=-c_{1,2j+1}$ and
$c_{2,2j+1}=c_{1,2j}$ for the unscaled coefficients
$c_{c,n}=a_{c,n}/J_{n+1/2}(k/2)$. The contact spinor points along
$(\cos\frac\pi8,\sin\frac\pi8)$, the $+1$ eigenvector of
$(\Xmat+\Zmat)/\sqrt2$. Near contact $\|\bm\Phi\|=1.86324\,r^{1/2}[1+O(r)]$,
and the next channel is $r^{3/2}e^{-2\ii\varphi}$.

\begin{figure}[htbp]
\centering
\includegraphics[width=\textwidth]{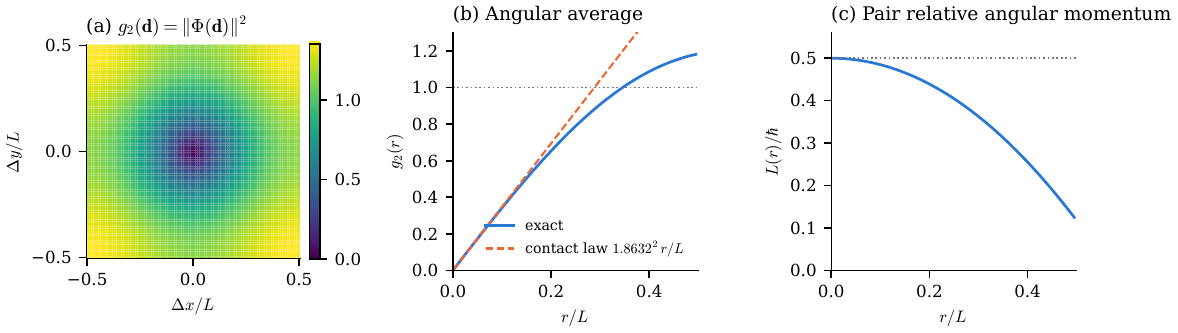}
\caption{Exact two-semion ground state on the unit torus. (a) Relative density
$\|\bm\Phi(\bm d)\|^2=g_2(\bm d)$. (b) Angular average with the contact law.
(c) Pair relative angular momentum; the dotted line is the isolated-pair value
$\alpha\hbar=\hbar/2$.}
\label{fig:n2-exact}
\end{figure}

\paragraph{Exact observables.}
Figure~\ref{fig:n2-exact} shows the exact pair correlation and the pair
relative angular momentum $L(r)$ of Eq.~\eqref{eq:pair-angular-momentum}.
For $N=2$ with $\bm K=0$, $\bm v_1-\bm v_2=(2\hbar/m)
\operatorname{Re}[\bm\Phi^\dagger(-\ii\nabla_w+\bm a)\bm\Phi]/\|\bm\Phi\|^2$.
The pair correlation rises as $1.8632^2\,r/L$ and exceeds one beyond
$r\simeq0.35L$. $L(r)$ starts at $\hbar/2$ and decreases monotonically to
$0.12\hbar$ at $r=L/2$. The radial part vanishes to $10^{-17}$. These curves
are exact references for the neural $N=2$ state.

\section{VMC estimators, optimization, and evaluation}
\label{app:vmc}

The ansatz is optimized and evaluated by variational Monte Carlo (VMC). For training through the projection, the same projected state,
Eq.~\eqref{eq:projected-ansatz}, is used for sampling, the energy, and its
gradient. Measurements use the frozen state specified for each result.

\paragraph{Sampling.}
Configurations are drawn from the Born density
$\|\Psi^{(M)}_\theta(\R)\|^2=|\Psi_1|^2+|\Psi_2|^2$ with the Metropolis
algorithm~\cite{Metropolis1953}. One sweep combines a Gaussian move of all
particles with a move of one randomly chosen particle to a uniformly drawn
position. During training, the Gaussian width is adjusted every 100 updates
from the measured acceptance, following~\cite{Pfau2020}, and is capped at
the shortest box length. Ten sweeps separate
consecutive optimization steps. Positions are wrapped into the cell only for
sampling. The density is periodic, while the amplitude itself obeys
Eqs.~\eqref{eq:boundary}.

\paragraph{Energy estimator.}
Write $\Psi=J_\eta\bm f$, where $\bm f$ is the smooth, momentum-projected
amplitude, and $u=\ln J_\eta$. Integrating the cross terms between $J_\eta$
and $\bm f$ by parts turns the form~\eqref{eq:quadratic-form} into an average
of first derivatives only,
with $e_{\rm W}$ as given in Eq.~\eqref{eq:weak-energy}.
The boundary terms at coincidences vanish for the $r^{1/2}$ cusp. Near a
collision, both $\|D_i\bm f\|^2/\|\bm f\|^2$ and $|\nabla_iu|^2$ contain a
$1/(4r^2)$ pole. The statistical connection and the rotated gradient of $u$
share the same singular part, so we subtract it analytically before summing.
The local estimator $e_{\rm W}$ has no leading isolated-pair $1/r^2$ divergence. It has the same expectation as the kinetic form, but unlike the
operator estimator $\operatorname{Re}[\Psi^\dagger H\Psi]/\|\Psi\|^2$ it does
not have zero variance at an eigenstate. It needs only first derivatives,
avoiding spatial second derivatives of the neural amplitude.

\paragraph{Optimization.}
The gradient differentiates the sampled energy expectation, numerator and norm
together, with the Born measure held fixed. It is therefore the exact gradient
of the variational energy, for the full complex two-component state. Updates
use Kronecker-factored approximate curvature
(K-FAC)~\cite{MartensGrosse2015,Pfau2020} in its kfac-jax implementation,
with the learning rate $\epsilon_t=0.05/(1+t/10^4)$, damping $10^{-3}$ and a
norm constraint $10^{-3}$, reduced to $10^{-4}$ in the final
$N=16$ stage described below. Every coefficient, attention, embedding and
displacement parameter is registered with the curvature estimator. Training
starts from random initialization with a fixed seed. The displacement head starts at zero,
so the first state uses the analytic starting frame.

\paragraph{Production training protocol.}
The pair range is fixed at $\eta=0.4a$, and the initial Gaussian width
parameter is $0.32a^2$, with $a=(\Acell/N)^{1/2}$ and $\Acell=1$.
All production states use two attention layers, four heads, two Fourier
harmonics, displacement scale $s=0.1$, and the same optimizer family
and learning-rate schedule. Training starts from random initialization
for $2\times10^4$ updates, continued to $4\times10^4$ at $N=10,12$ and to
$1.2\times10^5$ at $N=16$. The selected $N=16$ state then receives
$2\times10^4$ further updates with the K-FAC norm constraint reduced
from $10^{-3}$ to $10^{-4}$, fresh curvature estimates, and training
seed 55. This stage keeps the 120\,000-update wavefunction and walkers,
reinitializes the sampling stream with a new burn-in, and continues the
learning-rate schedule with offset 120\,000. The training grid is the smallest
$M\ge2$ coprime to $N/2$: $M=2,3,2,3,2,5,3$ for
$N=2,4,6,8,10,12,16$. The hidden width is 64 through $N=12$ and 256
for the selected $N=16$ state. We use 2048 walkers through $N=10$ and
1024 at $N=12,16$. Walkers are split over the available GPUs; the
parameters and optimizer arrays are replicated. The $N=16$ width controls
use the same seed, training grid, walker count, and $2\times10^4$ updates.
At $N=6,8$, width-128 controls lower the $\mathcal P_{2M}$ energies by
$0.27$ and $1.05$, respectively. Independent seed-53 runs through
$N=8$, evaluated on their training grids, reproduce the corresponding
seed-52 energies within $0.15$; the larger production states use one
initialization seed; the final $N=16$ stage uses the new sampling seed
stated above.

\paragraph{Evaluation.}
Energy, density, and current estimates come from a separate VMC run with frozen
parameters, a new random stream, and the sampling width of the final
checkpoint. States through $N=12$ are evaluated through $\mathcal P_{2M}$
on 4000 measurements, with evaluation seed 2 and the walker counts stated
above. The selected $N=16$ state is evaluated through $\mathcal P_{12}$
on 1000 measurements with 512 walkers. Each measurement follows ten sweeps. Statistical errors come from
blocking the measurement series~\cite{FlyvbjergPetersen1989}. We quote the
largest blocked error among block sizes that leave at least 32 blocks.
Spatial maps and profiles use errors from 32 contiguous groups of blocks. The same run measures the
center-of-mass energy~\eqref{eq:com-energy}, total momentum, and the
diagonal observables and currents of Sec.~\ref{sec:beyond-ed}.
Pair coherence is measured separately as described in
Sec.~\ref{app:pair-coherence}.

\section{Continuum exact diagonalization}
\label{app:ed}

The ED calculation is a linear Rayleigh--Ritz expansion in explicitly known
continuum functions. It uses the Hamiltonian, two global components,
and boundary conditions of Sec.~\ref{sec:model}, with the Friedrichs
collision condition described in Sec.~\ref{sec:jastrow}.
All energies below are for $L_x=L_y=1$, $\beta_x = \beta_y = 0$, and $\kappa=1$.
The basis is constructed independently of the neural wavefunction: its
sections and collision factors are fixed, and only linear coefficients are
determined by diagonalization.

\subsection{Boundary sections, collision factor, and Fourier cutoff}

An unordered configuration $\mu$ consists of $N$ integer vectors
$\bm k_i=(k_{ix},k_{iy})$. Let $\mathcal O_\mu$ be its distinct assignments
to the labeled particles. The symmetric Fourier polynomial is
\begin{equation}
 F_\mu(\R)=\frac{1}{\sqrt{|\mathcal O_\mu|}}
 \sum_{\bm k\in\mathcal O_\mu}
 \exp\!\left[2\pi\ii\sum_{i=1}^N
 \left(\frac{k_{ix}x_i}{L_x}+\frac{k_{iy}y_i}{L_y}\right)\right],
 \qquad \lambda_\mu=\sum_i(k_{ix}^2+k_{iy}^2)\le\Lambda.
 \label{eq:ed-fourier}
\end{equation}
Every configuration satisfying this inequality is included. Thus $\Lambda$
truncates coefficient functions.
The fixed factors below also carry spatial structure, so $\Lambda$ is not
a sharp Fourier cutoff on the full physical wavefunction.

To supply the boundary transport, periodize a product Gaussian with the
matrices $\Xmat$ and $\Zmat$ of Eq.~\eqref{eq:transport-matrices}. With
$s_i=Nx_i-\sum_jx_j$, define the $2\times2$ matrix
\begin{align}
 U_\sigma(\R)&=Y_\sigma(\R)X_\sigma(\R),\nonumber\\
 X_\sigma(\R)&=\prod_{i=1}^N\left[
 \sum_{n\in\mathbb Z}e^{-(x_i-nL_x)^2/(2\sigma^2L_x^2)}\Xmat^n\right],\nonumber\\
 Y_\sigma(\R)&=\prod_{i=1}^N\left[
 \sum_{m\in\mathbb Z}e^{-(y_i-mL_y)^2/(2\sigma^2L_y^2)}
 e^{\ii\pi m s_i/L_x}\Zmat^m\right].
 \label{eq:ed-sections}
\end{align}
Each column is permutation symmetric and obeys the required two-component
transport. Relabeling the Gaussian images proves the $y$ rule directly;
for the $x$ rule, $\Xmat\Zmat=-\Zmat\Xmat$ cancels the change in the
$y$-image phases. The product form evaluates the many-particle image sum
through one-dimensional sums and provides analytic first derivatives.
The reported bases use fixed width $\sigma=0.3$, images $|m|,|n|\le6$,
and 16 trapezoidal nodes per direction for the projection below.

The ED collision factor is the range-free function
\begin{equation}
 J_{\rm ED}(\R)=\prod_{i<j}d_{ij}^{1/2},\qquad
 \bm\phi_{\mu b}(\R)=J_{\rm ED}(\R)\,
 P_{\bm K}\!\left[F_\mu(\R)U_\sigma(\R)\bm e_b\right],
 \quad b=1,2,
 \label{eq:ed-basis}
\end{equation}
where $d_{ij}$ is the chord distance of Sec.~\ref{sec:jastrow} and $\bm e_b$
is a component unit vector. This factor has the same isolated-pair
$r^{1/2}$ cusp as the neural factor $J_\eta$, but has no range parameter.
It places the basis functions in the same Friedrichs form domain.
The collective translation projection is
\begin{equation}
 P_{\bm K}\bm f(\R)=\frac{1}{\Acell}\int_{T^2}d^2t\,
 e^{-\ii\bm K\cdot\bm t}\bm f(\R+\bm t),\qquad
 K_a=\frac{2\pi n_a}{L_a}.
 \label{eq:ed-projection}
\end{equation}
The comparisons use $\bm K=0$. Since $F_\mu$ has known total Fourier
momentum and $s_i$ is invariant under collective translation, this
projection separates into two one-dimensional integrals over the Gaussian
blocks. The translation nodes approximate these smooth integrals.
Increasing their number from 16 to 24 and the image cutoff from 6 to 7
changes tested $N=4$ columns and first derivatives through $\Lambda=8$
by less than $2\times10^{-11}$ in relative norm.

There are two nominal columns per Fourier configuration. At $N=4$, the
$\Lambda=4,5,6$ spaces contain 288, 576, and 992 columns. At $N=6$,
$\Lambda=6,7,8,9$ gives 1552, 3072, 5526, and 9558 columns. These counts precede
the removal of overlap dependencies.

The $N=4$ entry marked $+T$ in Table~\ref{tab:validation} also includes the
regular $\ell=-2$ channel of Sec.~\ref{app:collisions}. For a minimum-image
pair displacement $z_{ij}=\delta x_{ij}+\ii\delta y_{ij}$, let
$r_{ij}=|z_{ij}|$ and
\begin{equation}
 t_{ij}=\chi(r_{ij})\frac{\overline z_{ij}^{\,2}}{r_0r_{ij}},\qquad
 T(\R)=\sum_{i<j}t_{ij},\qquad r_0=0.4L.
 \label{eq:ed-channel}
\end{equation}
Here $\chi(r)=1$ for $r\le r_0/2$, $\chi(r)=0$ for $r\ge r_0$, and
$\chi(r)=1-h(2r/r_0-1)$ in between, with
$h(v)=e^{-1/v}/[e^{-1/v}+e^{-1/(1-v)}]$ for $0<v<1$.
The support lies inside the minimum-image cell, making $T$ periodic;
it is also symmetric and invariant under collective translation.
The added columns are $T\bm\phi_{\mu b}$ with their own coefficient cutoff
$\lambda_\mu\le\Lambda_T$. Their pair contribution has the leading form
$r^{3/2}e^{-2\ii\varphi}$, giving this channel an independent amplitude.
The bounded first derivatives of $t_{ij}$ keep the columns in the form domain.
The quoted $N=4$ result uses $\Lambda=6$ and $\Lambda_T=5$, for
$992+576=1568$ nominal columns. The reported $N=6$ calculations use the
base family of Eq.~\eqref{eq:ed-basis}, without $T$.

\subsection{Continuum matrices and numerical validation}

Writing $\Psi=\sum_a c_a\bm\phi_a$, assemble
\begin{equation}
 S_{ab}=\int d\R\,\bm\phi_a^\dagger\bm\phi_b,\qquad
 H_{ab}=\kappa\int d\R\,\sum_i
 (D_i\bm\phi_a)^\dagger\!\cdot(D_i\bm\phi_b),\qquad Hc=ESc.
 \label{eq:ed-matrices}
\end{equation}
All component contractions use the full two-component state. Derivatives of
the Gaussian sums, Fourier polynomials, and collision factors are analytic.
At fixed total momentum the integrands are invariant under collective
translation. Fixing $\rvec_N=0$ therefore removes one position integral;
the remaining $2(N-1)$ coordinates are integrated continuously, with the
common area factor restored.

For $\bm\phi_a=J_{\rm ED}\bm f_a$, integration by parts gives
\begin{equation}
 \frac{H_{ab}}{\kappa}=\int d\R\,J_{\rm ED}^2\left[
 \sum_i(D_i\bm f_a)^\dagger\!\cdot(D_i\bm f_b)
 -\frac{\Delta J_{\rm ED}}{J_{\rm ED}}\bm f_a^\dagger\bm f_b\right],
 \qquad \Delta=\sum_i\nabla_i^2.
 \label{eq:ed-reduced}
\end{equation}
Torus-face terms cancel by boundary covariance, and the collision terms
vanish with the regular pair cusp. This is the original kinetic form,
with no additional scalar interaction. With $\bm g_i=\nabla_i\ln J_{\rm ED}$,
the scalar term in the expanded kernel is
$\sum_i(|\avec_i|^2-|\bm g_i|^2)-\Delta\ln J_{\rm ED}$.
The equal leading $1/(4r^2)$ pair poles of $|\avec_i|^2$ and
$|\bm g_i|^2$ cancel before quadrature. This reduces integration scatter
substantially compared with directly integrating the positive form.

The reduced matrices use uniform scrambled Sobol points, assembled
independently for each scramble and averaged before diagonalization.
Independent scrambles and increasing point counts test the matrix integrals.
We also freeze a vector fitted without a given scramble and evaluate its
Rayleigh quotient with the excluded matrices. A separate check freezes the
averaged-matrix vector and evaluates the original positive form on new
points, without another eigensolve. For this latter check, half of the
integration measure is uniform and half samples close pairs. The pair
component has uniform angle and radial density proportional to
$(1-r/R)^2$ on $0<r<R=\min(L_x,L_y)/2$; exact mixture weights restore the
continuum integral. Permutation symmetry allows separate integration of
the uniform term and one representative pair term.

The basis is nonorthogonal and has exact and near dependencies. First scale
with $B_{aa}=S_{aa}^{-1/2}$ and diagonalize $\widetilde S=BSB$.
Keep overlap eigenvalues $s_j>\epsilon s_{\max}$ and canonically
orthogonalize with $W=BV_{\rm keep}s_{\rm keep}^{-1/2}$. Diagonalizing
$W^\dagger HW$ gives the retained Ritz spectrum and $c=Wv$.
Both the retained rank and the sensitivity to $\epsilon$ are recorded.

\begin{table}[htbp]
\begin{ruledtabular}
\begin{tabular}{rrrrrr}
$\Lambda$ & Columns & Rank & $2^{16}$ points & $2^{18}$ points & Excluded-scramble checks\\
\hline
4 & 288 & 178 & 73.85105 & 73.83746 & 73.83481--73.84084\\
5 & 576 & 320 & 73.18609 & 73.17819 & 73.17577--73.18012\\
\end{tabular}
\end{ruledtabular}
\caption{Refined $N=4$, $\bm K=0$ base-basis results. Each point count uses
four independent reduced-kernel scrambles; the energies diagonalize their
averaged matrices with $\epsilon=10^{-12}$. The last column evaluates vectors
selected using only the other three fine scrambles. Its range describes
integration variation, rather than a statistical confidence interval.}
\label{tab:ed-n4}
\end{table}

Table~\ref{tab:ed-n4} separates fixed-basis integration refinement from the
larger change on increasing $\Lambda$. Changing $\epsilon$ from $10^{-9}$
to $10^{-12}$ changes the fine energies from $73.87399$ to $73.83746$ at
$\Lambda=4$ and from $73.17858$ to $73.17819$ at $\Lambda=5$.
Thresholds $10^{-12}$ and $10^{-13}$ retain identical spaces.

Original-positive-form checks on independently frozen $10^{-9}$ vectors
agree at the few-hundredths level. Deterministic two-particle Gauss
quadrature also verifies agreement between the original and reduced kernels.

The $N=4$, $\Lambda=6+T(5)$ calculation uses two $2^{16}$-point scrambles
and $\epsilon=10^{-12}$, retaining 819 of 1568 columns. The averaged-matrix
value is $72.4283$; vectors fitted to one scramble and evaluated on the
other give $72.4807$ and $72.5017$. These checks place the integration
variation on the scale of a few hundredths. The remaining Fourier-cutoff
dependence is larger: at the same integration settings, the $\Lambda=5+T(5)$
value is $72.7924$.

\begin{table}[htbp]
\begin{ruledtabular}
\begin{tabular}{rrrrr}
$\Lambda$ & Matrix points & Fitted Ritz value & Excluded-scramble checks & Positive-form checks\\
\hline
8 & $2^{15}$ & 177.0216 & 179.4134, 180.2734 & 178.2115, 179.1292\\
8 & $2^{16}$ & 177.5448 & 178.6667, 179.3083 & 178.2370, 178.7453\\
9 & $2^{15}$ & 174.1903 & 177.9728, 178.9372 & 176.5430, 176.7334\\
9 & $2^{16}$ & 174.9042 & 176.7166, 177.3904 & 176.2013, 176.3074\\
\end{tabular}
\end{ruledtabular}
\caption{$N=6$, $\bm K=0$ base-basis calculations. The first two columns
identify the Fourier cutoff and the point count per reduced-matrix scramble.
Two independent matrix scrambles are used. Excluded-scramble checks evaluate
each single-scramble vector with the other matrices. The last column gives
two independent original-positive-form evaluations of the frozen
averaged-matrix vector; no coefficients are refitted for these checks.}
\label{tab:ed-n6}
\end{table}

For $N=6$, Table~\ref{tab:ed-n6} reports the completed $\Lambda=8,9$ runs.
Their matrix scrambles use seeds 9101 and 9102. The mean of the two fine
$\Lambda=9$ positive-form evaluations is $176.2543$, quoted as about
$176.3$ in Table~\ref{tab:validation}. Averaging the independently evaluated
Rayleigh quotients here gives a practical finite-state reference. The
roughly $1.35$ difference from its fitted value $174.9042$ demonstrates
why matrix fitting and state evaluation must be distinguished. The
coarser run gives a positive-form mean $176.6382$, so integration refinement
also changes the selected state on the scale of a few tenths. The two
validation values alone do not determine a confidence interval.

The Fourier cutoff, numerical integration, overlap threshold, Gaussian
width, and momentum sector are separate convergence controls. In particular,
the present $N=6$ sequence still changes with both cutoff and integration.
A sharper ED reference requires their joint refinement, followed by
independent positive-form and overlap-threshold checks at the desired
energy resolution. Establishing the global ground sector additionally
requires comparison with the inequivalent nonzero-momentum sectors.
For exact integrals the conforming trial spaces give variational upper
bounds; the numerical values above are quadrature estimates of those
finite-basis energies.

\section{Compact-pair coherence: torus string, channels, and controls}
\label{app:pair-coherence}

This section gives the compact-pair estimator used in
Sec.~\ref{sec:pair-coherence} and its numerical checks. The construction
adapts the local-pair prescription of Ref.~\cite{ZhangIdealAnyonGas2026}
to the two-component torus state. Its normalization depends on the stated
internal-pair selection and does not define an absolute condensate fraction.

\subsection{Local statistical string in unwrapped coordinates}

For a pair $(i,j)$, choose the minimum-image internal vector
$\bm\varepsilon=\rvec_i-\rvec_j$ and use the periodic copy of
particle $j$ nearest to $i$.
This gives $\widetilde\R$, with center $\bm c=\rvec_i-\bm\varepsilon/2$.
The moved configuration $\R'$ puts the members at
$\bm c+\bm d\pm\mathsf R_\phi\bm\varepsilon/2$, where
$\mathsf R_\phi$ is a planar rotation, and leaves all spectators fixed.
Both endpoint wavefunctions are evaluated in unwrapped coordinates,
without wrapping the moved members back into the primitive cell.

Use $f(z)$ from Eq.~\eqref{eq:theta}, and write
$t(z)=f(z)/|f(z)|$ and $c=c_x+\ii c_y$. The pair--spectator phase is
\begin{equation}
 \mathcal E_{ij}(\R)=\prod_{k\ne i,j}t(c-z_k)\,
 \exp\!\left\{\frac{\ii}{2}\sum_{k\ne i,j}\sum_{m=i,j}
       \operatorname{Arg}\frac{f(z_m-z_k)}{f(c-z_k)}\right\}.
 \label{eq:supp-pair-string}
\end{equation}
Here $\operatorname{Arg}$ is the principal branch. Splitting each
pair--spectator contribution to $\chi$ as in
Eq.~\eqref{eq:pair-phase-split} gives one center term with integer
coefficient and two fractional member-to-center ratios. The fractional
branch is confined near the compact pair rather than extended along two
separate particle trajectories. The internal term of $\chi$ is omitted;
it is unchanged by translation, and its omission for rotations defines
the bosonic-gauge angular labels of Eq.~\eqref{eq:coherence-channels}.

Theta quasi-periodicity compensates the scalar boundary phases in the
endpoint overlap. The component transport matrices cancel when the same
periodic copies are used at both endpoints, so no additional $2\times2$ matrix is
needed in Eq.~\eqref{eq:pair-overlap}. Identity moves give $F_{ij}=1$,
and full-cycle pair translations give
$C_2(\bm d+L_\mu\hat{\bm e}_\mu,0)=C_2(\bm d,0)$ at zero twists.
This revival is a transport identity; the physical evidence comes from
the correlations within the cell and their size dependence.

Continuing the statistical phase member by member along a straight move
adds $(-1)$ for each branch crossing of the member ratios. For sufficiently
compact pairs these are spectators passing through the internal segment.
We measure this continuation as a control. In the outer shell, the
farthest diagonal correlator at $N=12$ changes from about $0.105$ with
the local-pair string to $0.031$ with member-by-member continuation;
at $N=16$ (80\,000 updates) it changes from $0.030$ to $0.008$.
The statistical dressing of the operator therefore matters quantitatively.

\subsection{Sampling and independent checks}

We sample $\|\Psi\|^2$ of the full evaluated state
$\Psi=J\mathcal P_g f_\theta$. For each configuration and shell,
one pair is drawn uniformly among the shell's pairs and weighted by their
number, giving an unbiased estimate of the sums in
Eq.~\eqref{eq:pair-coherence}. Shells are $[0.25,0.45]a$ and
$[0.45,0.65]a$. Translations use six equally spaced steps to
$(L/2,0)$, $(0,L/2)$, and $(L/2,L/2)$.
The angular transform uses eight angles $\phi_j=\pi j/8$ at the origin
and at these farthest points. The reported weights thus resolve even
harmonics modulo 16; the resolved weight is dominated by $\ell=0,-2,+2$.

A shell-preserving translation or rotation satisfies the density-ratio
identity
\begin{equation}
 \frac{\left\langle\sum_{i<j}\omega(r_{ij})
       \|\Psi(\R')\|^2/\|\Psi(\widetilde\R)\|^2\right\rangle}
      {\left\langle\sum_{i<j}\omega(r_{ij})\right\rangle}=1.
 \label{eq:pair-density-check}
\end{equation}
This checks normalization independently of the component overlap.
The ratio has a slowly converging tail from close pair--spectator
configurations. Some inner-shell rotated moves of the 100\,000-update
$\mathcal P_{12}$ evaluation give values below one by up to $0.16$,
while the corresponding channel coherences agree with $\mathcal P_6$
to $0.007$. Blocked statistical errors are therefore assessed alongside
independent integration and repeated-state comparisons.

Structural tests cover identity, exchange, full-cycle revival,
periodicity, and invariance under alternative periodic copies of members and spectators.
Finite differences of the theta phase recover the production connection,
and path integration verifies the branch-crossing sign.
At $N=4$, four scrambles of $2^{14}$ uniform Sobol points independently
reproduce the Monte Carlo overlaps: the outer-shell $x$-translation by
$L/2$ gives $0.4250(23)$ by quadrature and $0.4253(28)$ by Monte Carlo.
The largest discrepancy among the checked moves is about three combined
standard errors.

The $N=4,6$ measurements use 256 walkers and 200 measurements;
$N=8,10$ use 2048 walkers and 1000 measurements; $N=12$ uses
1024 walkers and 1000 measurements; the selected $N=16$ state uses
512 walkers and 1000 measurements at $\mathcal P_{12}$. Errors use 32 contiguous groups,
including numerator--denominator covariance. The $N=16$ controls at
$\mathcal P_6$ use 1024 walkers; the 60\,000-update control has
955 measurements. All wavefunctions are frozen;
no pairing objective enters their optimization.

For the magnitude in Fig.~\ref{fig:pair-coherence}(b,c), we first average
the complex overlap and then take its modulus. Its statistical error uses
a delete-one-group jackknife over the same 32 contiguous groups,
recomputing the complex ratio and its magnitude after omitting each group.
This retains correlations between the real and imaginary parts and the
normalization. The channel panels retain the signed real parts.

\subsection{Chiral structure and variational accuracy}

The contact channel contains most compact-pair weight, but the
$\ell=-2$ population greatly exceeds $\ell=+2$. The physical labels
$-3/2$ and $+5/2$ have different centrifugal costs. Independent references
reproduce this imbalance. The exact $N=2$ state has inner/outer-shell
weights $0.054/0.125$ for $\ell=-2$ and $0.0011/0.0051$ for $\ell=+2$.
At $N=4$, continuum ED with $\Lambda=6$ and the additional collision
channel gives $0.084/0.192$ and $0.0052/0.0160$, respectively, compared
with neural values $0.108/0.202$ and $0.0052/0.0159$.
For the outer shell, ED retains
$\operatorname{Re}\Gamma_{-2}\simeq0.80/0.77$ at the farthest
$x$/diagonal points, close to the neural values $0.85/0.80$.
Thus independently constructed finite-system states share the same chiral
pair structure and channel dependence of coherence.

\begin{figure}[t]
\centering
\includegraphics[width=0.95\textwidth]{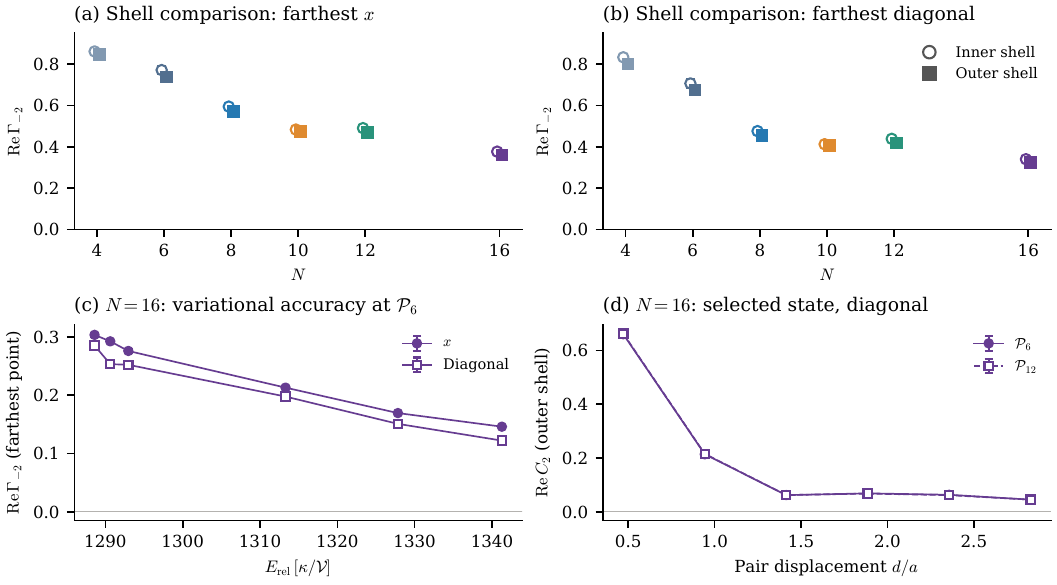}
\caption{Pair-coherence controls. (a,b) Normalized $\ell=-2$ coherence
at the farthest $x$ and diagonal points for the states in
Fig.~\ref{fig:pair-coherence}, comparing inner-shell open circles
and outer-shell filled squares. (c) Outer-shell coherence of $N=16$
states against their frozen relative-motion energy at $\mathcal P_6$:
widths 64 and 128, and width 256 at 20\,000, 60\,000, 80\,000,
and 100\,000 updates. Circles denote $x$ and squares the diagonal;
lines guide the eye. (d) Total outer-shell diagonal correlator of the
selected $N=16$ checkpoint at $\mathcal P_6$ and $\mathcal P_{12}$. Error bars show
two blocked statistical standard errors.}
\label{fig:pair-coherence-controls}
\end{figure}

Shell independence of $\Gamma_{-2}$ holds across the size series
[Fig.~\ref{fig:pair-coherence-controls}(a,b)]. At $N=8$, a second random
start gives farthest-point outer-shell coherence $0.58/0.45$
($x$/diagonal), and width 128 gives $0.58/0.47$, compared with
$0.57/0.45$ in the main state. The signal survives both controls.

At $N=16$, improving the state through wider networks and longer
training raises the coherence at fixed evaluation grid
[Fig.~\ref{fig:pair-coherence-controls}(c)]. From 80\,000 to
100\,000 updates in the width-256 run, the farthest-point values
increase from $0.292/0.253$ to $0.303/0.285$.
The subsequent 120\,000-update state gives $0.311/0.281$ at the same
pair-coherence grid. Another 20\,000 updates with norm constraint
$10^{-4}$ raise these values to $0.358/0.320$ in the selected state at
$\mathcal P_6$.
At $\mathcal P_{12}$, this stage lowers the energy from
$1275.83(103)$ to $1266.91(109)$, or $E/(N\pi n\kappa)=1.5753$.
The main energy and observable results use the same final checkpoint
at $\mathcal P_{12}$.

On the selected checkpoint, increasing the evaluation grid from
$\mathcal P_6$ to $\mathcal P_{12}$ changes the outer-shell channel
coherence from $0.358/0.320$ to $0.360/0.323$. The real part of
$C_2$ changes by at most $0.003$ at the measured translations in either shell
[Fig.~\ref{fig:pair-coherence-controls}(d)]. The main $N=16$ results use
$\mathcal P_{12}$ throughout. The earlier 100\,000-update checkpoint
also showed stable pair coherence under this grid change, while its
energy fell from $1293.88(63)$ to $1279.83(105)$ and its center-of-mass
energy from $5.28$ to $1.62$. Finite-size scaling should track both
displacement in spacing units and variational accuracy, using larger
cells to determine the long-distance limit.


\begin{thebibliography}{84}%
\makeatletter
\providecommand \@ifxundefined [1]{%
 \@ifx{#1\undefined}
}%
\providecommand \@ifnum [1]{%
 \ifnum #1\expandafter \@firstoftwo
 \else \expandafter \@secondoftwo
 \fi
}%
\providecommand \@ifx [1]{%
 \ifx #1\expandafter \@firstoftwo
 \else \expandafter \@secondoftwo
 \fi
}%
\providecommand \natexlab [1]{#1}%
\providecommand \enquote  [1]{``#1''}%
\providecommand \bibnamefont  [1]{#1}%
\providecommand \bibfnamefont [1]{#1}%
\providecommand \citenamefont [1]{#1}%
\providecommand \href@noop [0]{\@secondoftwo}%
\providecommand \href [0]{\begingroup \@sanitize@url \@href}%
\providecommand \@href[1]{\@@startlink{#1}\@@href}%
\providecommand \@@href[1]{\endgroup#1\@@endlink}%
\providecommand \@sanitize@url [0]{\catcode `\\12\catcode `\$12\catcode
  `\&12\catcode `\#12\catcode `\^12\catcode `\_12\catcode `\%12\relax}%
\providecommand \@@startlink[1]{}%
\providecommand \@@endlink[0]{}%
\providecommand \url  [0]{\begingroup\@sanitize@url \@url }%
\providecommand \@url [1]{\endgroup\@href {#1}{\urlprefix }}%
\providecommand \urlprefix  [0]{URL }%
\providecommand \Eprint [0]{\href }%
\providecommand \doibase [0]{https://doi.org/}%
\providecommand \selectlanguage [0]{\@gobble}%
\providecommand \bibinfo  [0]{\@secondoftwo}%
\providecommand \bibfield  [0]{\@secondoftwo}%
\providecommand \translation [1]{[#1]}%
\providecommand \BibitemOpen [0]{}%
\providecommand \bibitemStop [0]{}%
\providecommand \bibitemNoStop [0]{.\EOS\space}%
\providecommand \EOS [0]{\spacefactor3000\relax}%
\providecommand \BibitemShut  [1]{\csname bibitem#1\endcsname}%
\let\auto@bib@innerbib\@empty
\bibitem [{\citenamefont {Leinaas}\ and\ \citenamefont
  {Myrheim}(1977)}]{LeinaasMyrheim1977}%
  \BibitemOpen
  \bibfield  {author} {\bibinfo {author} {\bibfnamefont {J.~M.}\ \bibnamefont
  {Leinaas}}\ and\ \bibinfo {author} {\bibfnamefont {J.}~\bibnamefont
  {Myrheim}},\ }\bibfield  {title} {\bibinfo {title} {On the theory of
  identical particles},\ }\href {https://doi.org/10.1007/BF02727953} {\bibfield
   {journal} {\bibinfo  {journal} {Il Nuovo Cimento B (1971-1996)}\ }\textbf
  {\bibinfo {volume} {37}},\ \bibinfo {pages} {1} (\bibinfo {year}
  {1977})}\BibitemShut {NoStop}%
\bibitem [{\citenamefont {Wilczek}(1982{\natexlab{a}})}]{Wilczek1982}%
  \BibitemOpen
  \bibfield  {author} {\bibinfo {author} {\bibfnamefont {F.}~\bibnamefont
  {Wilczek}},\ }\bibfield  {title} {\bibinfo {title} {Quantum {Mechanics} of
  {Fractional}-{Spin} {Particles}},\ }\href
  {https://doi.org/10.1103/PhysRevLett.49.957} {\bibfield  {journal} {\bibinfo
  {journal} {Physical Review Letters}\ }\textbf {\bibinfo {volume} {49}},\
  \bibinfo {pages} {957} (\bibinfo {year} {1982}{\natexlab{a}})}\BibitemShut
  {NoStop}%
\bibitem [{\citenamefont {Wu}(1984)}]{Wu1984}%
  \BibitemOpen
  \bibfield  {author} {\bibinfo {author} {\bibfnamefont {Y.-S.}\ \bibnamefont
  {Wu}},\ }\bibfield  {title} {\bibinfo {title} {General {Theory} for {Quantum}
  {Statistics} in {Two} {Dimensions}},\ }\href
  {https://doi.org/10.1103/PhysRevLett.52.2103} {\bibfield  {journal} {\bibinfo
   {journal} {Physical Review Letters}\ }\textbf {\bibinfo {volume} {52}},\
  \bibinfo {pages} {2103} (\bibinfo {year} {1984})}\BibitemShut {NoStop}%
\bibitem [{\citenamefont {Kalmeyer}\ and\ \citenamefont
  {Laughlin}(1987)}]{KalmeyerLaughlin87}%
  \BibitemOpen
  \bibfield  {author} {\bibinfo {author} {\bibfnamefont {V.}~\bibnamefont
  {Kalmeyer}}\ and\ \bibinfo {author} {\bibfnamefont {R.~B.}\ \bibnamefont
  {Laughlin}},\ }\bibfield  {title} {\bibinfo {title} {Equivalence of the
  resonating-valence-bond and fractional quantum hall states},\ }\href
  {https://doi.org/10.1103/PhysRevLett.59.2095} {\bibfield  {journal} {\bibinfo
   {journal} {Phys. Rev. Lett.}\ }\textbf {\bibinfo {volume} {59}},\ \bibinfo
  {pages} {2095} (\bibinfo {year} {1987})}\BibitemShut {NoStop}%
\bibitem [{\citenamefont {Wen}\ \emph {et~al.}(1989)\citenamefont {Wen},
  \citenamefont {Wilczek},\ and\ \citenamefont {Zee}}]{WWZ89}%
  \BibitemOpen
  \bibfield  {author} {\bibinfo {author} {\bibfnamefont {X.~G.}\ \bibnamefont
  {Wen}}, \bibinfo {author} {\bibfnamefont {F.}~\bibnamefont {Wilczek}},\ and\
  \bibinfo {author} {\bibfnamefont {A.}~\bibnamefont {Zee}},\ }\bibfield
  {title} {\bibinfo {title} {Chiral spin states and superconductivity},\ }\href
  {https://doi.org/10.1103/PhysRevB.39.11413} {\bibfield  {journal} {\bibinfo
  {journal} {Phys. Rev. B}\ }\textbf {\bibinfo {volume} {39}},\ \bibinfo
  {pages} {11413} (\bibinfo {year} {1989})}\BibitemShut {NoStop}%
\bibitem [{\citenamefont {Laughlin}\ and\ \citenamefont
  {Zou}(1990)}]{LaughlinZou90}%
  \BibitemOpen
  \bibfield  {author} {\bibinfo {author} {\bibfnamefont {R.~B.}\ \bibnamefont
  {Laughlin}}\ and\ \bibinfo {author} {\bibfnamefont {Z.}~\bibnamefont {Zou}},\
  }\bibfield  {title} {\bibinfo {title} {Properties of the chiral-spin-liquid
  state},\ }\href {https://doi.org/10.1103/PhysRevB.41.664} {\bibfield
  {journal} {\bibinfo  {journal} {Phys. Rev. B}\ }\textbf {\bibinfo {volume}
  {41}},\ \bibinfo {pages} {664} (\bibinfo {year} {1990})}\BibitemShut
  {NoStop}%
\bibitem [{\citenamefont {Bauer}\ \emph {et~al.}(2014)\citenamefont {Bauer},
  \citenamefont {Cincio}, \citenamefont {Keller}, \citenamefont {Dolfi},
  \citenamefont {Vidal}, \citenamefont {Trebst},\ and\ \citenamefont
  {Ludwig}}]{bauer2014chiral}%
  \BibitemOpen
  \bibfield  {author} {\bibinfo {author} {\bibfnamefont {B.}~\bibnamefont
  {Bauer}}, \bibinfo {author} {\bibfnamefont {L.}~\bibnamefont {Cincio}},
  \bibinfo {author} {\bibfnamefont {B.~P.}\ \bibnamefont {Keller}}, \bibinfo
  {author} {\bibfnamefont {M.}~\bibnamefont {Dolfi}}, \bibinfo {author}
  {\bibfnamefont {G.}~\bibnamefont {Vidal}}, \bibinfo {author} {\bibfnamefont
  {S.}~\bibnamefont {Trebst}},\ and\ \bibinfo {author} {\bibfnamefont {A.~W.}\
  \bibnamefont {Ludwig}},\ }\bibfield  {title} {\bibinfo {title} {Chiral spin
  liquid and emergent anyons in a kagome lattice mott insulator},\ }\href@noop
  {} {\bibfield  {journal} {\bibinfo  {journal} {Nature communications}\
  }\textbf {\bibinfo {volume} {5}},\ \bibinfo {pages} {5137} (\bibinfo {year}
  {2014})}\BibitemShut {NoStop}%
\bibitem [{\citenamefont {Laughlin}(1983)}]{Laughlin1983}%
  \BibitemOpen
  \bibfield  {author} {\bibinfo {author} {\bibfnamefont {R.~B.}\ \bibnamefont
  {Laughlin}},\ }\bibfield  {title} {\bibinfo {title} {Anomalous {Quantum}
  {Hall} {Effect}: {An} {Incompressible} {Quantum} {Fluid} with {Fractionally}
  {Charged} {Excitations}},\ }\href
  {https://doi.org/10.1103/PhysRevLett.50.1395} {\bibfield  {journal} {\bibinfo
   {journal} {Physical Review Letters}\ }\textbf {\bibinfo {volume} {50}},\
  \bibinfo {pages} {1395} (\bibinfo {year} {1983})}\BibitemShut {NoStop}%
\bibitem [{\citenamefont {Arovas}\ \emph {et~al.}(1984)\citenamefont {Arovas},
  \citenamefont {Schrieffer},\ and\ \citenamefont {Wilczek}}]{Arovas1984}%
  \BibitemOpen
  \bibfield  {author} {\bibinfo {author} {\bibfnamefont {D.}~\bibnamefont
  {Arovas}}, \bibinfo {author} {\bibfnamefont {J.~R.}\ \bibnamefont
  {Schrieffer}},\ and\ \bibinfo {author} {\bibfnamefont {F.}~\bibnamefont
  {Wilczek}},\ }\bibfield  {title} {\bibinfo {title} {Fractional {Statistics}
  and the {Quantum} {Hall} {Effect}},\ }\href
  {https://doi.org/10.1103/PhysRevLett.53.722} {\bibfield  {journal} {\bibinfo
  {journal} {Physical Review Letters}\ }\textbf {\bibinfo {volume} {53}},\
  \bibinfo {pages} {722} (\bibinfo {year} {1984})}\BibitemShut {NoStop}%
\bibitem [{\citenamefont {Halperin}(1984)}]{Halperin84}%
  \BibitemOpen
  \bibfield  {author} {\bibinfo {author} {\bibfnamefont {B.~I.}\ \bibnamefont
  {Halperin}},\ }\bibfield  {title} {\bibinfo {title} {Statistics of
  quasiparticles and the hierarchy of fractional quantized hall states},\
  }\href {https://doi.org/10.1103/PhysRevLett.52.1583} {\bibfield  {journal}
  {\bibinfo  {journal} {Phys. Rev. Lett.}\ }\textbf {\bibinfo {volume} {52}},\
  \bibinfo {pages} {1583} (\bibinfo {year} {1984})}\BibitemShut {NoStop}%
\bibitem [{\citenamefont {Wen}\ and\ \citenamefont {Zee}(1992)}]{WenZee92}%
  \BibitemOpen
  \bibfield  {author} {\bibinfo {author} {\bibfnamefont {X.~G.}\ \bibnamefont
  {Wen}}\ and\ \bibinfo {author} {\bibfnamefont {A.}~\bibnamefont {Zee}},\
  }\bibfield  {title} {\bibinfo {title} {Shift and spin vector: New topological
  quantum numbers for the hall fluids},\ }\href
  {https://doi.org/10.1103/PhysRevLett.69.953} {\bibfield  {journal} {\bibinfo
  {journal} {Phys. Rev. Lett.}\ }\textbf {\bibinfo {volume} {69}},\ \bibinfo
  {pages} {953} (\bibinfo {year} {1992})}\BibitemShut {NoStop}%
\bibitem [{\citenamefont {Kallin}(1993)}]{Callin1993}%
  \BibitemOpen
  \bibfield  {author} {\bibinfo {author} {\bibfnamefont {C.}~\bibnamefont
  {Kallin}},\ }\bibfield  {title} {\bibinfo {title} {Flux quantization for
  semions on a torus},\ }\href {https://doi.org/10.1103/PhysRevB.48.13742}
  {\bibfield  {journal} {\bibinfo  {journal} {Phys. Rev. B}\ }\textbf {\bibinfo
  {volume} {48}},\ \bibinfo {pages} {13742} (\bibinfo {year}
  {1993})}\BibitemShut {NoStop}%
\bibitem [{\citenamefont {Laughlin}(1988)}]{Laughlin88}%
  \BibitemOpen
  \bibfield  {author} {\bibinfo {author} {\bibfnamefont {R.~B.}\ \bibnamefont
  {Laughlin}},\ }\bibfield  {title} {\bibinfo {title} {Superconducting ground
  state of noninteracting particles obeying fractional statistics},\ }\href
  {https://doi.org/10.1103/PhysRevLett.60.2677} {\bibfield  {journal} {\bibinfo
   {journal} {Phys. Rev. Lett.}\ }\textbf {\bibinfo {volume} {60}},\ \bibinfo
  {pages} {2677} (\bibinfo {year} {1988})}\BibitemShut {NoStop}%
\bibitem [{\citenamefont {Fetter}\ \emph {et~al.}(1989)\citenamefont {Fetter},
  \citenamefont {Hanna},\ and\ \citenamefont {Laughlin}}]{FetterHannaLaughlin}%
  \BibitemOpen
  \bibfield  {author} {\bibinfo {author} {\bibfnamefont {A.~L.}\ \bibnamefont
  {Fetter}}, \bibinfo {author} {\bibfnamefont {C.~B.}\ \bibnamefont {Hanna}},\
  and\ \bibinfo {author} {\bibfnamefont {R.~B.}\ \bibnamefont {Laughlin}},\
  }\bibfield  {title} {\bibinfo {title} {Random-phase approximation in the
  fractional-statistics gas},\ }\href
  {https://doi.org/10.1103/PhysRevB.39.9679} {\bibfield  {journal} {\bibinfo
  {journal} {Phys. Rev. B}\ }\textbf {\bibinfo {volume} {39}},\ \bibinfo
  {pages} {9679(R)} (\bibinfo {year} {1989})}\BibitemShut {NoStop}%
\bibitem [{\citenamefont {Chen}\ \emph {et~al.}(1989)\citenamefont {Chen},
  \citenamefont {Wilczek}, \citenamefont {Witten},\ and\ \citenamefont
  {Halperin}}]{ChenWilczekWittenHalperin1989}%
  \BibitemOpen
  \bibfield  {author} {\bibinfo {author} {\bibfnamefont {Y.-H.}\ \bibnamefont
  {Chen}}, \bibinfo {author} {\bibfnamefont {F.}~\bibnamefont {Wilczek}},
  \bibinfo {author} {\bibfnamefont {E.}~\bibnamefont {Witten}},\ and\ \bibinfo
  {author} {\bibfnamefont {B.~I.}\ \bibnamefont {Halperin}},\ }\bibfield
  {title} {\bibinfo {title} {On anyon superconductivity},\ }\href
  {https://doi.org/10.1142/S0217979289000725} {\bibfield  {journal} {\bibinfo
  {journal} {Int. J. Mod. Phys. B}\ }\textbf {\bibinfo {volume} {3}},\ \bibinfo
  {pages} {1001} (\bibinfo {year} {1989})}\BibitemShut {NoStop}%
\bibitem [{\citenamefont {Cai}\ \emph {et~al.}(2023)\citenamefont {Cai},
  \citenamefont {Anderson}, \citenamefont {Wang}, \citenamefont {Zhang},
  \citenamefont {Liu}, \citenamefont {Holtzmann}, \citenamefont {Zhang},
  \citenamefont {Fan}, \citenamefont {Taniguchi}, \citenamefont {Watanabe}
  \emph {et~al.}}]{cai2023signatures}%
  \BibitemOpen
  \bibfield  {author} {\bibinfo {author} {\bibfnamefont {J.}~\bibnamefont
  {Cai}}, \bibinfo {author} {\bibfnamefont {E.}~\bibnamefont {Anderson}},
  \bibinfo {author} {\bibfnamefont {C.}~\bibnamefont {Wang}}, \bibinfo {author}
  {\bibfnamefont {X.}~\bibnamefont {Zhang}}, \bibinfo {author} {\bibfnamefont
  {X.}~\bibnamefont {Liu}}, \bibinfo {author} {\bibfnamefont {W.}~\bibnamefont
  {Holtzmann}}, \bibinfo {author} {\bibfnamefont {Y.}~\bibnamefont {Zhang}},
  \bibinfo {author} {\bibfnamefont {F.}~\bibnamefont {Fan}}, \bibinfo {author}
  {\bibfnamefont {T.}~\bibnamefont {Taniguchi}}, \bibinfo {author}
  {\bibfnamefont {K.}~\bibnamefont {Watanabe}}, \emph {et~al.},\ }\bibfield
  {title} {\bibinfo {title} {Signatures of fractional quantum anomalous hall
  states in twisted mote2},\ }\href@noop {} {\bibfield  {journal} {\bibinfo
  {journal} {Nature}\ }\textbf {\bibinfo {volume} {622}},\ \bibinfo {pages}
  {63} (\bibinfo {year} {2023})}\BibitemShut {NoStop}%
\bibitem [{\citenamefont {Park}\ \emph {et~al.}(2023)\citenamefont {Park},
  \citenamefont {Cai}, \citenamefont {Anderson}, \citenamefont {Zhang},
  \citenamefont {Zhu}, \citenamefont {Liu}, \citenamefont {Wang}, \citenamefont
  {Holtzmann}, \citenamefont {Hu}, \citenamefont {Liu} \emph
  {et~al.}}]{park2023observation}%
  \BibitemOpen
  \bibfield  {author} {\bibinfo {author} {\bibfnamefont {H.}~\bibnamefont
  {Park}}, \bibinfo {author} {\bibfnamefont {J.}~\bibnamefont {Cai}}, \bibinfo
  {author} {\bibfnamefont {E.}~\bibnamefont {Anderson}}, \bibinfo {author}
  {\bibfnamefont {Y.}~\bibnamefont {Zhang}}, \bibinfo {author} {\bibfnamefont
  {J.}~\bibnamefont {Zhu}}, \bibinfo {author} {\bibfnamefont {X.}~\bibnamefont
  {Liu}}, \bibinfo {author} {\bibfnamefont {C.}~\bibnamefont {Wang}}, \bibinfo
  {author} {\bibfnamefont {W.}~\bibnamefont {Holtzmann}}, \bibinfo {author}
  {\bibfnamefont {C.}~\bibnamefont {Hu}}, \bibinfo {author} {\bibfnamefont
  {Z.}~\bibnamefont {Liu}}, \emph {et~al.},\ }\bibfield  {title} {\bibinfo
  {title} {Observation of fractionally quantized anomalous hall effect},\
  }\href@noop {} {\bibfield  {journal} {\bibinfo  {journal} {Nature}\ }\textbf
  {\bibinfo {volume} {622}},\ \bibinfo {pages} {74} (\bibinfo {year}
  {2023})}\BibitemShut {NoStop}%
\bibitem [{\citenamefont {Zeng}\ \emph {et~al.}(2023)\citenamefont {Zeng},
  \citenamefont {Xia}, \citenamefont {Kang}, \citenamefont {Zhu}, \citenamefont
  {Kn{\"u}ppel}, \citenamefont {Vaswani}, \citenamefont {Watanabe},
  \citenamefont {Taniguchi}, \citenamefont {Mak},\ and\ \citenamefont
  {Shan}}]{zeng2023thermodynamic}%
  \BibitemOpen
  \bibfield  {author} {\bibinfo {author} {\bibfnamefont {Y.}~\bibnamefont
  {Zeng}}, \bibinfo {author} {\bibfnamefont {Z.}~\bibnamefont {Xia}}, \bibinfo
  {author} {\bibfnamefont {K.}~\bibnamefont {Kang}}, \bibinfo {author}
  {\bibfnamefont {J.}~\bibnamefont {Zhu}}, \bibinfo {author} {\bibfnamefont
  {P.}~\bibnamefont {Kn{\"u}ppel}}, \bibinfo {author} {\bibfnamefont
  {C.}~\bibnamefont {Vaswani}}, \bibinfo {author} {\bibfnamefont
  {K.}~\bibnamefont {Watanabe}}, \bibinfo {author} {\bibfnamefont
  {T.}~\bibnamefont {Taniguchi}}, \bibinfo {author} {\bibfnamefont {K.~F.}\
  \bibnamefont {Mak}},\ and\ \bibinfo {author} {\bibfnamefont {J.}~\bibnamefont
  {Shan}},\ }\bibfield  {title} {\bibinfo {title} {Thermodynamic evidence of
  fractional chern insulator in moir{\'e} mote2},\ }\href@noop {} {\bibfield
  {journal} {\bibinfo  {journal} {Nature}\ }\textbf {\bibinfo {volume} {622}},\
  \bibinfo {pages} {69} (\bibinfo {year} {2023})}\BibitemShut {NoStop}%
\bibitem [{\citenamefont {Xu}\ \emph {et~al.}(2023)\citenamefont {Xu},
  \citenamefont {Sun}, \citenamefont {Jia}, \citenamefont {Liu}, \citenamefont
  {Xu}, \citenamefont {Li}, \citenamefont {Gu}, \citenamefont {Watanabe},
  \citenamefont {Taniguchi}, \citenamefont {Tong}, \citenamefont {Jia},
  \citenamefont {Shi}, \citenamefont {Jiang}, \citenamefont {Zhang},
  \citenamefont {Liu},\ and\ \citenamefont {Li}}]{XuPRX2023}%
  \BibitemOpen
  \bibfield  {author} {\bibinfo {author} {\bibfnamefont {F.}~\bibnamefont
  {Xu}}, \bibinfo {author} {\bibfnamefont {Z.}~\bibnamefont {Sun}}, \bibinfo
  {author} {\bibfnamefont {T.}~\bibnamefont {Jia}}, \bibinfo {author}
  {\bibfnamefont {C.}~\bibnamefont {Liu}}, \bibinfo {author} {\bibfnamefont
  {C.}~\bibnamefont {Xu}}, \bibinfo {author} {\bibfnamefont {C.}~\bibnamefont
  {Li}}, \bibinfo {author} {\bibfnamefont {Y.}~\bibnamefont {Gu}}, \bibinfo
  {author} {\bibfnamefont {K.}~\bibnamefont {Watanabe}}, \bibinfo {author}
  {\bibfnamefont {T.}~\bibnamefont {Taniguchi}}, \bibinfo {author}
  {\bibfnamefont {B.}~\bibnamefont {Tong}}, \bibinfo {author} {\bibfnamefont
  {J.}~\bibnamefont {Jia}}, \bibinfo {author} {\bibfnamefont {Z.}~\bibnamefont
  {Shi}}, \bibinfo {author} {\bibfnamefont {S.}~\bibnamefont {Jiang}}, \bibinfo
  {author} {\bibfnamefont {Y.}~\bibnamefont {Zhang}}, \bibinfo {author}
  {\bibfnamefont {X.}~\bibnamefont {Liu}},\ and\ \bibinfo {author}
  {\bibfnamefont {T.}~\bibnamefont {Li}},\ }\bibfield  {title} {\bibinfo
  {title} {Observation of integer and fractional quantum anomalous hall effects
  in twisted bilayer ${\mathrm{mote}}_{2}$},\ }\href
  {https://doi.org/10.1103/PhysRevX.13.031037} {\bibfield  {journal} {\bibinfo
  {journal} {Phys. Rev. X}\ }\textbf {\bibinfo {volume} {13}},\ \bibinfo
  {pages} {031037} (\bibinfo {year} {2023})}\BibitemShut {NoStop}%
\bibitem [{\citenamefont {Lu}\ \emph {et~al.}(2024)\citenamefont {Lu},
  \citenamefont {Han}, \citenamefont {Yao}, \citenamefont {Reddy},
  \citenamefont {Yang}, \citenamefont {Seo}, \citenamefont {Watanabe},
  \citenamefont {Taniguchi}, \citenamefont {Fu},\ and\ \citenamefont
  {Ju}}]{lu2024fractional}%
  \BibitemOpen
  \bibfield  {author} {\bibinfo {author} {\bibfnamefont {Z.}~\bibnamefont
  {Lu}}, \bibinfo {author} {\bibfnamefont {T.}~\bibnamefont {Han}}, \bibinfo
  {author} {\bibfnamefont {Y.}~\bibnamefont {Yao}}, \bibinfo {author}
  {\bibfnamefont {A.~P.}\ \bibnamefont {Reddy}}, \bibinfo {author}
  {\bibfnamefont {J.}~\bibnamefont {Yang}}, \bibinfo {author} {\bibfnamefont
  {J.}~\bibnamefont {Seo}}, \bibinfo {author} {\bibfnamefont {K.}~\bibnamefont
  {Watanabe}}, \bibinfo {author} {\bibfnamefont {T.}~\bibnamefont {Taniguchi}},
  \bibinfo {author} {\bibfnamefont {L.}~\bibnamefont {Fu}},\ and\ \bibinfo
  {author} {\bibfnamefont {L.}~\bibnamefont {Ju}},\ }\bibfield  {title}
  {\bibinfo {title} {Fractional quantum anomalous hall effect in multilayer
  graphene},\ }\href@noop {} {\bibfield  {journal} {\bibinfo  {journal}
  {Nature}\ }\textbf {\bibinfo {volume} {626}},\ \bibinfo {pages} {759}
  (\bibinfo {year} {2024})}\BibitemShut {NoStop}%
\bibitem [{\citenamefont {Xu}\ \emph {et~al.}(2025)\citenamefont {Xu},
  \citenamefont {Sun}, \citenamefont {Li}, \citenamefont {Zheng}, \citenamefont
  {Xu}, \citenamefont {Gao}, \citenamefont {Jia}, \citenamefont {Su},
  \citenamefont {Watanabe}, \citenamefont {Taniguchi} \emph
  {et~al.}}]{xu2025signatures}%
  \BibitemOpen
  \bibfield  {author} {\bibinfo {author} {\bibfnamefont {F.}~\bibnamefont
  {Xu}}, \bibinfo {author} {\bibfnamefont {Z.}~\bibnamefont {Sun}}, \bibinfo
  {author} {\bibfnamefont {J.}~\bibnamefont {Li}}, \bibinfo {author}
  {\bibfnamefont {C.}~\bibnamefont {Zheng}}, \bibinfo {author} {\bibfnamefont
  {C.}~\bibnamefont {Xu}}, \bibinfo {author} {\bibfnamefont {J.}~\bibnamefont
  {Gao}}, \bibinfo {author} {\bibfnamefont {T.}~\bibnamefont {Jia}}, \bibinfo
  {author} {\bibfnamefont {Y.}~\bibnamefont {Su}}, \bibinfo {author}
  {\bibfnamefont {K.}~\bibnamefont {Watanabe}}, \bibinfo {author}
  {\bibfnamefont {T.}~\bibnamefont {Taniguchi}}, \emph {et~al.},\ }\bibfield
  {title} {\bibinfo {title} {Signatures of unconventional superconductivity
  near reentrant and fractional quantum anomalous hall insulators},\
  }\href@noop {} {\bibfield  {journal} {\bibinfo  {journal} {arXiv preprint
  arXiv:2504.06972}\ } (\bibinfo {year} {2025})}\BibitemShut {NoStop}%
\bibitem [{\citenamefont {Schleith}\ \emph {et~al.}(2026)\citenamefont
  {Schleith}, \citenamefont {Soejima},\ and\ \citenamefont
  {Khalaf}}]{schleith2026anyon}%
  \BibitemOpen
  \bibfield  {author} {\bibinfo {author} {\bibfnamefont {M.-L.}\ \bibnamefont
  {Schleith}}, \bibinfo {author} {\bibfnamefont {T.}~\bibnamefont {Soejima}},\
  and\ \bibinfo {author} {\bibfnamefont {E.}~\bibnamefont {Khalaf}},\
  }\bibfield  {title} {\bibinfo {title} {Anyon dispersion from nonuniform
  magnetic field on the sphere},\ }\href@noop {} {\bibfield  {journal}
  {\bibinfo  {journal} {Physical Review Letters}\ }\textbf {\bibinfo {volume}
  {137}},\ \bibinfo {pages} {046506} (\bibinfo {year} {2026})}\BibitemShut
  {NoStop}%
\bibitem [{\citenamefont {Yan}\ \emph {et~al.}(2026)\citenamefont {Yan},
  \citenamefont {Li}, \citenamefont {Soejima},\ and\ \citenamefont
  {Khalaf}}]{yan2026anyon}%
  \BibitemOpen
  \bibfield  {author} {\bibinfo {author} {\bibfnamefont {Z.}~\bibnamefont
  {Yan}}, \bibinfo {author} {\bibfnamefont {Q.}~\bibnamefont {Li}}, \bibinfo
  {author} {\bibfnamefont {T.}~\bibnamefont {Soejima}},\ and\ \bibinfo {author}
  {\bibfnamefont {E.}~\bibnamefont {Khalaf}},\ }\bibfield  {title} {\bibinfo
  {title} {Anyon dispersion in aharonov-casher bands and implications for
  twisted mote 2},\ }\href@noop {} {\bibfield  {journal} {\bibinfo  {journal}
  {Physical Review B}\ }\textbf {\bibinfo {volume} {113}},\ \bibinfo {pages}
  {235155} (\bibinfo {year} {2026})}\BibitemShut {NoStop}%
\bibitem [{\citenamefont {Iyer}\ \emph {et~al.}(2026)\citenamefont {Iyer},
  \citenamefont {Feuerpfeil}, \citenamefont {Cr{\'e}pel}, \citenamefont
  {Regnault},\ and\ \citenamefont {Mora}}]{iyer2026dispersion}%
  \BibitemOpen
  \bibfield  {author} {\bibinfo {author} {\bibfnamefont {K.}~\bibnamefont
  {Iyer}}, \bibinfo {author} {\bibfnamefont {A.}~\bibnamefont {Feuerpfeil}},
  \bibinfo {author} {\bibfnamefont {V.}~\bibnamefont {Cr{\'e}pel}}, \bibinfo
  {author} {\bibfnamefont {N.}~\bibnamefont {Regnault}},\ and\ \bibinfo
  {author} {\bibfnamefont {C.}~\bibnamefont {Mora}},\ }\bibfield  {title}
  {\bibinfo {title} {Dispersion of anyon bloch bands},\ }\href@noop {}
  {\bibfield  {journal} {\bibinfo  {journal} {arXiv preprint arXiv:2604.24859}\
  } (\bibinfo {year} {2026})}\BibitemShut {NoStop}%
\bibitem [{\citenamefont {Lu}\ \emph {et~al.}(2009)\citenamefont {Lu},
  \citenamefont {Gao}, \citenamefont {G{\"u}hne}, \citenamefont {Zhou},
  \citenamefont {Chen},\ and\ \citenamefont {Pan}}]{lu2009demonstrating}%
  \BibitemOpen
  \bibfield  {author} {\bibinfo {author} {\bibfnamefont {C.-Y.}\ \bibnamefont
  {Lu}}, \bibinfo {author} {\bibfnamefont {W.-B.}\ \bibnamefont {Gao}},
  \bibinfo {author} {\bibfnamefont {O.}~\bibnamefont {G{\"u}hne}}, \bibinfo
  {author} {\bibfnamefont {X.-Q.}\ \bibnamefont {Zhou}}, \bibinfo {author}
  {\bibfnamefont {Z.-B.}\ \bibnamefont {Chen}},\ and\ \bibinfo {author}
  {\bibfnamefont {J.-W.}\ \bibnamefont {Pan}},\ }\bibfield  {title} {\bibinfo
  {title} {Demonstrating anyonic fractional statistics with a six-qubit quantum
  simulator},\ }\href@noop {} {\bibfield  {journal} {\bibinfo  {journal}
  {Physical review letters}\ }\textbf {\bibinfo {volume} {102}},\ \bibinfo
  {pages} {030502} (\bibinfo {year} {2009})}\BibitemShut {NoStop}%
\bibitem [{\citenamefont {{Google Quantum AI and
  Collaborators}}(2023)}]{google2023non}%
  \BibitemOpen
  \bibfield  {author} {\bibinfo {author} {\bibnamefont {{Google Quantum AI and
  Collaborators}}},\ }\bibfield  {title} {\bibinfo {title} {Non-abelian
  braiding of graph vertices in a superconducting processor},\ }\href@noop {}
  {\bibfield  {journal} {\bibinfo  {journal} {Nature}\ }\textbf {\bibinfo
  {volume} {618}},\ \bibinfo {pages} {264} (\bibinfo {year}
  {2023})}\BibitemShut {NoStop}%
\bibitem [{\citenamefont {Iqbal}\ \emph {et~al.}(2024)\citenamefont {Iqbal},
  \citenamefont {Tantivasadakarn}, \citenamefont {Verresen}, \citenamefont
  {Campbell}, \citenamefont {Dreiling}, \citenamefont {Figgatt}, \citenamefont
  {Gaebler}, \citenamefont {Johansen}, \citenamefont {Mills}, \citenamefont
  {Moses} \emph {et~al.}}]{iqbal2024non}%
  \BibitemOpen
  \bibfield  {author} {\bibinfo {author} {\bibfnamefont {M.}~\bibnamefont
  {Iqbal}}, \bibinfo {author} {\bibfnamefont {N.}~\bibnamefont
  {Tantivasadakarn}}, \bibinfo {author} {\bibfnamefont {R.}~\bibnamefont
  {Verresen}}, \bibinfo {author} {\bibfnamefont {S.~L.}\ \bibnamefont
  {Campbell}}, \bibinfo {author} {\bibfnamefont {J.~M.}\ \bibnamefont
  {Dreiling}}, \bibinfo {author} {\bibfnamefont {C.}~\bibnamefont {Figgatt}},
  \bibinfo {author} {\bibfnamefont {J.~P.}\ \bibnamefont {Gaebler}}, \bibinfo
  {author} {\bibfnamefont {J.}~\bibnamefont {Johansen}}, \bibinfo {author}
  {\bibfnamefont {M.}~\bibnamefont {Mills}}, \bibinfo {author} {\bibfnamefont
  {S.~A.}\ \bibnamefont {Moses}}, \emph {et~al.},\ }\bibfield  {title}
  {\bibinfo {title} {Non-abelian topological order and anyons on a trapped-ion
  processor},\ }\href@noop {} {\bibfield  {journal} {\bibinfo  {journal}
  {Nature}\ }\textbf {\bibinfo {volume} {626}},\ \bibinfo {pages} {505}
  (\bibinfo {year} {2024})}\BibitemShut {NoStop}%
\bibitem [{\citenamefont {Divic}\ \emph {et~al.}(2025)\citenamefont {Divic},
  \citenamefont {Cr{\'e}pel}, \citenamefont {Soejima}, \citenamefont {Song},
  \citenamefont {Millis}, \citenamefont {Zaletel},\ and\ \citenamefont
  {Vishwanath}}]{divic2025anyon}%
  \BibitemOpen
  \bibfield  {author} {\bibinfo {author} {\bibfnamefont {S.}~\bibnamefont
  {Divic}}, \bibinfo {author} {\bibfnamefont {V.}~\bibnamefont {Cr{\'e}pel}},
  \bibinfo {author} {\bibfnamefont {T.}~\bibnamefont {Soejima}}, \bibinfo
  {author} {\bibfnamefont {X.-Y.}\ \bibnamefont {Song}}, \bibinfo {author}
  {\bibfnamefont {A.~J.}\ \bibnamefont {Millis}}, \bibinfo {author}
  {\bibfnamefont {M.~P.}\ \bibnamefont {Zaletel}},\ and\ \bibinfo {author}
  {\bibfnamefont {A.}~\bibnamefont {Vishwanath}},\ }\bibfield  {title}
  {\bibinfo {title} {Anyon superconductivity from topological criticality in a
  hofstadter--hubbard model},\ }\href@noop {} {\bibfield  {journal} {\bibinfo
  {journal} {Proceedings of the National Academy of Sciences}\ }\textbf
  {\bibinfo {volume} {122}},\ \bibinfo {pages} {e2426680122} (\bibinfo {year}
  {2025})}\BibitemShut {NoStop}%
\bibitem [{\citenamefont {Shi}\ \emph {et~al.}(2025)\citenamefont {Shi},
  \citenamefont {Zhang},\ and\ \citenamefont {Senthil}}]{shi2025doping}%
  \BibitemOpen
  \bibfield  {author} {\bibinfo {author} {\bibfnamefont {Z.~D.}\ \bibnamefont
  {Shi}}, \bibinfo {author} {\bibfnamefont {C.}~\bibnamefont {Zhang}},\ and\
  \bibinfo {author} {\bibfnamefont {T.}~\bibnamefont {Senthil}},\ }\bibfield
  {title} {\bibinfo {title} {Doping lattice non-abelian quantum hall states},\
  }\href@noop {} {\bibfield  {journal} {\bibinfo  {journal} {arXiv preprint
  arXiv:2505.02893}\ } (\bibinfo {year} {2025})}\BibitemShut {NoStop}%
\bibitem [{\citenamefont {Shi}\ and\ \citenamefont
  {Senthil}(2026)}]{shi2026superconductivity}%
  \BibitemOpen
  \bibfield  {author} {\bibinfo {author} {\bibfnamefont {Z.~D.}\ \bibnamefont
  {Shi}}\ and\ \bibinfo {author} {\bibfnamefont {T.}~\bibnamefont {Senthil}},\
  }\bibfield  {title} {\bibinfo {title} {Superconductivity and non-fermi liquid
  metals in a charge-1/3 anyon fluid},\ }\href@noop {} {\bibfield  {journal}
  {\bibinfo  {journal} {arXiv preprint arXiv:2606.20403}\ } (\bibinfo {year}
  {2026})}\BibitemShut {NoStop}%
\bibitem [{\citenamefont {Shi}\ and\ \citenamefont
  {Senthil}(2025{\natexlab{a}})}]{shi2025non}%
  \BibitemOpen
  \bibfield  {author} {\bibinfo {author} {\bibfnamefont {Z.~D.}\ \bibnamefont
  {Shi}}\ and\ \bibinfo {author} {\bibfnamefont {T.}~\bibnamefont {Senthil}},\
  }\bibfield  {title} {\bibinfo {title} {Non-abelian topological
  superconductivity from melting abelian fractional chern insulators},\
  }\href@noop {} {\bibfield  {journal} {\bibinfo  {journal} {arXiv preprint
  arXiv:2512.17996}\ } (\bibinfo {year} {2025}{\natexlab{a}})}\BibitemShut
  {NoStop}%
\bibitem [{\citenamefont {Senthil}(2026)}]{senthil2026fractionalized}%
  \BibitemOpen
  \bibfield  {author} {\bibinfo {author} {\bibfnamefont {T.}~\bibnamefont
  {Senthil}},\ }\bibfield  {title} {\bibinfo {title} {Fractionalized metals
  from doped anyons: Application to tmote2},\ }\href@noop {} {\bibfield
  {journal} {\bibinfo  {journal} {arXiv preprint arXiv:2607.11484}\ } (\bibinfo
  {year} {2026})}\BibitemShut {NoStop}%
\bibitem [{\citenamefont {Han}\ \emph {et~al.}(2026)\citenamefont {Han},
  \citenamefont {Vishwanath},\ and\ \citenamefont {Khalaf}}]{han2026single}%
  \BibitemOpen
  \bibfield  {author} {\bibinfo {author} {\bibfnamefont {Z.}~\bibnamefont
  {Han}}, \bibinfo {author} {\bibfnamefont {A.}~\bibnamefont {Vishwanath}},\
  and\ \bibinfo {author} {\bibfnamefont {E.}~\bibnamefont {Khalaf}},\
  }\bibfield  {title} {\bibinfo {title} {Single-component twisted $z_3$
  orthogonal metal in an $ e/3$-anyon fluid},\ }\href@noop {} {\bibfield
  {journal} {\bibinfo  {journal} {arXiv preprint arXiv:2607.15274}\ } (\bibinfo
  {year} {2026})}\BibitemShut {NoStop}%
\bibitem [{\citenamefont {Fan}\ \emph {et~al.}(2026)\citenamefont {Fan},
  \citenamefont {Vishwanath},\ and\ \citenamefont {Wang}}]{fan2026hidden}%
  \BibitemOpen
  \bibfield  {author} {\bibinfo {author} {\bibfnamefont {Z.-D.}\ \bibnamefont
  {Fan}}, \bibinfo {author} {\bibfnamefont {A.}~\bibnamefont {Vishwanath}},\
  and\ \bibinfo {author} {\bibfnamefont {Z.}~\bibnamefont {Wang}},\ }\bibfield
  {title} {\bibinfo {title} {Hidden weak-pairing superconductivity of
  non-interacting anyons obeying $ \frac{1}{3} $ statistics},\ }\href@noop {}
  {\bibfield  {journal} {\bibinfo  {journal} {arXiv preprint arXiv:2605.19036}\
  } (\bibinfo {year} {2026})}\BibitemShut {NoStop}%
\bibitem [{\citenamefont {Wang}\ and\ \citenamefont {Zhang}(2026)}]{Taige2026}%
  \BibitemOpen
  \bibfield  {author} {\bibinfo {author} {\bibfnamefont {T.}~\bibnamefont
  {Wang}}\ and\ \bibinfo {author} {\bibfnamefont {Y.-H.}\ \bibnamefont
  {Zhang}},\ }\bibfield  {title} {\bibinfo {title} {Anyon superfluid in
  trilayer quantum hall systems},\ }\href {https://doi.org/10.1103/96h3-z6r8}
  {\bibfield  {journal} {\bibinfo  {journal} {Phys. Rev. Lett.}\ }\textbf
  {\bibinfo {volume} {137}},\ \bibinfo {pages} {136502} (\bibinfo {year}
  {2026})}\BibitemShut {NoStop}%
\bibitem [{\citenamefont {Mehta}\ \emph {et~al.}(2026)\citenamefont {Mehta},
  \citenamefont {Nakajima},\ and\ \citenamefont {Goldman}}]{mehta2026coloring}%
  \BibitemOpen
  \bibfield  {author} {\bibinfo {author} {\bibfnamefont {U.}~\bibnamefont
  {Mehta}}, \bibinfo {author} {\bibfnamefont {Y.}~\bibnamefont {Nakajima}},\
  and\ \bibinfo {author} {\bibfnamefont {H.}~\bibnamefont {Goldman}},\
  }\bibfield  {title} {\bibinfo {title} {Coloring in anyon superconductivity},\
  }\href@noop {} {\bibfield  {journal} {\bibinfo  {journal} {arXiv preprint
  arXiv:2607.19470}\ } (\bibinfo {year} {2026})}\BibitemShut {NoStop}%
\bibitem [{\citenamefont {Cui}\ \emph {et~al.}(2026)\citenamefont {Cui},
  \citenamefont {Calvera}, \citenamefont {Mehta},\ and\ \citenamefont
  {Goldman}}]{cui2026anyonic}%
  \BibitemOpen
  \bibfield  {author} {\bibinfo {author} {\bibfnamefont {H.-R.}\ \bibnamefont
  {Cui}}, \bibinfo {author} {\bibfnamefont {V.}~\bibnamefont {Calvera}},
  \bibinfo {author} {\bibfnamefont {U.}~\bibnamefont {Mehta}},\ and\ \bibinfo
  {author} {\bibfnamefont {H.}~\bibnamefont {Goldman}},\ }\bibfield  {title}
  {\bibinfo {title} {Anyonic molecules through the trion looking glass},\
  }\href@noop {} {\bibfield  {journal} {\bibinfo  {journal} {arXiv preprint
  arXiv:2609.30380}\ } (\bibinfo {year} {2026})}\BibitemShut {NoStop}%
\bibitem [{\citenamefont {Zhang}(2026{\natexlab{a}})}]{zhang2026color}%
  \BibitemOpen
  \bibfield  {author} {\bibinfo {author} {\bibfnamefont {Y.-H.}\ \bibnamefont
  {Zhang}},\ }\bibfield  {title} {\bibinfo {title} {Color superconductors and
  holon metals from doping a fractional chern insulator},\ }\href@noop {}
  {\bibfield  {journal} {\bibinfo  {journal} {arXiv preprint arXiv:2607.18238}\
  } (\bibinfo {year} {2026}{\natexlab{a}})}\BibitemShut {NoStop}%
\bibitem [{\citenamefont {Li}\ \emph {et~al.}(2026)\citenamefont {Li},
  \citenamefont {Nosov}, \citenamefont {Wang},\ and\ \citenamefont
  {Khalaf}}]{li2026boundstatesanyonsgeometric}%
  \BibitemOpen
  \bibfield  {author} {\bibinfo {author} {\bibfnamefont {Q.}~\bibnamefont
  {Li}}, \bibinfo {author} {\bibfnamefont {P.~A.}\ \bibnamefont {Nosov}},
  \bibinfo {author} {\bibfnamefont {T.}~\bibnamefont {Wang}},\ and\ \bibinfo
  {author} {\bibfnamefont {E.}~\bibnamefont {Khalaf}},\ }\href
  {https://arxiv.org/abs/2603.24701} {\bibinfo {title} {Bound states of anyons:
  a geometric quantization approach}} (\bibinfo {year} {2026}),\ \Eprint
  {https://arxiv.org/abs/2603.24701} {arXiv:2603.24701 [cond-mat.str-el]}
  \BibitemShut {NoStop}%
\bibitem [{\citenamefont {Nosov}\ \emph {et~al.}(2026)\citenamefont {Nosov},
  \citenamefont {Han},\ and\ \citenamefont {Khalaf}}]{Nosov_2026}%
  \BibitemOpen
  \bibfield  {author} {\bibinfo {author} {\bibfnamefont {P.~A.}\ \bibnamefont
  {Nosov}}, \bibinfo {author} {\bibfnamefont {Z.}~\bibnamefont {Han}},\ and\
  \bibinfo {author} {\bibfnamefont {E.}~\bibnamefont {Khalaf}},\ }\bibfield
  {title} {\bibinfo {title} {Anyon superconductivity and plateau transitions in
  doped fractional quantum anomalous hall insulators},\ }\bibfield  {journal}
  {\bibinfo  {journal} {Physical Review Letters}\ }\textbf {\bibinfo {volume}
  {136}},\ \href {https://doi.org/10.1103/6bgj-bfdn} {10.1103/6bgj-bfdn}
  (\bibinfo {year} {2026})\BibitemShut {NoStop}%
\bibitem [{\citenamefont {Shi}\ and\ \citenamefont
  {Senthil}(2025{\natexlab{b}})}]{ShiPRX}%
  \BibitemOpen
  \bibfield  {author} {\bibinfo {author} {\bibfnamefont {Z.~D.}\ \bibnamefont
  {Shi}}\ and\ \bibinfo {author} {\bibfnamefont {T.}~\bibnamefont {Senthil}},\
  }\bibfield  {title} {\bibinfo {title} {Doping a fractional quantum anomalous
  hall insulator},\ }\href {https://doi.org/10.1103/kcm5-hx56} {\bibfield
  {journal} {\bibinfo  {journal} {Phys. Rev. X}\ }\textbf {\bibinfo {volume}
  {15}},\ \bibinfo {pages} {031069} (\bibinfo {year}
  {2025}{\natexlab{b}})}\BibitemShut {NoStop}%
\bibitem [{\citenamefont {Wagner}\ and\ \citenamefont
  {Neupert}(2026)}]{WagnerNeupert2026}%
  \BibitemOpen
  \bibfield  {author} {\bibinfo {author} {\bibfnamefont {G.}~\bibnamefont
  {Wagner}}\ and\ \bibinfo {author} {\bibfnamefont {T.}~\bibnamefont
  {Neupert}},\ }\bibfield  {title} {\bibinfo {title} {Sensing the binding and
  unbinding of anyons at impurities},\ }\href
  {https://doi.org/10.1103/cfm5-wgz1} {\bibfield  {journal} {\bibinfo
  {journal} {Phys. Rev. Research}\ }\textbf {\bibinfo {volume} {8}},\ \bibinfo
  {pages} {013263} (\bibinfo {year} {2026})}\BibitemShut {NoStop}%
\bibitem [{\citenamefont {Iengo}\ and\ \citenamefont
  {Lechne}(1992)}]{iengo1992anyon}%
  \BibitemOpen
  \bibfield  {author} {\bibinfo {author} {\bibfnamefont {R.}~\bibnamefont
  {Iengo}}\ and\ \bibinfo {author} {\bibfnamefont {K.}~\bibnamefont {Lechne}},\
  }\bibfield  {title} {\bibinfo {title} {Anyon quantum mechanics and
  chern-simons theory},\ }\href@noop {} {\bibfield  {journal} {\bibinfo
  {journal} {Physics Reports}\ }\textbf {\bibinfo {volume} {213}},\ \bibinfo
  {pages} {179} (\bibinfo {year} {1992})}\BibitemShut {NoStop}%
\bibitem [{\citenamefont {Einarsson}(1990)}]{Einarsson1990}%
  \BibitemOpen
  \bibfield  {author} {\bibinfo {author} {\bibfnamefont {T.}~\bibnamefont
  {Einarsson}},\ }\bibfield  {title} {\bibinfo {title} {Fractional statistics
  on a torus},\ }\href {https://doi.org/10.1103/PhysRevLett.64.1995} {\bibfield
   {journal} {\bibinfo  {journal} {Phys. Rev. Lett.}\ }\textbf {\bibinfo
  {volume} {64}},\ \bibinfo {pages} {1995} (\bibinfo {year}
  {1990})}\BibitemShut {NoStop}%
\bibitem [{\citenamefont {Hosotani}\ and\ \citenamefont
  {Ho}(1992)}]{HosotaniHo1992}%
  \BibitemOpen
  \bibfield  {author} {\bibinfo {author} {\bibfnamefont {Y.}~\bibnamefont
  {Hosotani}}\ and\ \bibinfo {author} {\bibfnamefont {C.-L.}\ \bibnamefont
  {Ho}},\ }\bibfield  {title} {\bibinfo {title} {Anyons on a torus},\ }\href
  {https://doi.org/10.1063/1.43444} {\bibfield  {journal} {\bibinfo  {journal}
  {AIP Conf. Proc.}\ }\textbf {\bibinfo {volume} {272}},\ \bibinfo {pages}
  {1466} (\bibinfo {year} {1992})},\ \Eprint
  {https://arxiv.org/abs/hep-th/9210112} {arXiv:hep-th/9210112} \BibitemShut
  {NoStop}%
\bibitem [{\citenamefont {Wilczek}(1982{\natexlab{b}})}]{Wilczek82}%
  \BibitemOpen
  \bibfield  {author} {\bibinfo {author} {\bibfnamefont {F.}~\bibnamefont
  {Wilczek}},\ }\bibfield  {title} {\bibinfo {title} {Magnetic flux, angular
  momentum, and statistics},\ }\href
  {https://doi.org/10.1103/PhysRevLett.48.1144} {\bibfield  {journal} {\bibinfo
   {journal} {Phys. Rev. Lett.}\ }\textbf {\bibinfo {volume} {48}},\ \bibinfo
  {pages} {1144} (\bibinfo {year} {1982}{\natexlab{b}})}\BibitemShut {NoStop}%
\bibitem [{\citenamefont {Correggi}\ \emph {et~al.}(2019)\citenamefont
  {Correggi}, \citenamefont {Duboscq}, \citenamefont {Lundholm},\ and\
  \citenamefont {Rougerie}}]{correggi2019vortex}%
  \BibitemOpen
  \bibfield  {author} {\bibinfo {author} {\bibfnamefont {M.}~\bibnamefont
  {Correggi}}, \bibinfo {author} {\bibfnamefont {R.}~\bibnamefont {Duboscq}},
  \bibinfo {author} {\bibfnamefont {D.}~\bibnamefont {Lundholm}},\ and\
  \bibinfo {author} {\bibfnamefont {N.}~\bibnamefont {Rougerie}},\ }\bibfield
  {title} {\bibinfo {title} {Vortex patterns in the almost-bosonic anyon gas},\
  }\href@noop {} {\bibfield  {journal} {\bibinfo  {journal} {Europhysics
  Letters}\ }\textbf {\bibinfo {volume} {126}},\ \bibinfo {pages} {20005}
  (\bibinfo {year} {2019})}\BibitemShut {NoStop}%
\bibitem [{\citenamefont {Girardot}\ and\ \citenamefont
  {Rougerie}(2021)}]{girardot2021semiclassical}%
  \BibitemOpen
  \bibfield  {author} {\bibinfo {author} {\bibfnamefont {T.}~\bibnamefont
  {Girardot}}\ and\ \bibinfo {author} {\bibfnamefont {N.}~\bibnamefont
  {Rougerie}},\ }\bibfield  {title} {\bibinfo {title} {Semiclassical limit for
  almost fermionic anyons},\ }\href@noop {} {\bibfield  {journal} {\bibinfo
  {journal} {Communications in Mathematical Physics}\ }\textbf {\bibinfo
  {volume} {387}},\ \bibinfo {pages} {427} (\bibinfo {year}
  {2021})}\BibitemShut {NoStop}%
\bibitem [{\citenamefont {Lundholm}\ and\ \citenamefont
  {Rougerie}(2015)}]{LundholmRougerie2015}%
  \BibitemOpen
  \bibfield  {author} {\bibinfo {author} {\bibfnamefont {D.}~\bibnamefont
  {Lundholm}}\ and\ \bibinfo {author} {\bibfnamefont {N.}~\bibnamefont
  {Rougerie}},\ }\bibfield  {title} {\bibinfo {title} {The average field
  approximation for almost bosonic extended anyons},\ }\href
  {https://doi.org/10.1007/s10955-015-1382-y} {\bibfield  {journal} {\bibinfo
  {journal} {J. Stat. Phys.}\ }\textbf {\bibinfo {volume} {161}},\ \bibinfo
  {pages} {1236} (\bibinfo {year} {2015})}\BibitemShut {NoStop}%
\bibitem [{\citenamefont {Lundholm}(2024)}]{Lundholm2024}%
  \BibitemOpen
  \bibfield  {author} {\bibinfo {author} {\bibfnamefont {D.}~\bibnamefont
  {Lundholm}},\ }\bibfield  {title} {\bibinfo {title} {Properties of {2D} anyon
  gas},\ }in\ \href {https://doi.org/10.1016/B978-0-323-90800-9.00260-2} {\emph
  {\bibinfo {booktitle} {Encyclopedia of Condensed Matter Physics}}},\ \bibinfo
  {editor} {edited by\ \bibinfo {editor} {\bibfnamefont {T.}~\bibnamefont
  {Chakraborty}}}\ (\bibinfo  {publisher} {Academic Press},\ \bibinfo {year}
  {2024})\ \bibinfo {edition} {2nd}\ ed.,\ pp.\ \bibinfo {pages} {450--484},\
  \Eprint {https://arxiv.org/abs/2303.09544} {arXiv:2303.09544} \BibitemShut
  {NoStop}%
\bibitem [{\citenamefont {Ataei}\ \emph {et~al.}(2025)\citenamefont {Ataei},
  \citenamefont {Ellingsen}, \citenamefont {Getzner}, \citenamefont {Girardot},
  \citenamefont {Lundholm},\ and\ \citenamefont {Nguyen}}]{AtaeiEtAl2025}%
  \BibitemOpen
  \bibfield  {author} {\bibinfo {author} {\bibfnamefont {A.}~\bibnamefont
  {Ataei}}, \bibinfo {author} {\bibfnamefont {A.}~\bibnamefont {Ellingsen}},
  \bibinfo {author} {\bibfnamefont {F.}~\bibnamefont {Getzner}}, \bibinfo
  {author} {\bibfnamefont {T.}~\bibnamefont {Girardot}}, \bibinfo {author}
  {\bibfnamefont {D.}~\bibnamefont {Lundholm}},\ and\ \bibinfo {author}
  {\bibfnamefont {D.-T.}\ \bibnamefont {Nguyen}},\ }\href
  {https://arxiv.org/abs/2510.14679v3} {\bibinfo {title} {Nonlinear {Landau}
  levels in the almost-bosonic anyon gas}} (\bibinfo {year} {2025}),\ \bibinfo
  {note} {version 3 (2026)},\ \Eprint {https://arxiv.org/abs/2510.14679}
  {arXiv:2510.14679 [cond-mat.quant-gas]} \BibitemShut {NoStop}%
\bibitem [{\citenamefont {Komargodski}\ \emph {et~al.}(2026)\citenamefont
  {Komargodski}, \citenamefont {Lou}, \citenamefont {Nagar}, \citenamefont
  {Orlando}, \citenamefont {Reffert},\ and\ \citenamefont
  {Sever}}]{komargodski2026anyon}%
  \BibitemOpen
  \bibfield  {author} {\bibinfo {author} {\bibfnamefont {Z.}~\bibnamefont
  {Komargodski}}, \bibinfo {author} {\bibfnamefont {X.}~\bibnamefont {Lou}},
  \bibinfo {author} {\bibfnamefont {I.}~\bibnamefont {Nagar}}, \bibinfo
  {author} {\bibfnamefont {D.}~\bibnamefont {Orlando}}, \bibinfo {author}
  {\bibfnamefont {S.}~\bibnamefont {Reffert}},\ and\ \bibinfo {author}
  {\bibfnamefont {A.}~\bibnamefont {Sever}},\ }\bibfield  {title} {\bibinfo
  {title} {Anyon crystallization by statistics},\ }\href@noop {} {\bibfield
  {journal} {\bibinfo  {journal} {arXiv preprint arXiv:2609.25249}\ } (\bibinfo
  {year} {2026})}\BibitemShut {NoStop}%
\bibitem [{\citenamefont {Sporre}\ \emph {et~al.}(1991)\citenamefont {Sporre},
  \citenamefont {Verbaarschot},\ and\ \citenamefont {Zahed}}]{Sporre1991}%
  \BibitemOpen
  \bibfield  {author} {\bibinfo {author} {\bibfnamefont {M.}~\bibnamefont
  {Sporre}}, \bibinfo {author} {\bibfnamefont {J.~J.~M.}\ \bibnamefont
  {Verbaarschot}},\ and\ \bibinfo {author} {\bibfnamefont {I.}~\bibnamefont
  {Zahed}},\ }\bibfield  {title} {\bibinfo {title} {Numerical solution of the
  three-anyon problem},\ }\href {https://doi.org/10.1103/PhysRevLett.67.1813}
  {\bibfield  {journal} {\bibinfo  {journal} {Physical Review Letters}\
  }\textbf {\bibinfo {volume} {67}},\ \bibinfo {pages} {1813} (\bibinfo {year}
  {1991})}\BibitemShut {NoStop}%
\bibitem [{\citenamefont {Sporre}\ \emph {et~al.}(1992)\citenamefont {Sporre},
  \citenamefont {Verbaarschot},\ and\ \citenamefont {Zahed}}]{Sporre1992}%
  \BibitemOpen
  \bibfield  {author} {\bibinfo {author} {\bibfnamefont {M.}~\bibnamefont
  {Sporre}}, \bibinfo {author} {\bibfnamefont {J.~J.~M.}\ \bibnamefont
  {Verbaarschot}},\ and\ \bibinfo {author} {\bibfnamefont {I.}~\bibnamefont
  {Zahed}},\ }\bibfield  {title} {\bibinfo {title} {Four anyons in a harmonic
  well},\ }\href {https://doi.org/10.1103/PhysRevB.46.5738} {\bibfield
  {journal} {\bibinfo  {journal} {Physical Review B}\ }\textbf {\bibinfo
  {volume} {46}},\ \bibinfo {pages} {5738} (\bibinfo {year}
  {1992})}\BibitemShut {NoStop}%
\bibitem [{\citenamefont {Wang}\ and\ \citenamefont
  {Zaletel}(2026)}]{wang2026anyon}%
  \BibitemOpen
  \bibfield  {author} {\bibinfo {author} {\bibfnamefont {T.}~\bibnamefont
  {Wang}}\ and\ \bibinfo {author} {\bibfnamefont {M.~P.}\ \bibnamefont
  {Zaletel}},\ }\bibfield  {title} {\bibinfo {title} {Anyon molecules in
  fractional quantum hall states},\ }\href@noop {} {\bibfield  {journal}
  {\bibinfo  {journal} {arXiv preprint arXiv:2604.09798}\ } (\bibinfo {year}
  {2026})}\BibitemShut {NoStop}%
\bibitem [{\citenamefont {Valentini}\ \emph {et~al.}(2026)\citenamefont
  {Valentini}, \citenamefont {Wang}, \citenamefont {Mohammadi}, \citenamefont
  {Samuelson}, \citenamefont {Li}, \citenamefont {Cohen}, \citenamefont
  {Zakharyan}, \citenamefont {Feng}, \citenamefont {Wang}, \citenamefont
  {Taniguchi} \emph {et~al.}}]{valentini2026single}%
  \BibitemOpen
  \bibfield  {author} {\bibinfo {author} {\bibfnamefont {M.}~\bibnamefont
  {Valentini}}, \bibinfo {author} {\bibfnamefont {W.}~\bibnamefont {Wang}},
  \bibinfo {author} {\bibfnamefont {A.~S.}\ \bibnamefont {Mohammadi}}, \bibinfo
  {author} {\bibfnamefont {N.}~\bibnamefont {Samuelson}}, \bibinfo {author}
  {\bibfnamefont {C.}~\bibnamefont {Li}}, \bibinfo {author} {\bibfnamefont
  {L.}~\bibnamefont {Cohen}}, \bibinfo {author} {\bibfnamefont
  {M.}~\bibnamefont {Zakharyan}}, \bibinfo {author} {\bibfnamefont
  {J.}~\bibnamefont {Feng}}, \bibinfo {author} {\bibfnamefont {T.}~\bibnamefont
  {Wang}}, \bibinfo {author} {\bibfnamefont {T.}~\bibnamefont {Taniguchi}},
  \emph {et~al.},\ }\bibfield  {title} {\bibinfo {title} {Single anyons and
  anyon binding in a quantum hall dot},\ }\href@noop {} {\bibfield  {journal}
  {\bibinfo  {journal} {arXiv preprint arXiv:2610.18817}\ } (\bibinfo {year}
  {2026})}\BibitemShut {NoStop}%
\bibitem [{\citenamefont {Carleo}\ and\ \citenamefont
  {Troyer}(2017)}]{CarleoTroyer2017}%
  \BibitemOpen
  \bibfield  {author} {\bibinfo {author} {\bibfnamefont {G.}~\bibnamefont
  {Carleo}}\ and\ \bibinfo {author} {\bibfnamefont {M.}~\bibnamefont
  {Troyer}},\ }\bibfield  {title} {\bibinfo {title} {Solving the {Quantum}
  {Many}-{Body} {Problem} with {Artificial} {Neural} {Networks}},\ }\href
  {https://doi.org/10.1126/science.aag2302} {\bibfield  {journal} {\bibinfo
  {journal} {Science}\ }\textbf {\bibinfo {volume} {355}},\ \bibinfo {pages}
  {602} (\bibinfo {year} {2017})}\BibitemShut {NoStop}%
\bibitem [{\citenamefont {Pfau}\ \emph {et~al.}(2020)\citenamefont {Pfau},
  \citenamefont {Spencer}, \citenamefont {Matthews},\ and\ \citenamefont
  {Foulkes}}]{Pfau2020}%
  \BibitemOpen
  \bibfield  {author} {\bibinfo {author} {\bibfnamefont {D.}~\bibnamefont
  {Pfau}}, \bibinfo {author} {\bibfnamefont {J.~S.}\ \bibnamefont {Spencer}},
  \bibinfo {author} {\bibfnamefont {A.~G. D.~G.}\ \bibnamefont {Matthews}},\
  and\ \bibinfo {author} {\bibfnamefont {W.~M.~C.}\ \bibnamefont {Foulkes}},\
  }\bibfield  {title} {\bibinfo {title} {Ab initio solution of the
  many-electron {S}chr\"odinger equation with deep neural networks},\ }\href
  {https://doi.org/10.1103/PhysRevResearch.2.033429} {\bibfield  {journal}
  {\bibinfo  {journal} {Phys. Rev. Research}\ }\textbf {\bibinfo {volume}
  {2}},\ \bibinfo {pages} {033429} (\bibinfo {year} {2020})}\BibitemShut
  {NoStop}%
\bibitem [{\citenamefont {von Glehn}\ \emph {et~al.}(2023)\citenamefont {von
  Glehn}, \citenamefont {Spencer},\ and\ \citenamefont {Pfau}}]{vonglehn2023}%
  \BibitemOpen
  \bibfield  {author} {\bibinfo {author} {\bibfnamefont {I.}~\bibnamefont {von
  Glehn}}, \bibinfo {author} {\bibfnamefont {J.~S.}\ \bibnamefont {Spencer}},\
  and\ \bibinfo {author} {\bibfnamefont {D.}~\bibnamefont {Pfau}},\ }\bibfield
  {title} {\bibinfo {title} {A self-attention ansatz for ab-initio quantum
  chemistry},\ }in\ \href@noop {} {\emph {\bibinfo {booktitle} {International
  Conference on Learning Representations}}}\ (\bibinfo {year}
  {2023})\BibitemShut {NoStop}%
\bibitem [{\citenamefont {Geier}\ \emph {et~al.}(2025)\citenamefont {Geier},
  \citenamefont {Nazaryan}, \citenamefont {Zaklama},\ and\ \citenamefont
  {Fu}}]{geier2025}%
  \BibitemOpen
  \bibfield  {author} {\bibinfo {author} {\bibfnamefont {M.}~\bibnamefont
  {Geier}}, \bibinfo {author} {\bibfnamefont {K.}~\bibnamefont {Nazaryan}},
  \bibinfo {author} {\bibfnamefont {T.}~\bibnamefont {Zaklama}},\ and\ \bibinfo
  {author} {\bibfnamefont {L.}~\bibnamefont {Fu}},\ }\bibfield  {title}
  {\bibinfo {title} {Self-attention neural network for solving correlated
  electron problems in solids},\ }\href {https://doi.org/10.1103/qxc3-bkc7}
  {\bibfield  {journal} {\bibinfo  {journal} {Phys. Rev. B}\ }\textbf {\bibinfo
  {volume} {112}},\ \bibinfo {pages} {045119} (\bibinfo {year}
  {2025})}\BibitemShut {NoStop}%
\bibitem [{\citenamefont {Luo}\ \emph {et~al.}(2024)\citenamefont {Luo},
  \citenamefont {Dai},\ and\ \citenamefont {Fu}}]{LuoDaiFu2024}%
  \BibitemOpen
  \bibfield  {author} {\bibinfo {author} {\bibfnamefont {D.}~\bibnamefont
  {Luo}}, \bibinfo {author} {\bibfnamefont {D.~D.}\ \bibnamefont {Dai}},\ and\
  \bibinfo {author} {\bibfnamefont {L.}~\bibnamefont {Fu}},\ }\href
  {https://doi.org/10.48550/arXiv.2406.17645} {\bibinfo {title} {Simulating
  moir\'e quantum matter with neural network}} (\bibinfo {year} {2024}),\
  \Eprint {https://arxiv.org/abs/2406.17645} {arXiv:2406.17645
  [cond-mat.str-el]} \BibitemShut {NoStop}%
\bibitem [{\citenamefont {Abouelkomsan}\ \emph {et~al.}(2025)\citenamefont
  {Abouelkomsan}, \citenamefont {Geier},\ and\ \citenamefont
  {Fu}}]{abouelkomsan2025topologicalorderdeepstate}%
  \BibitemOpen
  \bibfield  {author} {\bibinfo {author} {\bibfnamefont {A.}~\bibnamefont
  {Abouelkomsan}}, \bibinfo {author} {\bibfnamefont {M.}~\bibnamefont
  {Geier}},\ and\ \bibinfo {author} {\bibfnamefont {L.}~\bibnamefont {Fu}},\
  }\href {https://arxiv.org/abs/2512.01863} {\bibinfo {title} {Topological
  order in deep state}} (\bibinfo {year} {2025}),\ \Eprint
  {https://arxiv.org/abs/2512.01863} {arXiv:2512.01863 [cond-mat.mes-hall]}
  \BibitemShut {NoStop}%
\bibitem [{\citenamefont {Abouelkomsan}\ and\ \citenamefont
  {Fu}(2026)}]{abouelkomsan2026firstprinciplesaifindscrystallization}%
  \BibitemOpen
  \bibfield  {author} {\bibinfo {author} {\bibfnamefont {A.}~\bibnamefont
  {Abouelkomsan}}\ and\ \bibinfo {author} {\bibfnamefont {L.}~\bibnamefont
  {Fu}},\ }\href {https://arxiv.org/abs/2602.03927} {\bibinfo {title}
  {First-principles ai finds crystallization of fractional quantum hall
  liquids}} (\bibinfo {year} {2026}),\ \Eprint
  {https://arxiv.org/abs/2602.03927} {arXiv:2602.03927 [cond-mat.mes-hall]}
  \BibitemShut {NoStop}%
\bibitem [{\citenamefont {Hu}\ \emph {et~al.}(2026)\citenamefont {Hu},
  \citenamefont {Lin}, \citenamefont {Ran},\ and\ \citenamefont
  {Xiao}}]{hu2026hyperdet}%
  \BibitemOpen
  \bibfield  {author} {\bibinfo {author} {\bibfnamefont {X.}~\bibnamefont
  {Hu}}, \bibinfo {author} {\bibfnamefont {G.-L.}\ \bibnamefont {Lin}},
  \bibinfo {author} {\bibfnamefont {Y.}~\bibnamefont {Ran}},\ and\ \bibinfo
  {author} {\bibfnamefont {D.}~\bibnamefont {Xiao}},\ }\bibfield  {title}
  {\bibinfo {title} {Hyperdet wavefunction: A phase-agnostic ansatz for
  strongly correlated systems},\ }\href@noop {} {\bibfield  {journal} {\bibinfo
   {journal} {arXiv preprint arXiv:2609.04146}\ } (\bibinfo {year}
  {2026})}\BibitemShut {NoStop}%
\bibitem [{\citenamefont {Zhang}\ \emph {et~al.}(2026)\citenamefont {Zhang},
  \citenamefont {{\.I}mamo{\u{g}}lu}, \citenamefont {Shi}, \citenamefont
  {Demler},\ and\ \citenamefont {Morera}}]{zhang2026correlated}%
  \BibitemOpen
  \bibfield  {author} {\bibinfo {author} {\bibfnamefont {W.}~\bibnamefont
  {Zhang}}, \bibinfo {author} {\bibfnamefont {A.}~\bibnamefont
  {{\.I}mamo{\u{g}}lu}}, \bibinfo {author} {\bibfnamefont {T.}~\bibnamefont
  {Shi}}, \bibinfo {author} {\bibfnamefont {E.~A.}\ \bibnamefont {Demler}},\
  and\ \bibinfo {author} {\bibfnamefont {I.}~\bibnamefont {Morera}},\
  }\bibfield  {title} {\bibinfo {title} {Correlated metals, metamagnetism, and
  orbital nematic order in moir$\backslash$'e materials with neural quantum
  states},\ }\href@noop {} {\bibfield  {journal} {\bibinfo  {journal} {arXiv
  preprint arXiv:2610.03145}\ } (\bibinfo {year} {2026})}\BibitemShut {NoStop}%
\bibitem [{\citenamefont {Khalifa}\ and\ \citenamefont
  {Chatterjee}(2026)}]{khalifa2026continuum}%
  \BibitemOpen
  \bibfield  {author} {\bibinfo {author} {\bibfnamefont {A.}~\bibnamefont
  {Khalifa}}\ and\ \bibinfo {author} {\bibfnamefont {S.}~\bibnamefont
  {Chatterjee}},\ }\href {https://arxiv.org/abs/2610.08955} {\bibinfo {title}
  {From continuum neural quantum states to effective lattice models:
  Fluctuating wigner molecules in moir\'e superlattices}} (\bibinfo {year}
  {2026}),\ \Eprint {https://arxiv.org/abs/2610.08955} {arXiv:2610.08955
  [cond-mat.str-el]} \BibitemShut {NoStop}%
\bibitem [{\citenamefont {Nakahara}(2018)}]{nakahara2018geometry}%
  \BibitemOpen
  \bibfield  {author} {\bibinfo {author} {\bibfnamefont {M.}~\bibnamefont
  {Nakahara}},\ }\href@noop {} {\emph {\bibinfo {title} {Geometry, topology and
  physics}}}\ (\bibinfo  {publisher} {CRC press},\ \bibinfo {year}
  {2018})\BibitemShut {NoStop}%
\bibitem [{\citenamefont {Casazza}\ and\ \citenamefont
  {Chen}(2017)}]{CasazzaChen2017}%
  \BibitemOpen
  \bibfield  {author} {\bibinfo {author} {\bibfnamefont {P.~G.}\ \bibnamefont
  {Casazza}}\ and\ \bibinfo {author} {\bibfnamefont {X.}~\bibnamefont {Chen}},\
  }\bibfield  {title} {\bibinfo {title} {Frame scalings: A condition number
  approach},\ }\href {https://doi.org/10.1016/j.laa.2017.02.020} {\bibfield
  {journal} {\bibinfo  {journal} {Linear Algebra Appl.}\ }\textbf {\bibinfo
  {volume} {523}},\ \bibinfo {pages} {152} (\bibinfo {year} {2017})},\ \Eprint
  {https://arxiv.org/abs/1510.01653} {arXiv:1510.01653 [math.FA]} \BibitemShut
  {NoStop}%
\bibitem [{\citenamefont {Haldane}(1985)}]{Haldane1985}%
  \BibitemOpen
  \bibfield  {author} {\bibinfo {author} {\bibfnamefont {F.~D.~M.}\
  \bibnamefont {Haldane}},\ }\bibfield  {title} {\bibinfo {title}
  {Many-particle translational symmetries of two-dimensional electrons at
  rational {Landau}-level filling},\ }\href
  {https://doi.org/10.1103/PhysRevLett.55.2095} {\bibfield  {journal} {\bibinfo
   {journal} {Phys. Rev. Lett.}\ }\textbf {\bibinfo {volume} {55}},\ \bibinfo
  {pages} {2095} (\bibinfo {year} {1985})}\BibitemShut {NoStop}%
\bibitem [{\citenamefont {{NIST Digital Library of Mathematical
  Functions}}(2026)}]{DLMFTheta}%
  \BibitemOpen
  \bibfield  {author} {\bibinfo {author} {\bibnamefont {{NIST Digital Library
  of Mathematical Functions}}},\ }\href {https://dlmf.nist.gov/20.2} {\bibinfo
  {title} {Theta functions: Definitions and periodic properties, sec.~20.2}}
  (\bibinfo {year} {2026}),\ \bibinfo {note} {fourier series and zeros of the
  Jacobi theta functions; accessed October 26, 2026}\BibitemShut {NoStop}%
\bibitem [{\citenamefont {Belavin}\ and\ \citenamefont
  {Polyakov}(1975)}]{BelavinPolyakov1975}%
  \BibitemOpen
  \bibfield  {author} {\bibinfo {author} {\bibfnamefont {A.~A.}\ \bibnamefont
  {Belavin}}\ and\ \bibinfo {author} {\bibfnamefont {A.~M.}\ \bibnamefont
  {Polyakov}},\ }\bibfield  {title} {\bibinfo {title} {Metastable states of
  two-dimensional isotropic ferromagnets},\ }\href@noop {} {\bibfield
  {journal} {\bibinfo  {journal} {JETP Lett.}\ }\textbf {\bibinfo {volume}
  {22}},\ \bibinfo {pages} {245} (\bibinfo {year} {1975})}\BibitemShut
  {NoStop}%
\bibitem [{\citenamefont {Majorana}(1932)}]{Majorana1932}%
  \BibitemOpen
  \bibfield  {author} {\bibinfo {author} {\bibfnamefont {E.}~\bibnamefont
  {Majorana}},\ }\bibfield  {title} {\bibinfo {title} {Atomi orientati in campo
  magnetico variabile},\ }\href {https://doi.org/10.1007/BF02960953} {\bibfield
   {journal} {\bibinfo  {journal} {Nuovo Cimento}\ }\textbf {\bibinfo {volume}
  {9}},\ \bibinfo {pages} {43} (\bibinfo {year} {1932})}\BibitemShut {NoStop}%
\bibitem [{\citenamefont {Feynman}\ and\ \citenamefont
  {Cohen}(1956)}]{FeynmanCohen1956}%
  \BibitemOpen
  \bibfield  {author} {\bibinfo {author} {\bibfnamefont {R.~P.}\ \bibnamefont
  {Feynman}}\ and\ \bibinfo {author} {\bibfnamefont {M.}~\bibnamefont
  {Cohen}},\ }\bibfield  {title} {\bibinfo {title} {Energy spectrum of the
  excitations in liquid helium},\ }\href
  {https://doi.org/10.1103/PhysRev.102.1189} {\bibfield  {journal} {\bibinfo
  {journal} {Phys. Rev.}\ }\textbf {\bibinfo {volume} {102}},\ \bibinfo {pages}
  {1189} (\bibinfo {year} {1956})}\BibitemShut {NoStop}%
\bibitem [{\citenamefont {Pescia}\ \emph {et~al.}(2024)\citenamefont {Pescia},
  \citenamefont {Nys}, \citenamefont {Kim}, \citenamefont {Lovato},\ and\
  \citenamefont {Carleo}}]{PhysRevB.110.035108}%
  \BibitemOpen
  \bibfield  {author} {\bibinfo {author} {\bibfnamefont {G.}~\bibnamefont
  {Pescia}}, \bibinfo {author} {\bibfnamefont {J.}~\bibnamefont {Nys}},
  \bibinfo {author} {\bibfnamefont {J.}~\bibnamefont {Kim}}, \bibinfo {author}
  {\bibfnamefont {A.}~\bibnamefont {Lovato}},\ and\ \bibinfo {author}
  {\bibfnamefont {G.}~\bibnamefont {Carleo}},\ }\bibfield  {title} {\bibinfo
  {title} {Message-passing neural quantum states for the homogeneous electron
  gas},\ }\href {https://doi.org/10.1103/PhysRevB.110.035108} {\bibfield
  {journal} {\bibinfo  {journal} {Phys. Rev. B}\ }\textbf {\bibinfo {volume}
  {110}},\ \bibinfo {pages} {035108} (\bibinfo {year} {2024})}\BibitemShut
  {NoStop}%
\bibitem [{\citenamefont {Metropolis}\ \emph {et~al.}(1953)\citenamefont
  {Metropolis}, \citenamefont {Rosenbluth}, \citenamefont {Rosenbluth},
  \citenamefont {Teller},\ and\ \citenamefont {Teller}}]{Metropolis1953}%
  \BibitemOpen
  \bibfield  {author} {\bibinfo {author} {\bibfnamefont {N.}~\bibnamefont
  {Metropolis}}, \bibinfo {author} {\bibfnamefont {A.~W.}\ \bibnamefont
  {Rosenbluth}}, \bibinfo {author} {\bibfnamefont {M.~N.}\ \bibnamefont
  {Rosenbluth}}, \bibinfo {author} {\bibfnamefont {A.~H.}\ \bibnamefont
  {Teller}},\ and\ \bibinfo {author} {\bibfnamefont {E.}~\bibnamefont
  {Teller}},\ }\bibfield  {title} {\bibinfo {title} {Equation of state
  calculations by fast computing machines},\ }\href
  {https://doi.org/10.1063/1.1699114} {\bibfield  {journal} {\bibinfo
  {journal} {J. Chem. Phys.}\ }\textbf {\bibinfo {volume} {21}},\ \bibinfo
  {pages} {1087} (\bibinfo {year} {1953})}\BibitemShut {NoStop}%
\bibitem [{\citenamefont {Martens}\ and\ \citenamefont
  {Grosse}(2015)}]{MartensGrosse2015}%
  \BibitemOpen
  \bibfield  {author} {\bibinfo {author} {\bibfnamefont {J.}~\bibnamefont
  {Martens}}\ and\ \bibinfo {author} {\bibfnamefont {R.}~\bibnamefont
  {Grosse}},\ }\bibfield  {title} {\bibinfo {title} {Optimizing neural networks
  with {K}ronecker-factored approximate curvature},\ }in\ \href
  {https://proceedings.mlr.press/v37/martens15.html} {\emph {\bibinfo
  {booktitle} {Proceedings of the 32nd International Conference on Machine
  Learning}}},\ \bibinfo {series} {Proceedings of Machine Learning Research},
  Vol.~\bibinfo {volume} {37}\ (\bibinfo {year} {2015})\ pp.\ \bibinfo {pages}
  {2408--2417}\BibitemShut {NoStop}%
\bibitem [{\citenamefont {Flyvbjerg}\ and\ \citenamefont
  {Petersen}(1989)}]{FlyvbjergPetersen1989}%
  \BibitemOpen
  \bibfield  {author} {\bibinfo {author} {\bibfnamefont {H.}~\bibnamefont
  {Flyvbjerg}}\ and\ \bibinfo {author} {\bibfnamefont {H.~G.}\ \bibnamefont
  {Petersen}},\ }\bibfield  {title} {\bibinfo {title} {Error estimates on
  averages of correlated data},\ }\href {https://doi.org/10.1063/1.457480}
  {\bibfield  {journal} {\bibinfo  {journal} {J. Chem. Phys.}\ }\textbf
  {\bibinfo {volume} {91}},\ \bibinfo {pages} {461} (\bibinfo {year}
  {1989})}\BibitemShut {NoStop}%
\bibitem [{\citenamefont {Yang}(1962)}]{Yang1962}%
  \BibitemOpen
  \bibfield  {author} {\bibinfo {author} {\bibfnamefont {C.~N.}\ \bibnamefont
  {Yang}},\ }\bibfield  {title} {\bibinfo {title} {Concept of off-diagonal
  long-range order and the quantum phases of liquid {He} and of
  superconductors},\ }\href {https://doi.org/10.1103/RevModPhys.34.694}
  {\bibfield  {journal} {\bibinfo  {journal} {Rev. Mod. Phys.}\ }\textbf
  {\bibinfo {volume} {34}},\ \bibinfo {pages} {694} (\bibinfo {year}
  {1962})}\BibitemShut {NoStop}%
\bibitem [{\citenamefont {Elitzur}(1975)}]{Elitzur1975}%
  \BibitemOpen
  \bibfield  {author} {\bibinfo {author} {\bibfnamefont {S.}~\bibnamefont
  {Elitzur}},\ }\bibfield  {title} {\bibinfo {title} {Impossibility of
  spontaneously breaking local symmetries},\ }\href
  {https://doi.org/10.1103/PhysRevD.12.3978} {\bibfield  {journal} {\bibinfo
  {journal} {Phys. Rev. D}\ }\textbf {\bibinfo {volume} {12}},\ \bibinfo
  {pages} {3978} (\bibinfo {year} {1975})}\BibitemShut {NoStop}%
\bibitem [{\citenamefont {Zhang}(2026{\natexlab{b}})}]{ZhangIdealAnyonGas2026}%
  \BibitemOpen
  \bibfield  {author} {\bibinfo {author} {\bibfnamefont {Y.-H.}\ \bibnamefont
  {Zhang}},\ }\bibfield  {title} {\bibinfo {title} {Variational wavefunctions
  for ideal anyon gas: superfluid and skyrmion {Hall} crystal}\ }\href
  {https://doi.org/10.48550/arXiv.2610.40657} {10.48550/arXiv.2610.40657}
  (\bibinfo {year} {2026}{\natexlab{b}}),\ \Eprint
  {https://arxiv.org/abs/2610.40657} {arXiv:2610.40657 [cond-mat.str-el]}
  \BibitemShut {NoStop}%
\bibitem [{\citenamefont {Nayak}\ \emph {et~al.}(2008)\citenamefont {Nayak},
  \citenamefont {Simon}, \citenamefont {Stern}, \citenamefont {Freedman},\ and\
  \citenamefont {{Das Sarma}}}]{NayakEtAl2008}%
  \BibitemOpen
  \bibfield  {author} {\bibinfo {author} {\bibfnamefont {C.}~\bibnamefont
  {Nayak}}, \bibinfo {author} {\bibfnamefont {S.~H.}\ \bibnamefont {Simon}},
  \bibinfo {author} {\bibfnamefont {A.}~\bibnamefont {Stern}}, \bibinfo
  {author} {\bibfnamefont {M.}~\bibnamefont {Freedman}},\ and\ \bibinfo
  {author} {\bibfnamefont {S.}~\bibnamefont {{Das Sarma}}},\ }\bibfield
  {title} {\bibinfo {title} {Non-abelian anyons and topological quantum
  computation},\ }\href {https://doi.org/10.1103/RevModPhys.80.1083} {\bibfield
   {journal} {\bibinfo  {journal} {Rev. Mod. Phys.}\ }\textbf {\bibinfo
  {volume} {80}},\ \bibinfo {pages} {1083} (\bibinfo {year}
  {2008})}\BibitemShut {NoStop}%
\bibitem [{\citenamefont {Boerner}\ \emph {et~al.}(2023)\citenamefont
  {Boerner}, \citenamefont {Deems}, \citenamefont {Furlani}, \citenamefont
  {Knuth},\ and\ \citenamefont {Towns}}]{access}%
  \BibitemOpen
  \bibfield  {author} {\bibinfo {author} {\bibfnamefont {T.~J.}\ \bibnamefont
  {Boerner}}, \bibinfo {author} {\bibfnamefont {S.}~\bibnamefont {Deems}},
  \bibinfo {author} {\bibfnamefont {T.~R.}\ \bibnamefont {Furlani}}, \bibinfo
  {author} {\bibfnamefont {S.~L.}\ \bibnamefont {Knuth}},\ and\ \bibinfo
  {author} {\bibfnamefont {J.}~\bibnamefont {Towns}},\ }\bibfield  {title}
  {\bibinfo {title} {Access: Advancing innovation: Nsf's advanced
  cyberinfrastructure coordination ecosystem: Services \& support}\ }(\bibinfo
  {publisher} {Association for Computing Machinery},\ \bibinfo {address} {New
  York, NY, USA},\ \bibinfo {year} {2023})\ pp.\ \bibinfo {pages}
  {173--176}\BibitemShut {NoStop}%
\bibitem [{\citenamefont {{Perez Fadon}}\ \emph {et~al.}(2026)\citenamefont
  {{Perez Fadon}}, \citenamefont {Tam},\ and\ \citenamefont
  {Shackleton}}]{PerezFadonEtAl2026}%
  \BibitemOpen
  \bibfield  {author} {\bibinfo {author} {\bibfnamefont {A.}~\bibnamefont
  {{Perez Fadon}}}, \bibinfo {author} {\bibfnamefont {P.~M.}\ \bibnamefont
  {Tam}},\ and\ \bibinfo {author} {\bibfnamefont {L.}~\bibnamefont
  {Shackleton}},\ }\href {https://arxiv.org/abs/2610.23815v1} {\bibinfo {title}
  {Anyonic neural network wavefunctions}} (\bibinfo {year} {2026}),\ \Eprint
  {https://arxiv.org/abs/2610.23815} {arXiv:2610.23815 [cond-mat.str-el]}
  \BibitemShut {NoStop}%
\bibitem [{\citenamefont {Betcke}\ and\ \citenamefont
  {Trefethen}(2005)}]{BetckeTrefethen2005}%
  \BibitemOpen
  \bibfield  {author} {\bibinfo {author} {\bibfnamefont {T.}~\bibnamefont
  {Betcke}}\ and\ \bibinfo {author} {\bibfnamefont {L.~N.}\ \bibnamefont
  {Trefethen}},\ }\bibfield  {title} {\bibinfo {title} {Reviving the method of
  particular solutions},\ }\href {https://doi.org/10.1137/S0036144503437336}
  {\bibfield  {journal} {\bibinfo  {journal} {SIAM Rev.}\ }\textbf {\bibinfo
  {volume} {47}},\ \bibinfo {pages} {469} (\bibinfo {year} {2005})}\BibitemShut
  {NoStop}%
\end{thebibliography}
\end{document}